\documentclass[a4paper,11pt]{article}
\pdfoutput=1
\usepackage{jheppub} 
\usepackage{graphicx,color,rotating}
\usepackage{epsfig}
\usepackage{slashed}
\usepackage{amsfonts}
\usepackage{ulem}

\usepackage{tabularx}
\usepackage{adjustbox}
\usepackage{placeins}
\usepackage{threeparttablex}
\usepackage{multirow}
\usepackage{float}
\usepackage{subcaption}
\usepackage{hyperref}
\usepackage{siunitx}
\usepackage{comment}

\begin{document}
\title{Primordial Black Hole signatures from femtolensing and spectral fringe of Gamma-Ray Bursts}

\renewcommand{\thefootnote}{\arabic{footnote}}

\author{
Chang-Yu Dai$^{1}$, and Po-Yan Tseng$^{1,2}$}
\affiliation{
$^1$ Department of Physics, National Tsing Hua University,
101 Kuang-Fu Rd., Hsinchu 300044, Taiwan R.O.C. \\
$^2$ Physics Division, National Center for Theoretical Sciences, Taipei 106319, Taiwan R.O.C. \\
}

\date{\today}

\abstract{
Femtolensing of gamma-ray bursts (GRBs) is vastly studied to constrain primordial black holes lighter than $10^{-13}$ solar mass and may close the window for PBH dark matter. In this case, wave optics formalism is required and carefully implemented in our analysis. Incorporating the GRB observational data from Swift XRT, we perform the statistical analysis of PBH lensing, comparing it with the null hypothesis where the BAND model is used to parametrize the GRB spectrum. We found a few GRB data manifest the spectral fringe which characterizes the feature of femtolensing by PBHs, and the analysis shows moderate statistical preference in terms of goodness of fit. Conversely, since most of the fits to GRB spectral data do not improve with PBH lensing, we utilize this to obtain an upper bound on the PBH fractional abundance with respect to dark matter. However, the robust constraint cannot be achieved, unless the size of GRBs is smaller than $5\times 10^7$ m for PBH mass around $5\times 10^{-15}$ solar mass.
}

\maketitle

\section{Introduction}

Primordial Black Holes (PBHs) generated before star and galaxy formations serve as promising dark matter (DM) candidates~\cite{Zeldovich:1967lct,Hawking:1971ei,Chapline:1975ojl,Khlopov:2008qy,Carr:2016drx,Carr:2020gox,Carr:2020xqk,Green:2020jor,Bertone:2004pz}. Recently, LIGO/VIRGO gravitational wave (GW) signals implying BH binary mergers have increased investigations into PBHs~\cite{LIGOScientific:2016aoc}. Several PBH production mechanisms in the early Universe have been proposed: the collapse of an overdensity region stemming from primordial fluctuations after inflation~\cite{Carr:1974nx,Sasaki:2018dmp}; bubble wall collisions during cosmological first-order phase transition (FOPT) may induce sufficient energy density within the Schwarzschild radius~\cite{Hawking:1982ga,Kodama:1982sf,Moss:1994iq,Konoplich:1999qq}; and the attractive Yukawa interaction from dark FOPT catalyzes the intermediate state, a Fermi ball (FB), collapsing into a PBH~\cite{Baker:2021nyl,Gross:2021qgx,Kawana:2021tde,Marfatia:2021hcp}.

In the literature, efforts have been made to constrain the fractional abundance of PBHs, $f_{\rm PBH}$, depending on their mass and properties. When a PBH mass is lower than about $10^{-16}M_\odot$~\cite{Carr:2020gox}, it reveals its existence via Hawking radiation that evaporates particles and contributes to the cosmic gamma-ray or neutrino flux~\cite{Marfatia:2021hcp}. On the other hand, when a PBH is heavier than about $10^{-11}\, M_\odot$~\cite{Carr:2020gox}, the temporal enhancement of the luminosity of an electromagnetic source by the microlensing effect is expected when a PBH is transiting through the line of sight~\cite{Croon:2020ouk}. The mass window $10^{-16}\lesssim M_{\rm PBH}/M_\odot\lesssim 10^{-11}$ remains unconstrained, and thus the abundance of PBHs as a 100\% DM relic is still allowed. However, a method has been proposed in which the femtolensing of Gamma-ray Bursts (GRBs) with two-detector separation of two astronomical units has the potential to cover this PBH mass window~\cite{Jung:2019fcs}. Furthermore, an asteroid-mass PBH can also trigger Type Ia supernovae when it passes through a white dwarf and generates detonation ignition~\cite{Chen:2023oew}.

Femtolensing of distant GRBs was proposed in Refs.~\cite{Gould:1991td,Nemiroff:1995ak} to probe dark objects with a mass range between $\mathcal{O}(10^{-17})M_\odot$ and $\mathcal{O}(10^{-13})M_\odot$. This is based on the fact that, even though the two images of a GRB created by a lens are unresolved in space or time, their wavefronts will induce different phase shifts during propagation because of the different paths and gravitational potentials. We expect to observe interference fringes in the GRB frequency spectrum if the phase shift is $\mathcal{O}(1)$. Recent GRB observations provide more accurate measurements of redshift and frequency spectra, and they have been utilized for femtolensing analysis to constrain PBHs~\cite{Barnacka:2012bm,Katz:2018zrn}.

In this work, we scrutinize the signatures of interference fringes in the GRB energy spectra from Swift XRT data. The XRT data focuses on the energy range from 0.2 keV to 10 keV. The PBH-induced femtolensing is modeled from transiting PBHs, following the DM spacial distribution, with masses between $10^{-16}M_\odot$ and $10^{-13}M_\odot$. We adopt the formalism from Ref.~\cite{Katz:2018zrn}, in which the non-pointlike nature of GRBs is taken into account, and  predict the interference fringes in the GRB energy spectrum. For the null hypothesis, each GRB energy spectrum is modeled by the BAND parameterization after the minimization of the chi-square value. By comparing the statistical quantities of $\chi^2$ per degree of freedom and $P$-value between the PBH femtolensing and the BAND model, we highlight the 21 most likely GRB events in Table~\ref{tab:label1} that are imprinted with the signatures of interference fringes. On the other hand, 85 GRB events in Table~\ref{tab:label2},~\ref{tab:label3},~\ref{tab:label4},~\ref{tab:label5}, and~\ref{tab:label6} disfavor the interpretation of PBH femtolensing. They are used to constrain $f_{\rm PBH}$, the fractional abundance of PBHs, in the mass window $10^{-16} M_\odot \leq M_{\rm PBH} \leq 10^{-13} M_\odot$.

This paper is organized as follows. In Section~\ref{sec:femtolensing}, we review the formalism of femtolensing, adopting the wave optics approach and the finite source size. We emphasize that the oscillating characteristics of the 21 GRB spectra are statistically preferred under the interpretation of PBH femtolensing, while utilizing the other 85 GRBs to obtain an upper limit of $f_{\rm PBH}$ in Section~\ref{sec:result}. The results are discussed in Section~\ref{sec:conclusion}.

\bigskip

\section{Formalism of Femtolensing}
\label{sec:femtolensing}
\begin{figure}[t]
    \centering
        \includegraphics[width=0.8\linewidth]{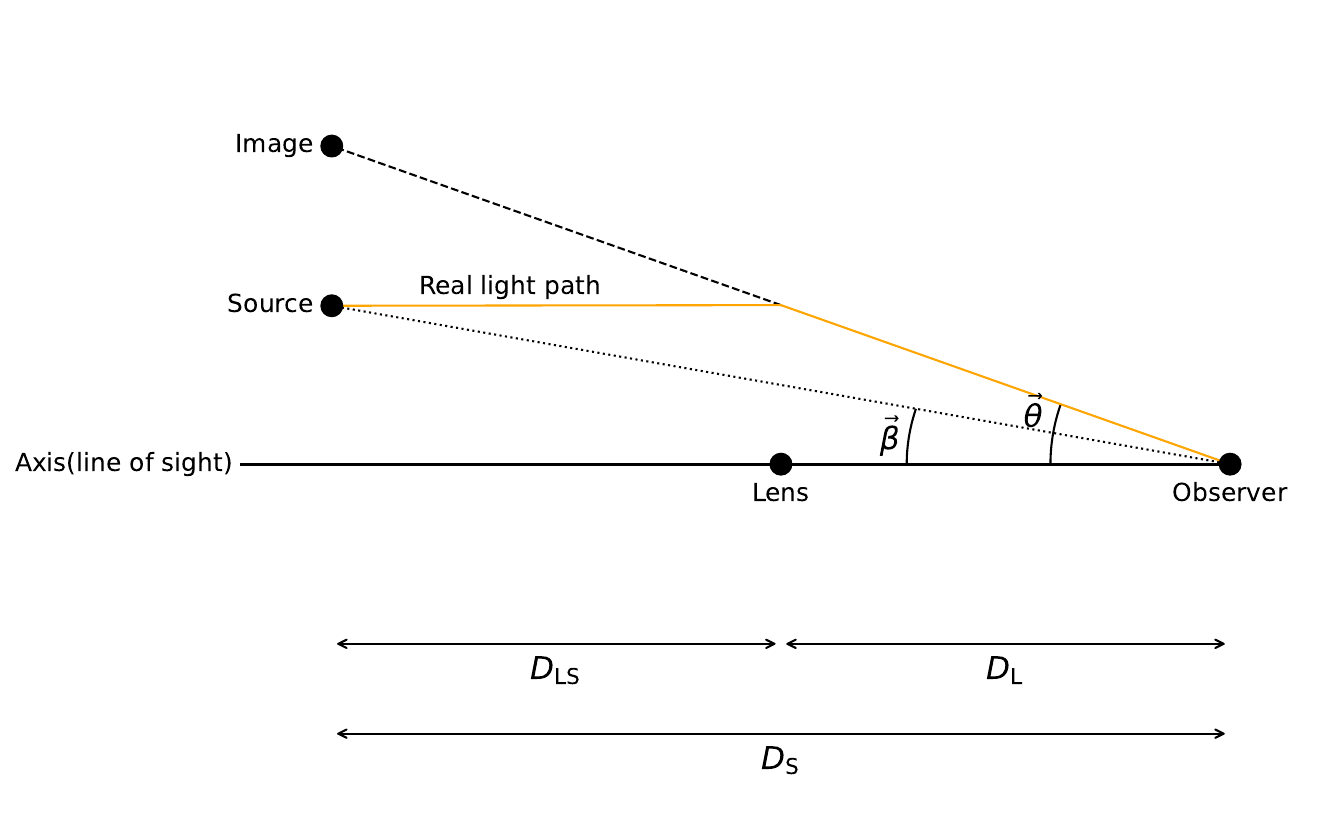}
    \caption{Geometric configuration of a gravitational lensing event. The diagram illustrates the deflection of light from a distant source by a lens. The relationships between the true source position $\vec{\beta}$, the observed image position $\vec{\theta}$, and the various cosmological distances ($D_{\rm{L}}$, $D_{\rm{S}}$, $D_{\rm{LS}}$) are shown.}
    \label{lensing}
\end{figure}
We review the femtolensing formalism by following Refs.~\cite{Katz:2018zrn,Tamta:2024pow,Hogg:1999ad}. We will start with the geometric optics approximation and then introduce the wave optics corrections. In the end, we discuss the physical condition under which the geometric approach deviates from the wave optics prediction.

We can set an axis from the observer to the lens; then, $\beta$ is the angle between the line from the observer to the source's real position and the axis. $\theta$ is the angle between the line from the observer to the source's lensed image and the axis. These are shown in Fig.~\ref{lensing}. From Fermat’s principle, $\nabla_{\theta}(\Delta t)=0$ in the geometric optics limit, where $\Delta t$ denotes the time delay which will be discussed later~\cite{Blandford:1986zz}. And one can use the lensing equation
\begin{equation}\label{eq:2.1}
    \vec{\theta}-\vec{\beta}=\nabla\psi(\vec{\theta}).
\end{equation}
to determine the positions of the images. The vector form means that those angles are projected to the lens plane, which is perpendicular to the observer-lens axis. $\psi(\vec{\theta})$ is the lensing potential given by
\begin{equation}
    \psi(\vec{\theta})=\frac{D_{LS}}{D_{L}D_{S}}\frac{2}{c^2}\int dz\,\Phi(D_{L}\vec{\theta},z)
\end{equation}
where $\Phi$ is the gravitational potential, and $D_{L}$, $D_{S}$, and $D_{LS}$ are angular diameter distances from the observer to the lens, from the observer to the source, and from the lens to the source, respectively~\cite{Hogg:1999ad}. One can relate $\psi$ to the lens mass distribution $\rho$ by the Poisson equation
\begin{equation}
    \nabla^{2}\psi(\vec{\theta})=\frac{8\pi G}{c^2}\frac{D_{LS}D_{L}}{D_{S}}\int^{\infty}_{-\infty} dz\,\rho(\sqrt{(D_{L}\theta)^{2}+z^{2}}).
\end{equation}

If a gamma-ray encounters a lens at redshift $z_{L}$,  it will be split into different rays. These rays will go through different paths, leading to a geometric time delay~\cite{Bartelmann:2010fz,Katz:2018zrn}
\begin{equation}
    \Delta t_{\rm{geom}}=\frac{1}{c}\frac{D_{L}D_{S}}{D_{LS}}(1+z_{L})\frac{|\vec{\theta}-\vec{\beta}|^2}{2}.
\end{equation}
In addition, the gravitational potential will cause another time delay compared to a non-lensed gamma ray. Because the distances between the source, lens, and observer are far more than the size of the lensing region, we can use the thin lens approximation to write the time delay as~\cite{Bartelmann:2010fz,Katz:2018zrn}
\begin{equation}
    \Delta t_{\rm{grav}}=\frac{1}{c}\frac{D_{L}D_{S}}{D_{LS}}(1+z_{L})\psi(\vec{\theta}).
\end{equation}
Therefore, they will have a total time delay
\begin{equation}
    \Delta t=\Delta t_{\rm{geom}}+\Delta t_{\rm{grav}}=\frac{1}{c}\frac{D_{L}D_{S}}{D_{LS}}(1+z_{L})\left(\frac{|\vec{\theta}-\vec{\beta}|^2}{2}-\psi(\vec{\theta})\right),
\end{equation}
which causes a phase shift $\Delta \phi=\omega\Delta t$, where $\omega$ is the angular frequency of the photon. If the arrival times of the split rays cannot be distinguished by the detector, then the detector will get an interference pattern.Consequently, the intensity of the signal received by the detector will be magnified. This phenomenon will be manifested in the spectrum.

We assume a point-like lens, so the lens potential is spherically symmetric
\begin{equation}
\psi(\vec{\theta})\rightarrow\psi(\theta)=\theta^{2}_{E}\,\rm{ln}(\theta),
\end{equation}
where $\theta_{E}$ is the Einstein angle
\begin{equation}\label{eq:3.9}
\theta_{E}\equiv\sqrt{\frac{4GM}{c^{2}}\frac{D_{LS}}{D_{L}D_{S}}},
\end{equation}
where $G$ is the gravitational constant and $M_{\rm{PBH}}$ is the PBH mass. We normalize $\vec{\theta}$ and $\vec{\beta}$ by the Einstein angle by defining $\vec{x}$ and $\vec{y}$ as $\vec{x} \equiv \vec{\theta}/\theta_{E}$ and $\vec{y} \equiv \vec{\beta}/\theta_{E}$, respectively. Therefore, the lensing equation Eq.~\ref{eq:2.1} becomes
\begin{equation}\label{eq.3.11}
    x-y=\frac{1}{x}.
\end{equation}
This equation has two solutions
\begin{equation}
    x_{\pm}=\frac{1}{2}(y\pm\sqrt{y^{2}+4}),
\end{equation}
which means we can observe two images. These two images lead to the intensity
\begin{equation}
    \mu_{\pm}=\frac{y^{2}+2}{2y\sqrt{y^{2}+4}}\pm\frac{1}{2}.
\end{equation}
If they can produce the interference pattern, then the total magnification will become
\begin{equation}\label{eq:3.14}
    \mu=\frac{y^{2}+2}{y\sqrt{y^{2}+4}}+\frac{2}{y\sqrt{y^{2}+4}}\text{sin}\left[\Omega\left(\frac{y\sqrt{y^{2}+4}}{2}+\text{ln}\left|\frac{y+\sqrt{y^{2}+4}}{y-\sqrt{y^{2}+4}}\right|\right)\right].
\end{equation}
Here we introduce the dimensionless frequency $\Omega$ defined as
\begin{equation}\label{eq:3.8}
    \Omega\equiv\frac{4GM_{\rm{PBH}}(1+z_{L})}{c^{3}}\omega.
\end{equation}
Because the energy range of GRB from Swift XRT is $10^{0}\sim10^{1}\text{keV}$, and the mass of the PBH of interest is $10^{-17}\sim10^{-12}\,M_{\odot}$, the wavelength of the GRB is comparable with the Schwarzschild radius of the PBH. Thus, we need to consider wave optics. The magnification for wave optics can be described by $\mu=|F|^{2}$. Here $F$ is the ratio of the wave amplitude $\phi_{L}/\phi_{0}$ ($\phi_{L}$ is the lensing amplitude and $\phi_{0}$ is the amplitude without lensing) and is given by~\cite{Tamta:2024pow,Nakamura:1999uwi}
\begin{equation}\label{eq:3.7}
    F(\vec{y},\Omega)=\frac{\Omega}{2\pi i}\int d^{2}\vec{x}\,e^{i\Omega\Delta t(\vec{x},\vec{y})}.
\end{equation}
Again, considering a point-like, spherically symmetric lens, Eq.~(\ref{eq:3.7}) becomes
\begin{equation}\label{eq:3.15}
    F(y,\Omega)=-i\Omega e^{i\Omega y^{2}/2}\int^{\infty}_{0}J_{0}(\Omega xy)xe^{i\Omega (\frac{1}{2}x^{2}-\psi(x))}\,dx,
\end{equation}
where $J_{0}(z)$ is the Bessel function of the first kind of zeroth order
\begin{equation}
    J_{0}(z)=\frac{1}{\pi}\int^{\pi}_{0}e^{iz\text{cos}\theta}\,d\theta.
\end{equation}
We can use the relation
\begin{equation}
    \int^{\infty}_{0}x^{m}e^{-\alpha x^{2}}J_{n}(\beta x)\,dx=\frac{\beta^{n}\Gamma(\frac{m+n+1}{2})}{2^{n+1}\alpha^{\frac{m+n+1}{2}}\Gamma(n+1)}\times\ _{1}F_{1}\left(\frac{m+n+1}{2},n+1,-\frac{\beta^{2}}{4\alpha}\right)
\end{equation}
and substitute it into Eq.~(\ref{eq:3.15}). After algebraic calculations, we get
\begin{equation}\label{eq:2.18}
    \mu(\Omega,y)=|F|^{2}=\frac{\pi \Omega}{1-e^{-\pi \Omega}}\left|_{1}F_{1}\left(\frac{i\Omega}{2},1,\frac{i\Omega y^{2}}{2}\right)\right|^{2},
\end{equation}
where $_{1}F_{1}$ is the confluent hypergeometric function of the first kind.

\begin{figure}[t]
    \centering
    \includegraphics[width=0.45\linewidth]{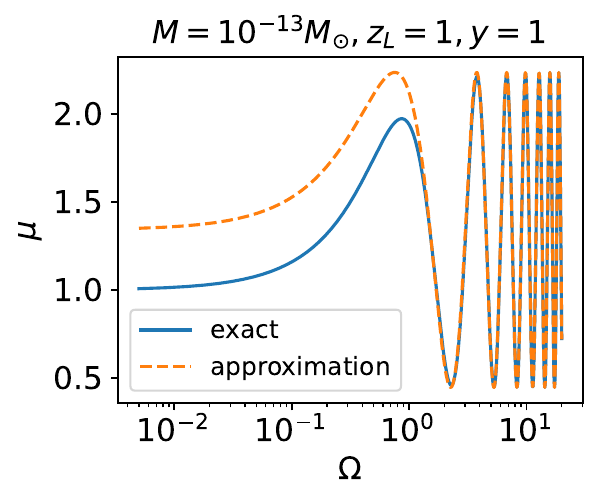}
    \includegraphics[width=0.45\linewidth]{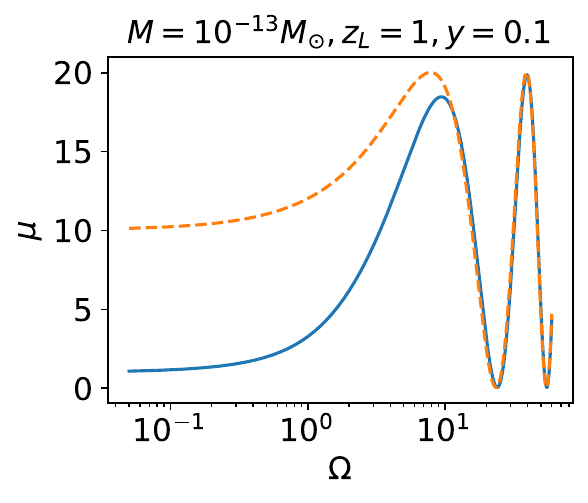}
    \caption{The orange curve represents the geometric optics approximation Eq.(\ref{eq:3.14}) and the blue curve represents the wave optics Eq.(\ref{eq:2.18}) for a point-like source and a point-like lens. $\Omega$ is calculated by Eq.(\ref{eq:3.8}). The geometric optics approximation is identical to wave optics when $\Omega\gg\frac{1}{y}$. $\mu$ will approach 1 when $\Omega\rightarrow0$.}
    \label{fig:enter-label}
\end{figure}

We can use the Schwarzschild radius $R_{s}=\frac{2GM}{c^{2}}$ and the light wavelength $\lambda$ to express $\Omega$ as
\begin{equation}\label{eq:3.10}
    \Omega=\frac{4\pi R_{s}(1+z_{L})}{\lambda}.
\end{equation}
Then we can use Eq.(\ref{eq:3.10}) to determine where the geometric optics approximation will break down and where we need to consider wave optics. Fig.~\ref{fig:enter-label} shows that the geometric optics approximation and wave optics are different when $\Omega\lesssim\frac{1}{y}$. When $\Omega\ll\frac{1}{y}$, the magnification is equal to 1. This is because $\Omega\ll\frac{1}{y}$ means the wavelength is much larger than the Schwarzschild radius of the lens. In this case, the wavelength cannot see the lens, which means no lensing. So the magnification is equal to 1.

Next, we take extended sources into account; thus, the source projected onto the lens plane will have an area. Therefore, we need to integrate over $y$ to calculate the magnification ~\cite{Katz:2018zrn,1993ApJ...413L...7S}
\begin{equation}\label{eq:3.19}
    \bar{\mu}=\frac{\int dy^{2}\,W(\vec{y},\sigma_{y})\mu(\vec{y},\Omega)}{\int\ dy^{2}\,W(\vec{y},\sigma_{y})}.
\end{equation}
$\sigma_{y}$ is a function of the source area $a_{s}=c\times T_{90}$
\begin{equation}\label{eq:3.20}
    \sigma_{y}\equiv\frac{a_{s}}{D_{s}\theta_{E}},
\end{equation}
where $T_{90}$ is the time interval over which a GRB emission ranges from $5\%$ to $95\%$ of its total measured counts. We require $\sigma_{y}\ll1$; otherwise, the interference fringes will disappear. The GRB emission region size is mostly located in $\sim 10^{9}$m, and these GRBs' redshifts are mostly $O(1)$, which means $D_{s}\sim \text{Gpc}$. If we require the lens' redshift to be of the same order, $z_{L}\sim O(1)$, then we will be forced to exclude most GRBs. Thus, we do not restrict the lens redshift to be of the same order as the source redshift; otherwise, the condition $\sigma_{y}\ll1$ cannot be satisfied. $W(\vec{y},\sigma_{y})$ is the emission intensity weight of the source. Following Ref.~\cite{Katz:2018zrn}, we set it to be Gaussian
\begin{equation}
    W(\vec{y},\sigma_{y})=e^{\frac{-|\vec{y}-\vec{y_{0}}|^{2}}{2\sigma_{y}^{2}}},
\end{equation}
where $\vec{y_{0}}$ is the center of the emission region. Substituting this into Eq.(\ref{eq:3.19}) and considering the spherically symmetric lens mentioned before, we get
\begin{equation}\label{eq:2.22}
    \bar{\mu}=\frac{e^{-y_{0}^{2}/2\sigma_{y}^{2}}}{\sigma_{y}^{2}}\int^{\infty}_{0}dy\,ye^{-y^{2}/2\sigma_{y}^{2}}I_{0}\left(\frac{y_{0}y}{\sigma_{y}^{2}}\right)\mu(y,\Omega),
\end{equation}
where $I_{0}$ is the modified Bessel function of the first kind of zeroth order
\begin{equation}
    I_{0}(z)=\sum^{\infty}_{k=0}\frac{(\frac{1}{4}z^{2})^{k}}{(k!)^{2}}.
\end{equation}

We choose the BAND model as the GRB model~\cite{BAND:1993eg}
\begin{equation}
f_{\text{BAND}}(E)=\begin{cases}A(\frac{E}{E_{0}})^{\alpha_{1}}e^{-\frac{E}{E_{0}}},\ E\leq(\alpha_{1}-\alpha_{2})E_{0} \\ A(\alpha_{1}-\alpha_{2})^{\alpha_{1}-\alpha_{2}}(\frac{E}{E_{0}})^{\alpha_{2}}e^{\alpha_{2}-\alpha_{1}},\ \text{otherwise}  \end{cases}
\end{equation}
where $\alpha_{1}$ and $\alpha_{2}$ are the power-law indices, $E_{0}$ is the break energy, and $A$ is the normalization constant with units of counts/$\text{cm}^{2}$/sec/keV.

\subsection{Numerical Calculations}

\begin{figure}[t]
    \centering
    \includegraphics[height=2in]{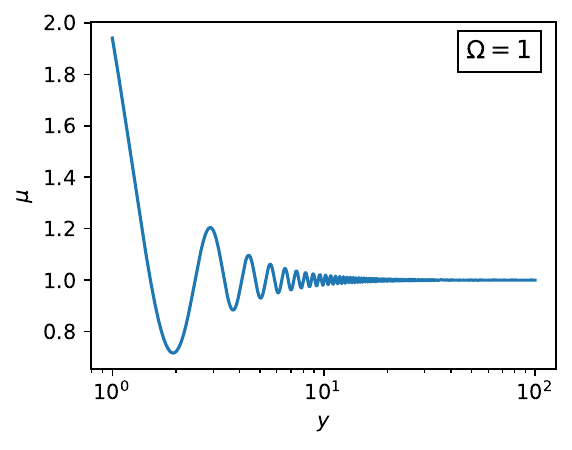}
    \includegraphics[height=2in]{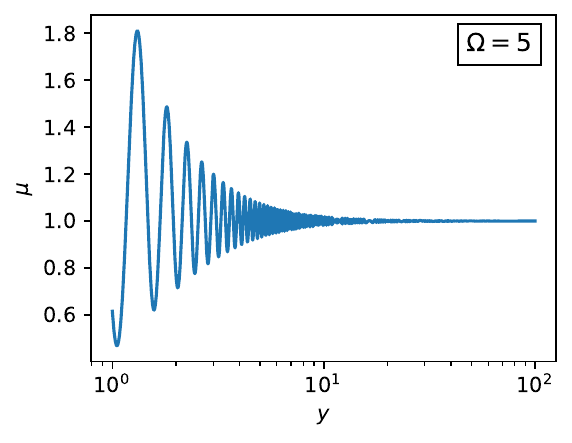}
    \caption{The magnification $\mu$ of Eq.(\ref{eq:2.18}) is shown as a function of $y$ with different $\Omega$. We can see that as $y$ increases, $\mu$ will approach 1 regardless of the value of $\Omega$.}
    \label{u plot to y}
\end{figure}

We have seven parameters: three related to the magnification function—$M_{\rm{PBH}}$, $z_{L}$, and $y_{0}$-and four related to the BAND’s function—$A$, $\alpha_{1}$, $\alpha_{2}$, and $E_{0}$. First, we will use Xspec to find the best-fitting BAND’s function. Then, our fitting parameters for the BAND’s function will be varied within a certain range around the values obtained from Xspec. We will focus on a mass range from $10^{-17}M_{\odot}$ to $10^{-12}M_{\odot}$. Since there is no strong evidence requires the lens redshift to be of the same order as that of the source, we impose no special constraints on $z_{L}$. Because the conversion from redshift to angular diameter distances is too sensitive at low $z_{L}$, we divide $z_{L}$ into two parts. In the first part, we use redshift from $z_{S}$ to $z_{L}=10^{-5}$. We change redshift from $z_{L}=10^{-5}$ to $z_{L}=10^{-10}$ into parsec units in the second part. For the last parameter $y_{0}$, we impose specific constraints to facilitate numerical calculations.

In the numerical calculation of Eq.(\ref{eq:2.22}), we can analyze the integrand to have a preliminary understanding of Eq.(\ref{eq:2.22}). First, notice that the integrand can be decomposed into four components. From Fig.~\ref{u plot to y}, $\mu(y,\Omega)$ will approach  1 at large $y$, which means we can ignore it when looking for the upper limit. For large $x$, we can use the approximation
\begin{equation}
    I_{0}(x)\sim\frac{e^{x}}{\sqrt{2\pi x}}.
\end{equation}
Consequently, we can rewrite the integrand in Eq.(\ref{eq:2.22}) as
\begin{equation}\label{eq:3.26}
    ye^{-y^{2}/2\sigma_{y}^{2}}I_{0}\left(\frac{y_{0}y}{\sigma_{y}^{2}}\right)\mu(y,\Omega)\sim\sqrt{\frac{y\sigma_{y}^{2}}{2\pi y_{0}}}e^{\frac{-(y-y_{0})^{2}+y_{0}^{2}}{2\sigma_{y}^{2}}}\mu(y,\Omega).
\end{equation}
Take Eq.(\ref{eq:3.26}) into Eq.(\ref{eq:2.22}). Then we get
\begin{equation}\label{eq:2.27}
    \bar{\mu}=\frac{1}{\sigma_{y}\sqrt{2\pi y_{0}}}\int^{\infty}_{0}dy\,\sqrt{y}e^{\frac{-(y-y_{0})^{2}}{2\sigma_{y}^{2}}}\mu(y,\Omega).
\end{equation}

\begin{figure}[t]
    \centering
    \includegraphics[width=0.3\linewidth]{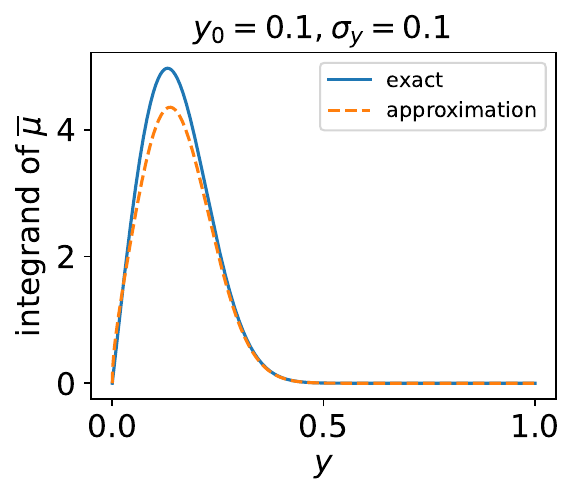}
    \includegraphics[width=0.3\linewidth]{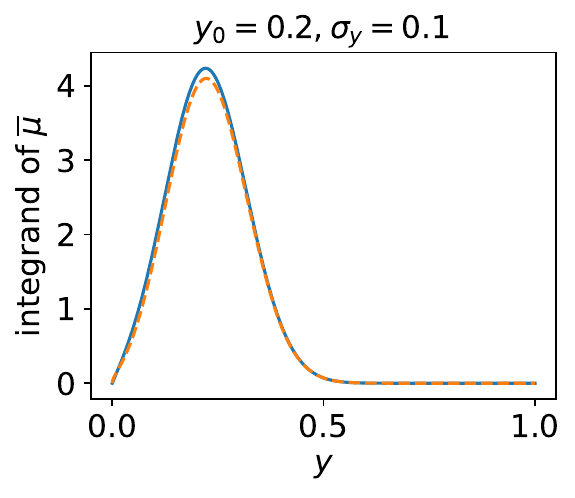}
    \includegraphics[width=0.3\linewidth]{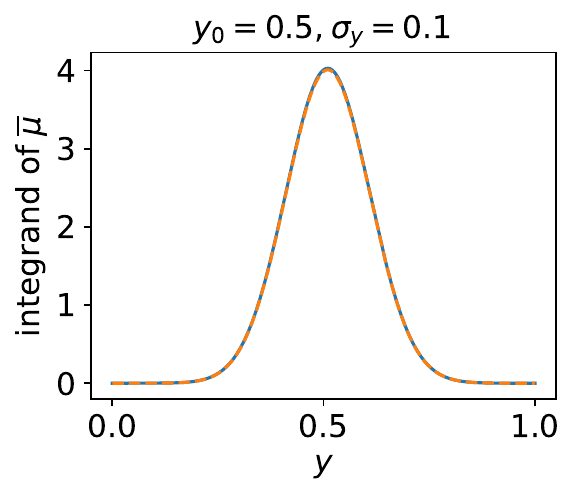}
    \includegraphics[width=0.3\linewidth]{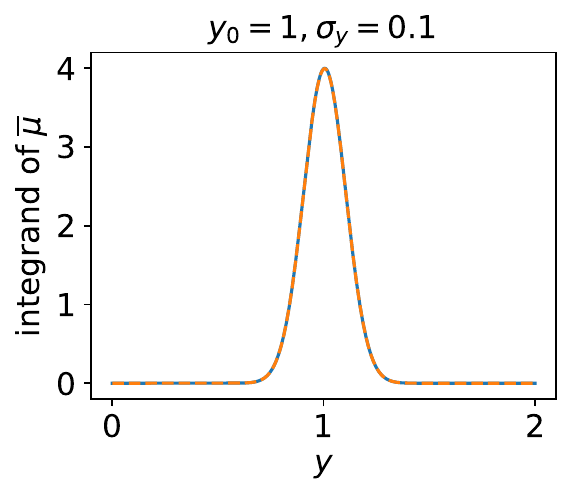}
    \includegraphics[width=0.3\linewidth]{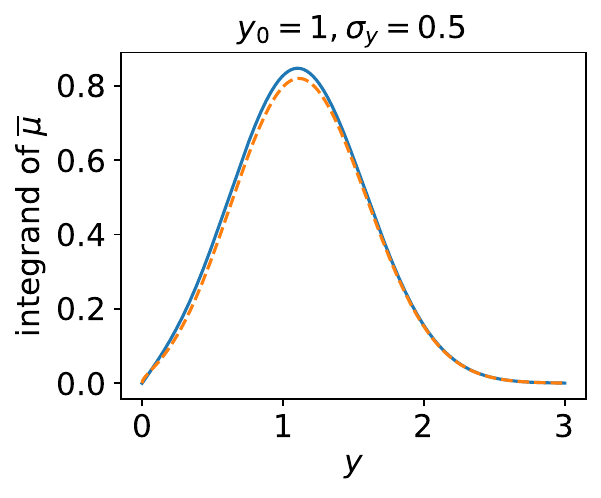}
    \includegraphics[width=0.3\linewidth]{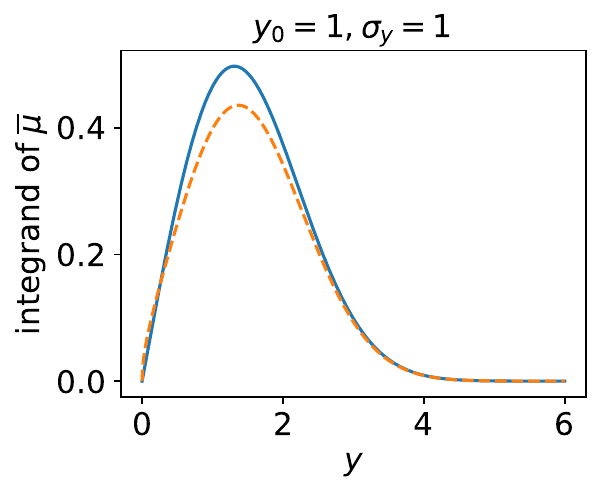}
    \caption{The comparison between the exact form Eq.(\ref{eq:2.22}) and the approximation form Eq.(\ref{eq:2.27}) with different $y_{0}$ and $\sigma_{y}$.  The curves are the integrands of Eq.(\ref{eq:2.22}) and Eq.(\ref{eq:2.27}) that include the factors appearing before the integration. We can see that when $yy_{0}\gg\sigma_{y}^{2}$, the approximation holds. Therefore, considering our requirement that $\sigma_{y}\ll1$, the upper limit of $y_{0}+3\sigma_{y}$ will not lose accuracy. The peak of each curve is approximately located at $y_{0}$ for different situations.}
    \label{fig:3}
\end{figure}

Thus, the integrand in Eq.(\ref{eq:2.27}) is suppressed by the Gaussian $e^{\frac{-(y-y_{0})^{2}}{2\sigma_{y}^{2}}}$ at large $y$; therefore, we can choose $y_{0}+3\sigma_{y}$ to be the upper limit of Eq.(\ref{eq:2.22}) without any significant loss of accuracy. Fig.~\ref{fig:3} shows that the approximation can well describe Eq.(\ref{eq:2.22}) at large $y$ regardless of which values of $y_{0}$ and $\sigma_{y}$ are chosen. If $y_{0}\gg\sigma_{y}$, the whole function can be described by the approximation. For $\Omega\gg\frac{1}{y}$, we can use the geometric optics approximation. From Eq.(\ref{eq:3.14}), when $y$ becomes too large, the sin will oscillate too fast. That will make numerical calculation difficult. So we will restrict $y_{0}$ to satisfy the following constraint
\begin{equation}\label{eq:3.28}
    \Omega\left(\frac{\widetilde{y}\sqrt{\widetilde{y}^{2}+4}}{2}+\text{ln}\left|\frac{\widetilde{y}+\sqrt{\widetilde{y}^{2}+4}}{\widetilde{y}-\sqrt{\widetilde{y}^{2}+4}}\right| \right)<2\pi\times10,
\end{equation}
\begin{equation}
    \widetilde{y}=y_{0}+3\sigma_{y}.
\end{equation}
Here we choose $\Omega$ at $E=10\,{\rm keV}$, which is the max energy of the GRB data. From a physical interpretation, the detector can only measure the average value of rapid oscillations, resulting in a smooth curve. Therefore, if there really are interference fringes in the spectrum, our requirements for $y_{0}$ should be reasonable. Finally, we require
\begin{equation}
    \lim_{\Omega\to0} \bar{\mu}=1.
\end{equation}
Since $\Omega$ is inversely proportional to $\lambda$, gamma-rays cannot be lensed at large wavelengths. Then the magnification will equal $1$ at small $\Omega$.

\subsection{Source Size}

In the above calculation, we used $a_{s}=c\times T_{90}$ to estimate the source radius. But according to Ref.~\cite{Barnacka:2014yja}, the maximum source size could be estimated by
\begin{equation}\label{eq:2.32}
    a_{s}^{max}\simeq ct_{var}'\simeq\frac{c\mathcal{D}t_{var}}{1+z_{s}}\simeq\frac{3\times10^{11}\text{cm}}{1+z_{s}}\left(\frac{t_{var}}{0.1\text{sec}}\right)\left(\frac{\mathcal{D}}{100}\right).
\end{equation}
$t_{var}$ is the observed minimum variability time scale, and the prime denotes it is in the rest frame. $\mathcal{D}$ is the Doppler factor
\begin{equation}
    \mathcal{D}=\frac{1}{\Gamma(1-\beta\cos\theta_{obs})},
\end{equation}
where $\theta_{obs}$ is the angle between the source's direction of motion and the line of sight to the source, $\Gamma=\frac{1}{\sqrt{1-\beta^{2}}}$, and $\beta=\frac{v}{c}$. Since the Doppler factor varies within a certain range, the source size of long GRBs actually has lower and upper limits. For short GRBs, the maximum source size is estimated by Eq.(\ref{eq:2.32}) as for long GRBs. But the minimum source size is estimated by Ref.~\cite{Barnacka:2014yja}
\begin{equation}\label{eq:2.34}
    a_{s}^{min}\simeq\frac{d^{2}_{L}}{\mathcal{D}^{4}}\Upsilon
\end{equation}
where
\begin{equation}\label{eq:2.35}
    \Upsilon=\int^{1}_{-1}\frac{1-\text{cos}\theta}{2}\,d\text{cos}\theta\int^{\infty}_{\epsilon_{th}}\frac{\Phi(\epsilon)}{m_{e}c^{3}\epsilon}\sigma_{\gamma\gamma}\,d\epsilon.
\end{equation}
Consider the high-energy photon emitted from the source. Then we can distinguish these emitted photons from the ambient photons. $\theta$ is the angle between the emitted photon and the ambient photon; $\epsilon$ is the energy of the ambient photon; the threshold energy $\epsilon_{th}=\frac{2}{\epsilon(1-\text{cos}\theta)}$ is the condition for positron-electron pair production~\cite{Gould:1967zzb}, where these energies are normalized by the electron rest mass $m_{e}$. $\Phi$ is the energy flux, and $\sigma_{\gamma\gamma}$ is the polarization-averaged cross section for pair production~\cite{Jauch:1976ava,Boettcher:2014csa}.

However, this range includes the estimation we have adopted for the source size. If further research indicates that the source size needs to be adjusted, because the source size only appears in $\sigma_{y}$ Eq.(\ref{eq:3.19}), we can change the Einstein angle to keep $\sigma_{y}$ unchanged. The Einstein angle depends on the lens mass and redshift by Eq.(\ref{eq:3.9}), but the lens mass also occurs in the dimensionless frequency Eq.(\ref{eq:3.8}). Therefore, we can simply adjust the lens redshift to get the same magnification.

\subsection{GRB data of Swift XRT}
\label{sec:PBH_511keV}

Swift was launched on November 20, 2004~\cite{SwiftScience:2004ykd}. It is equipped with three instruments: the Burst Alert Telescope (BAT)~\cite{Barthelmy:2005hs}, XRT~\cite{SWIFT:2005ngz}, and the Ultraviolet/Optical Telescope (UVOT)~\cite{Roming:2005hv}. It is used to observe gamma-ray bursts (GRBs). BAT can be considered a GRB trigger. When it receives a signal, it sends the position to the spacecraft, which then slews to allow XRT and UVOT to observe the GRB. XRT focuses on the $0.2\sim10\,\text{keV}$ energy range. It is designed to measure light curves and spectra, and it can determine locations with greater accuracy than BAT.

Swift XRT uses a CCD spectrometer to measure the spectrum. The spectrum is not the real spectrum, but rather photon counts $\mathcal{C}_{\text{fold},I}$ within a specific instrument channel $I$, \textit{i.e.}, folded observation raw data from Swift XRT. The relation between $\mathcal{C}_{\text{fold},I}$ and the real spectrum $\mathcal{M}(E)$ is
\begin{equation}\label{eq:2.36}
    \mathcal{C}_{\text{fold},I}=\int \mathcal{M}(E)R(I,E)A(E)\,dE
\end{equation}
where $R(I,E)$ is the instrumental response that gives the probability that an incident photon with energy $E$ is recorded in channel $I$. $A(E)$ accounts for the detection efficiency of a real detector. Although the function $R(I,E)$ is continuous with respect to $E$, it is stored in matrix form for use in XSPEC~\cite{Xspec}. The transition is 
\begin{equation}
    R_{I,J}=\frac{\int^{E_{J}}_{E_{J-1}}R(I,E)\,dE}{E_{J}-E_{J-1}}
\end{equation}
The spectra $\mathcal{M}(E)$ and $A(E)$ have a similar transition
\begin{equation}\label{eq:2.37}
    \mathcal{M}_{J}=\int^{E_{J}}_{E_{J-1}}\mathcal{M}(E)\,dE,~~~~~
    A_{J}=\int^{E_{J}}_{E_{J-1}}A(E)\,dE.
\end{equation}
Thus, the real process should be
\begin{equation}\label{eq:2.38}
    \mathcal{F}_{\text{fold},I}=\sum^{N}_{J=1}R_{I,J}A_{J}\mathcal{F}_{\text{unfold},J}.
\end{equation}
Here $\mathcal{F}$ is the general symbol {representing the raw data or the model function. When we choose a model function to fit the observed data, XSPEC performs the above process to transform the model function $X_{\text{unfold},J}$ into the folded model function $X_{\text{fold},I}$. XSPEC could also transform $X_{\text{fold}}$ from the channel domain into the energy domain $X_{\text{fold},I}\rightarrow X_{\text{unfold},J}$. Then we extract the unfolded data. XSPEC uses
\begin{equation}\label{eq:2.39}
    O_{\text{unfold},J}=\mathcal{C}_{\text{fold},J}\times\frac{X_{\text{unfold},J}}{X_{\text{fold},J}}
\end{equation}
to calculate the unfolded data. The model function $X_{\text{unfold},J}$ is chosen to represent the expected source spectrum. Taking $X_{\text{unfold},J}$ into the process mentioned in Eq.(\ref{eq:2.38}) above calculates $X_{\text{fold}}$.

We use the data of spectra~\cite{Evans:2008wp} from the UK Swift Science Data Centre (UKSSDC). There are two kinds of default modes for GRB spectra: Windowed Timing (WT) and Photon Counting (PC). Because WT mode has a higher time resolution ($1.7\,\text{ms}$) than PC mode ($2.5\,\text{s}$), we choose WT mode.

\bigskip

\section{Results}
\label{sec:result}
\begin{figure}[t]
    \centering
        \includegraphics[width=0.45\linewidth]{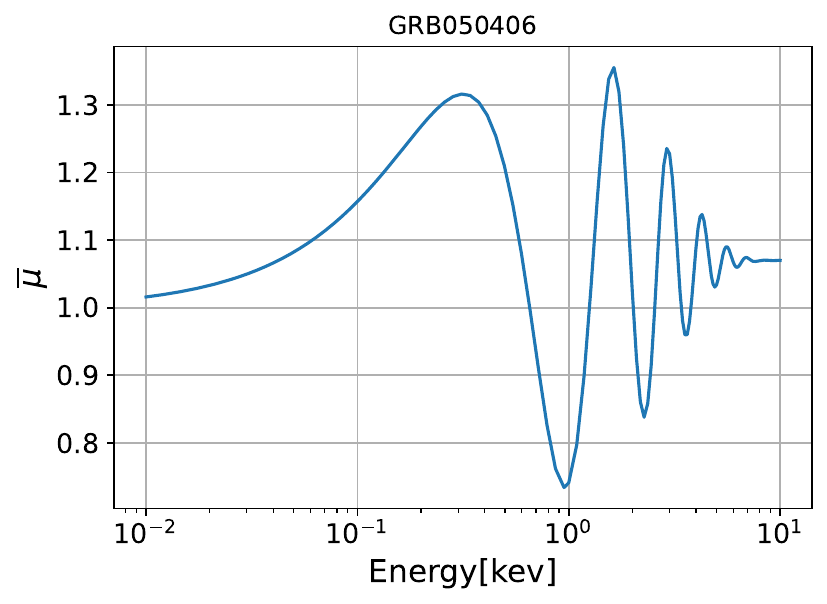}
        \includegraphics[width=0.45\linewidth]{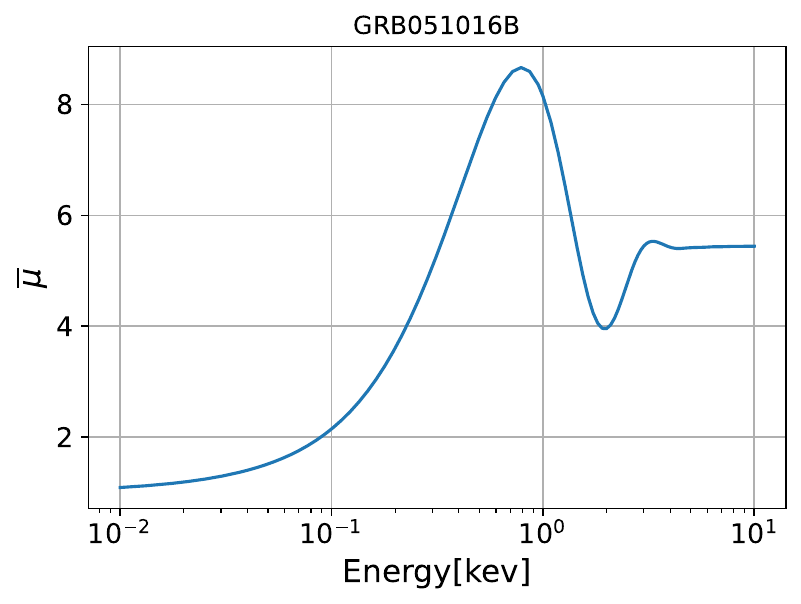}
    \caption{The first local minimum of $\bar{\mu}$ is used to determine the oscillation property of $\bar{\mu}$. The left panel is classified as oscillation because its local minimum is smaller than $1$. The local minimum of the right panel exceeds $1$; therefore, it is classified as no oscillation. The value of $y_{0}$ is $1.9$ for GRB050406 and $0.2$ for GRB051016.}
    \label{fig:6}
\end{figure}

Here we list the parameter ranges for the fitting. Here the Xspec in the superscript indicates that the parameters are obtained from the Xspec fitting program. The following are the parameters used to fit the GRB data:
\begin{eqnarray}
    0.1 & \leq & A/A^{\text{Xspec}} \leq 10 \,,~~
    0.1\, {\rm keV} \leq E_{0} \leq 2E^{\text{Xspec}}_{0}\, \nonumber\\
    -1  & \leq  &  \alpha_{1}\leq 1\,,~~
    2\alpha^{\text{Xspec}}_{2}\leq \alpha_{2}\leq -0.5 \nonumber \\
    10^{-10} & \leq  & z_{L} \leq z_{S}  \,, ~~
    10^{-17} \leq M_{\text{PBH}}/M_{\odot} \leq 10^{-12}\,, ~~ 
    0.1\leq y_{0} \leq 5 \nonumber
\end{eqnarray}

The detailed results are presented in Tables~\ref{tab:label1}-\ref{tab:label6} and Figs.~\ref{fig:PBH_femtolensing1}-\ref{fig:PBH_femtolensing6} in Appendix~\ref{app_fitting}, where a total of 106 GRBs are classified into six categories: 
i) $21$ GRBs in Table~\ref{tab:label1} with spectra in Fig.~\ref{fig:PBH_femtolensing1}, 
ii) $11$ GRBs in Table~\ref{tab:label2} with Fig.~\ref{fig:PBH_femtolensing2}, 
iii) $14$ GRBs in Table~\ref{tab:label3} with Fig.~\ref{fig:PBH_femtolensing3}, 
iv) $41$ GRBs in Table~\ref{tab:label4} with Fig.~\ref{fig:PBH_femtolensing4}, 
v) $13$ GRBs in Table~\ref{tab:label5} with Fig.~\ref{fig:PBH_femtolensing5}, 
and vi) $5$ GRBs in Table~\ref{tab:label6} with Fig.~\ref{fig:PBH_femtolensing6}. 

We classify the fitted GRB data according to the following characteristics. First, we determine if the GRB data has an oscillation pattern or not. Because the oscillation is caused by the lensing effect, it should be possible to determine by $\bar{\mu}$. We use the first local minimum of $\bar{\mu}$ to determine whether there is an oscillation in the GRB lensing signal. Since $\bar{\mu} \to 1$ at low energies, if the local minimum is less than $1$, we classify it as an oscillation pattern shown in the left panel of Fig.~\ref{fig:6}. The oscillation pattern is associated with $y_{0}$. Those GRBs with an oscillation pattern have a larger $y_{0}$. The $y_{0}$ parameter characterizes the geometric configuration of the source, the lens, and the observer. For example, if $y_{0}=0$, these three objects are collinear. Thus, an oscillation pattern corresponding to a larger $y_{0}$ implies that the lens must lie slightly off the line of sight between the observer and the source. Conversely, a smaller $y_{0}$ leads to a larger $\bar{\mu}$ at high energies. Because a smaller $y_{0}$ indicates that the source, the lens, and the observer are nearly aligned, the lens can deflect a larger fraction of photons toward the observer in a coherent phase. Once we classify the GRB data with (without) an oscillation pattern in Tables~\ref{tab:label1}-\ref{tab:label3} (Tables~\ref{tab:label4}-\ref{tab:label6}), we further divide the GRB data into three categories based on the goodness of fit, which will be discussed in a later section. We classify GRBs using the difference between two goodness-of-fit values: one obtained using the BAND model only, and the other obtained by considering the PBH lensing effect. We consider three scenarios based on the difference in goodness of fit between the two models:
\begin{itemize}
\item [1.] The goodness of fit of the BAND model exceeds that of the PBH lensing by more than $0.05$, \textit{i.e.}, as shown in Table~\ref{tab:label1} and Table~\ref{tab:label4}.
\item [2.] The goodness of fit of the BAND model exceeds that of the PBH lensing by less than $0.05$, \textit{i.e.},  as shown in Table~\ref{tab:label2} and Table~\ref{tab:label5}.
\item [3.] The goodness of fit of the BAND model is smaller than the goodness of fit of the PBH lensing, \textit{i.e.}, as shown in Table~\ref{tab:label3} and Table~\ref{tab:label6}.
\end{itemize}
\begin{figure}[t]
    \centering
        \includegraphics[width=0.45\linewidth]{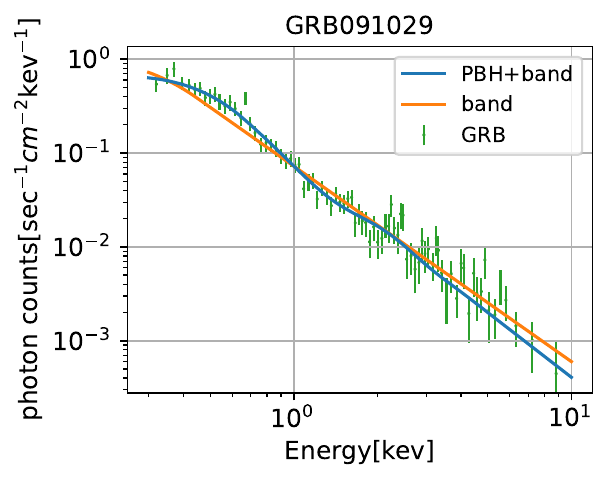}
        \includegraphics[width=0.45\linewidth]{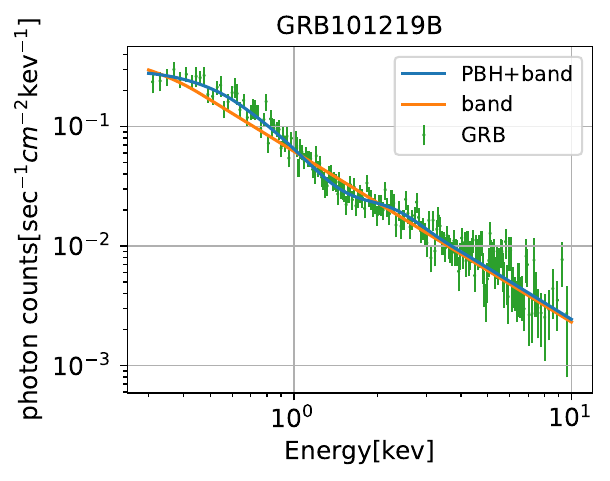}
    \caption{Two of the GRBs in Table~\ref{tab:label1} with an oscillation pattern appearing in the spectrum. Green error bars are the observed GRB data. Blue and orange curves are the fitted BAND model with and without the lensing effect, respectively. GRB091029 (GRB101219B) has $T_{90}=39.2\,s$ ($T_{90}=34\,s$) and is located $1.6\times10^{6}\,\rm{kpc}$ ($1.3\times10^{6}\,\rm{kpc}$) away from us. According to our best fit, these are lensed by a PBH with a mass of $2.33\times10^{-14}M_{\odot}$ ($2.95\times10^{-14}M_{\odot}$) at $11\,\rm{kpc}$ ($11\,\rm{kpc}$) from us and $y_{0}=2.1$ ($y_{0}=1.6$).}
    \label{example GRB}
\end{figure}
Fig.~\ref{example GRB} shows two examples of GRBs in Table~\ref{tab:label1}. They manifest an oscillation pattern. GRB091029 (GRB101219B) has $T_{90}=39.2\,s$ ($T_{90}=34\,s$) and is located $1.6\times10^{6}\,\rm{kpc}$ ($1.3\times10^{6}\,\rm{kpc}$) away from us. The signals are lensed by a PBH with mass $2.33\times10^{-14}M_{\odot}$ ($2.95\times10^{-14}M_{\odot}$) located at $11\,\rm{kpc}$ ($11\,\rm{kpc}$) from Earth and $y_{0}=2.1$ ($y_{0}=1.6$).

The relevant mass range is from $10^{-16\,}M_{\odot}$ to $10^{-13}\,M_{\odot}$ in our fitting results. This mass range matches the window of PBHs DM which remains unconstrained. The order of the redshift of lens is between $10^{-7}$ and $10^{-5}$, which indicates that PBHs are within the Milky Way.

We use~\cite{Katz:2018zrn}
\begin{equation}\label{eq:4.1}
    \tau=\int^{z_{s}}_{0}\frac{dz_{L}}{H(z_{L})}\sigma(D_{L})\frac{\rho_{PBH}}{M_{\rm{PBH}}}(1+z_{L})^{2}
\end{equation}
to calculate the optical depth $\tau$. Here $H(z_{L})$ is the Hubble parameter, Hubble constant and $f_{\text{PBH}}\equiv \rho_{\rm PBH}/\rho_{\rm DM}$ is the ratio of PBH abundance to $\rho_{\text{DM}}$, density of DM. We separate $\rho_{\text{DM}}$ into two parts, depending on whether it is inside the Milky Way ($z\approx 10^{-5}$) or not~\cite{Croon:2020ouk}.
\begin{eqnarray}
    \rho_{\text{DM}}&=&\begin{cases}\frac{\rho_{\text{s}}}{(r/r_{s})(1+r/r_{s})^{2}},~~ \text{inside the Milky Way}\\
    1.3\times10^{-6}\ \text{GeV}/\text{cm}^{3}, ~~\text{extragalactic}
    \end{cases}\\
    r&\equiv& \sqrt{R^{2}_{\text{Sol}}-2R_{\text{Sol}}D_{\text{L}}\text{cos}(l)\text{cos}(b)+D^{2}_{\text{L}}}
\end{eqnarray}
$\rho_{\text{s}}=0.184 \ \text{GeV}/\text{cm}^{3}$, $r_{\text{s}}=21.5\ \text{kpc}$, $R_{\text{Sol}}=8.5\ \text{kpc}$ is the distance from the Sun to the center of the Milky Way, and $l$, $b$ are the galactic coordinates. The lensing cross section is given by $\sigma(D_{L})=\pi(y_{\text{max}}\theta_{E}D_{L})^{2}$, where $y_{\text{max}}$ denotes the value obtained by solving
\begin{equation}
    -2\text{ln}\left(\frac{L_{0}(\vec{\mu_{s}}|y=y_{\text{max}})}{L_{0}(\vec{\mu_{s}}|y=\infty)}\right)=\alpha,
\end{equation}
$y_{\text{max}}$ is the maximal distance at which the lens still produces a distinguishable lensing effect from the BAND model to a given confidence level (CL). We use $\alpha=2.7$ for $90\%$ CL. Scanning over the parameters of the BAND model $\vec{\mu}_b$, $L_{0}$ is the maximum value of likelihood function given by~\cite{Katz:2018zrn}
\begin{equation}
\label{eq:L0}
    -2\text{ln}L_{0}(\vec{\mu_{s}})\equiv \min_{\vec{\mu_{b}}}\left[\sum_{J=1}^{\# \text{of bins}}\left(\frac{O_{\text{unfold},J}-X_{\text{unfold},J}(\vec{\mu}_b,\vec{\mu}_s)}{\sigma_{J}}\right)^2\right]+\text{const},
\end{equation}
where $O_{\text{unfold},J}$ from Eq.(\ref{eq:2.39}) and $\sigma_{J}$ are the observed values of the GRB and the uncertainty extracted from Xspec, respectively. $X_{\text{unfold},J}$ is the theoretical values of the model with/without PBH lensing, \textit{i.e.}, $X_{\text{unfold}}= f_{\text{BAND}}$ or $\bar{\mu}\times f_{\text{BAND}}$. The $\vec{\mu}_b$ and $\vec{\mu}_s$} denote the parameters of the BAND model ($E_{0},\alpha_{1},\alpha_{2},A$) and the lens ($z_{L},M_{\rm{PBH}},y_{0}$), respectively. The subscript $J$ indicates the $J\text{-}\text{th}$ energy bin as Eq.(\ref{eq:2.37}). The term in square brackets in Eq.(\ref{eq:L0}) is our definition of the chi-square $\chi^{2}$, where the minimum values of $\chi^2$ are denoted as $\chi^{2}_{\text{BAND}}$ for the BAND model and $\chi^{2}_{\text{PBH}}$ for including PBH lensing, obtained by scanning the parameters $\vec{\mu}_b$ and $\vec{\mu}_b+\vec{\mu}_s$, respectively. We use $\chi^{2}_{\rm BAND/PBH}/\text{d.o.f}$ to calculate the goodness of fit, where d.o.f (degrees of freedom) is defined as the number of data points in each GRB spectrum minus the number of fitted parameters (four for $\chi^{2}_{\text{BAND}}$ and seven for $\chi^{2}_{\text{PBH}}$).
\begin{figure}[t]
    \centering
        \includegraphics[width=0.8\linewidth]{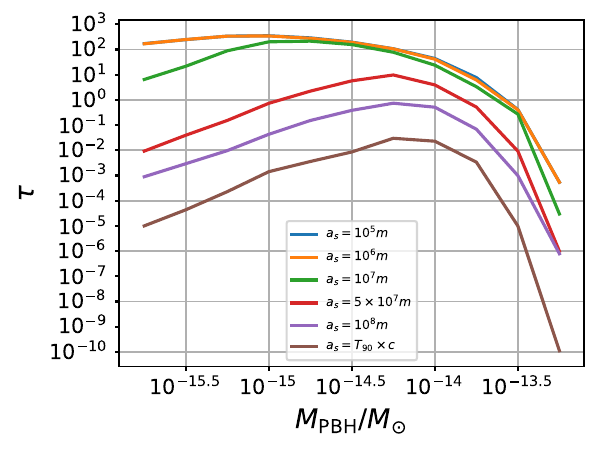}
    \caption{Optical depth $\tau$ from Eq.(\ref{eq:4.1}) varies with different source  sizes $a_{\text{s}}$ and $M_{\rm PBH}$. Each curve approaches a similar asymptotic slope on both sides, i.e., $M_{\rm PBH}\lesssim 3\times 10^{-16} M_\odot$ and $M_{\rm PBH}\gtrsim 3\times 10^{-14} M_\odot$. When $a_{\text{s}}$ decreases to $10^5$ m, $\tau$ will saturate to an upper limit. Notice, the curves for $a_s=10^5$ m and $a_s=10^6$ m almost overlap.}
    \label{fig:4}
\end{figure}
Following Ref.~\cite{Jung:2019fcs}, we assume the Poisson distribution for lensing events, and the no-lensing probability for one GRB is given by $P_{1}=e^{-\tau}$. The total probability of no-lensing events for $N$ GRBs is given by $P_{\text{null}} (f_{\text{PBH}})=\Pi^{N}_{i=1}P_{1,i}=\Pi^{N}_{i=1}e^{-\textstyle\tau_{i}}$, where $i$ is the GRB index. We set $P_{\text{null}}=0.05$ to find the upper limits of $f_{\text{PBH}}$. The GRBs in Table~\ref{tab:label1} exhibit an oscillation pattern in $\bar\mu$ with a decrease in the goodness of fit greater than $0.05$. We therefore consider these GRBs in Table~\ref{tab:label1} to have higher confidence for lensing.

\begin{figure}[t]
    \centering
        \includegraphics[width=0.8\linewidth]{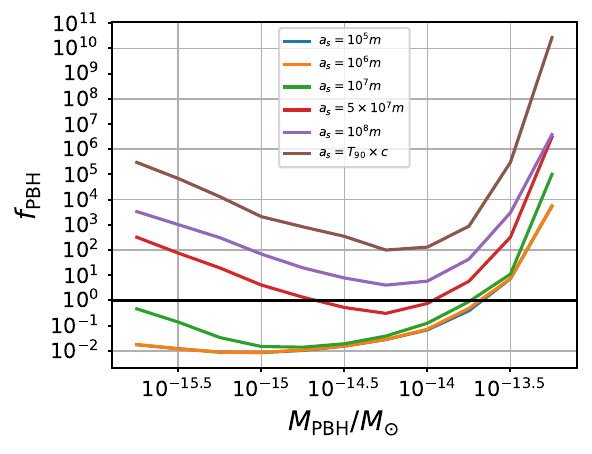}
    \caption{The sensitivity of PBH fraction in DM varying with source size $a_\text{S}$ and $M_{\rm PBH}$. The peak of the sensitivity is also dependent on the $a_\text{S}$. We adopted GRB data without an oscillating pattern, included in Table~\ref{tab:label2},~\ref{tab:label3},~\ref{tab:label4},~\ref{tab:label5}, and~\ref{tab:label6}, to calculate the sensitivities at 95\% C.L. The solid-black horizontal line represents the fraction of PBH equal to $1$. Notice, the curves for $a_s=10^5$ m and $a_s=10^6$ m almost overlap.}
    \label{fig:5}
\end{figure}

Alternatively, we use the GRBs from Table~\ref{tab:label2} to Table~\ref{tab:label6} to calculate the optical depth $\tau$ and find the upper limit of $f_{\text{PBH}}$. The results are presented in Fig.~\ref{fig:4} and Fig.~\ref{fig:5}, assuming various source sizes for GRBs. For a fixed value of $P_{\rm{null}}$, $f_{\rm PBH}$ decreases as $\tau$ increases. There are three terms in $\tau$ that depend on $M_{\text{PBH}}$. The first term is the integrand in Eq.(\ref{eq:4.1}). The remaining two terms are implicit in the constraint on $\sigma_{y}\ll1$ and Eq.(\ref{eq:3.28}), respectively. Next, we examine the effect of each of these three terms on $\tau$ individually. In the following, we highlight three points that dictate these curves:
\begin{itemize}
\item [1.] $M_{\text{PBH}}$ appears in the denominator of the integrand in Eq.(\ref{eq:4.1}), so $\tau$ is inversely proportional to $M_{\text{PBH}}$.
\item [2.] Since we require $\sigma_{y}\ll1$, $\theta_{\text{E}}$ has a lower limit ($\sigma_{y}$ is defined by Eq.(\ref{eq:3.20}), while $a_{\text{s}}$ and $D_{\text{s}}$ are fixed). $\theta_{\text{E}}$ is defined by Eq.(\ref{eq:3.9}). As $D_{LS}/D_{L}$ decreases with increasing $z_{L}$, the lens redshift has an upper limit for each fixed value of $M_{\text{PBH}}$. The upper limit of $z_{L}$ indicates that the upper bound in Eq.(\ref{eq:4.1}) might not actually extend to $z_{s}$. Because this upper limit on $z_{L}$ will increase as $M_{\text{PBH}}$ increases, $\tau$ correspondingly increases with increasing $M_{\text{PBH}}$. As $M_{\text{PBH}}$ continues to increase, the upper limit on $z_{L}$ eventually reaches $z_{s}$. Then the constraint will become ineffective for larger $M_{\text{PBH}}$.
\item [3.] The effect of the constraint of Eq.(\ref{eq:3.28}) is similar but opposite to the constraint of $\sigma_{y}\ll1$. For a fixed value of $M_{\text{PBH}}$, $y_{0}$ has an upper limit. However, the upper limit will increase as $M_{\text{PBH}}$ decreases. We confine the range of $y_{0}$ from $0$ to $10$, which means that this constraint cannot be satisfied when $M_{\text{PBH}}$ is less than a certain value.
\end{itemize}

Fig.~\ref{fig:4} is the plot of ($\tau$, $M_{\rm{PBH}}$) for different source sizes $a_{s}$. Those curves approach a similar slope for the lighter regime of $M_{\rm PBH}$; this is due to point-2 mentioned above. As $M_{\rm{PBH}}$ continues to increase, this effect disappears at a certain $M_{\rm{PBH}}$. The effect of point-3 requires these curves to have a similar slope in the heavier regime of $M_{\rm PBH}$. However, the almost identical slope of the curves corresponding to $a_{s}\leq10^7\,\text{m}$ in the intermediate mass region, $10^{-15}<{M_{\rm{PBH}}}/{M_{\odot}}<10^{-14}$, differs from others of larger $a_{s}$ values. This is because the effects of points 2 and 3 become irrelevant and are dictated only by the effect of point 1, {\it i.e.}, inversely proportional to $M_{\rm PBH}$.

As the source size $a_{S}$ decreases, $\tau$ saturates to an upper limit. This is because when $a_{S}$ decreases, the lower limit of $\theta_{\text{E}}$ also decreases, which means that the entire integration interval of Eq.(\ref{eq:4.1}) can satisfy the condition of the point 2 mentioned above. At this point, further decreasing $a_{S}$ will no longer increase $\tau$.

\newpage

\section{Conclusion}
\label{sec:conclusion}

We used 106 GRB data observed by Swift XRT to examine PBH gravitational femtolensing, which occurs when light propagates through the vicinity of a PBH; the gravitational field of a PBH causes light to deflect and induces phase shifts along different paths. The interference among different paths causes an oscillating pattern, which is imprinted on the energy spectrum of detected photon data.

According to our analysis, some of the GRBs display the feature of lensing oscillation caused by PBHs, rather than the simple exponential power-law spectrum predicted from the BAND model. Those potential PBHs are inferred to be localized within the Milky Way. This is because most GRBs have relatively large $a_{s}$ by assuming $a_{s}=c\times T_{90}$. Under the constraint of $\sigma_{y}\ll1$, they can only be lensed by PBHs at small lens redshifts $z_{L}$.

Whether there is an oscillation pattern cannot be determined directly from the GRB data. Therefore, we define the oscillation feature using $\bar{\mu}$ as a function of energy, in which the first local minimum is less than one. Some of the GRB data display the oscillation but are not eligible for our oscillation requirement with $\bar{\mu}$. The oscillation pattern occurs in $\bar{\mu}$ when the source, the lens, and the observer are not collinear. So $y_{0}$ is critical for oscillation; in particular, the oscillation cannot occur when $y_{0}$ is smaller than one. We highlight two GRBs, GRB091029 and GRB101219B, which satisfy the aforementioned oscillation requirement and prefer the PBH lensing interpretation.

Finally, we use the $85$ GRBs from Table~\ref{tab:label2} $\sim$ Table~\ref{tab:label6} to calculate the constraint on $f_{\rm{PBH}}$. The constraint of $f_{\text{PBH}}\leq 1$ can only be achieved if we set the $a_{\text{s}}$ smaller than $10^{7}$ m. For $a_{\rm{s}}=c\times T_{90}$, more GRB observation data are needed to reach a constraint with $f_{\text{PBH}}\leq 1$. We assume the monochromatic mass of PBHs when calculating the constraint on $f_{\text{PBH}}$. The mass interval with $f_{\text{PBH}}\leq 1$ extends from $5\times10^{-14}M_{\odot}$ to $10^{-16}M_{\odot}$. The quantities $f_{\rm{PBH}}$ and $\tau$ are inversely proportional under the requirement $P_{\text{null}}=0.05$. $\tau$ is proportional to the number of GRB observations. For projection sensitivity, we assume that GRBs observed in the future have the same distribution as our current sample of $85$ GRBs. Then a sample of $10^{7}$ GRBs would be sufficient to achieve $f_{\text{PBH}}\leq 1$ for $a_{\rm{s}}=c\times T_{90}$.

\bigskip

\section*{Acknowledgment}
We acknowledge the kind support of the National Science and Technology Council of Taiwan R.O.C., with grant number NSTC 111-2811-M-007-018-MY3.
P.Y.Tseng acknowledges support from the Physics Division of the National Center for Theoretical Sciences of Taiwan R.O.C. with grant NSTC 114-2124-M-002-003.
C.Y.Dai is supported in part by the Ministry of Education with Grant No. 111J0382I4.
This work made use of data supplied by the UK Swift Science Data Centre at the University of Leicester.

\bigskip
\newpage

\appendix
\section{Fitting GRB spectra data}
\label{app_fitting}

\begin{table}[H]
\centering
\resizebox{\textwidth}{!}{
\begin{tabular}{c|c|c|cccc|ccc|cc|cc|c}
\hline
\hline
GRB&$T_{90}$ & $z_{S}$ & A\hyperlink{A}{\textsuperscript{$a$}} & $E_0$\hyperlink{$E_0$}{\textsuperscript{$b$}} & $\alpha_1$ & $\alpha_2$ & $M_{\rm PBH}/M_\odot$ & $z_L$ & $y_0$  & $\chi^2_{\rm BAND/PBH}$ & $\chi^2_{\rm BAND/PBH}/{\rm d.o.f}$ & $P$-value~\cite{Cheung:2018ave} \\

\hline
\multirow{2}{4cm}{050406} & \multirow{2}{4em}{5.4} & \multirow{2}{4em}{2.44} & 67.76 & 0.16 & 1.00 & -2.31 & - & - & - & 51.57 & 0.77 & \multirow{2}{4em}{0.13}\\
&&& 71.02 & 0.19 & 1.10 & -2.31 & \num{3.73e-14} & \num{3.36e-5} & 1.90 & 45.90 & 0.72\\

\hline
\multirow{2}{4cm}{051221A} & \multirow{2}{4em}{1.4} & \multirow{2}{4em}{0.547} & 24.52 & 0.35 & 0.96 & -1.89 & - & - & - & 129.31 & 0.97 & \multirow{2}{4em}{\num{7.95e-3}}\\
&&& 25.70 & 0.42 & 1.06 & -1.89 & \num{2.33e-14} & \num{5.82e-5} & 1.90 & 117.47 & 0.90\\

\hline
\multirow{2}{4cm}{060604} & \multirow{2}{4em}{95} & \multirow{2}{4em}{2.1357} & 1.71 & 0.79 & 0.10 & -2.34 & - & - & - & 225.62 & 1.03 & \multirow{2}{4em}{\num{2.66e-5}}\\
&&& 1.23 & 0.91 & \num{9.18e-2} & -2.58 & \num{1.84e-14} & \num{5.95e-7} & 2.10 & 201.75 & 0.93\\

\hline
\multirow{2}{4cm}{060707} & \multirow{2}{4em}{66.2} & \multirow{2}{4em}{3.43} & \num{3.98e-6} & 15000.00 & -1.91 & -13.39 & - & - & - & 21.47 & 0.93 & \multirow{2}{4em}{0.23}\\
&&& \num{9.73e-6} & 30000.00 & -1.72 & -10.71 & \num{2.95e-14} & \num{5.95e-7} & 1.90 & 17.21 & 0.86\\

\hline
\multirow{2}{4cm}{061021} & \multirow{2}{4em}{46.2} & \multirow{2}{4em}{0.3463} & 995.89 & 0.18 & 1.00 & -1.97 & - & - & - & 144.01 & 1.27 & \multirow{2}{4em}{\num{7.49e-5}}\\
&&& 492.14 & 0.18 & 0.90 & -1.78 & \num{1.84e-14} & \num{5.95e-7} & 2.60 & 122.30 & 1.11\\

\hline
\multirow{2}{4cm}{080604} & \multirow{2}{4em}{82} & \multirow{2}{4em}{1.416} & 456.02 & 0.19 & 1.00 & -1.86 & - & - & - & 97.15 & 0.96 & \multirow{2}{4em}{\num{5.03e-2}}\\
&&& 1113.64 & 0.20 & 1.20 & -1.86 & \num{1.46e-14} & \num{1.09e-6} & 2.60 & 89.34 & 0.91\\

\hline
\multirow{2}{4cm}{091020} & \multirow{2}{4em}{34.6} & \multirow{2}{4em}{1.71} & 68.35 & 0.39 & 0.88 & -1.89 & - & - & - & 141.59 & 0.88 & \multirow{2}{4em}{\num{1.19e-3}}\\
&&& 151.93 & 0.39 & 1.05 & -1.89 & \num{1.84e-14} & \num{4.06e-6} & 2.60 & 125.70 & 0.80\\

\hline
\multirow{2}{4cm}{091029} & \multirow{2}{4em}{39.2} & \multirow{2}{4em}{2.752} & 1946.13 & 0.14 & 1.00 & -2.08 & - & - & - & 103.53 & 1.33 & \multirow{2}{4em}{\num{1.38e-7}}\\
&&& 2704.14 & 0.19 & 1.20 & -2.29 & \num{2.33e-14} & \num{2.57e-6} & 2.10 & 68.80 & 0.92\\

\hline
\multirow{2}{4cm}{101219B} & \multirow{2}{4em}{34} & \multirow{2}{4em}{0.5519} & 11.87 & 0.25 & 0.43 & -1.43 & - & - & - & 230.94 & 1.26 & \multirow{2}{4em}{\num{2.15e-15}}\\
&&& 10.31 & 0.33 & 0.51 & -1.43 & \num{2.95e-14} & \num{2.57e-6} & 1.60 & 159.56 & 0.89\\

\hline
\multirow{2}{4cm}{110731A} & \multirow{2}{4em}{38.8} & \multirow{2}{4em}{2.83} & 32.10 & 0.54 & 0.84 & -1.83 & - & - & - & 323.23 & 1.18 & \multirow{2}{4em}{\num{3.94e-7}}\\
&&& 36.96 & 0.60 & 0.92 & -2.01 & \num{2.33e-14} & \num{3.07e-6} & 2.00 & 290.64 & 1.07\\

\hline
\multirow{2}{4cm}{110808A} & \multirow{2}{4em}{48} & \multirow{2}{4em}{1.348} & 1890.42 & 0.16 & 1.00 & -2.96 & - & - & - & 47.57 & 1.29 & \multirow{2}{4em}{\num{5.46e-3}}\\
&&& 1981.37 & 0.17 & 1.10 & -2.66 & \num{2.95e-14} & \num{5.95e-7} & 1.60 & 34.92 & 1.03\\

\hline
\multirow{2}{4cm}{120326A} & \multirow{2}{4em}{69.6} & \multirow{2}{4em}{1.798} & 877.27 & 0.26 & 1.00 & -3.40 & - & - & - & 72.22 & 1.05 & \multirow{2}{4em}{\num{2.65e-3}}\\
&&& 574.72 & 0.26 & 1.00 & -3.06 & \num{2.33e-14} & \num{5.95e-7} & 1.30 & 58.03 & 0.88\\

\hline
\multirow{2}{4cm}{130831A} & \multirow{2}{4em}{32.5} & \multirow{2}{4em}{0.4791} & \num{3.72e-7} & 14999.04 & -2.60 & -12.71 & - & - & - & 81.43 & 1.63 & \multirow{2}{4em}{\num{7.05e-5}}\\
&&& \num{3.72e-6} & 104.23 & -2.08 & -10.17 & \num{3.73e-14} & \num{1.58e-6} & 1.40 & 59.59 & 1.27\\

\hline
\multirow{2}{4cm}{141220A} & \multirow{2}{4em}{7.21} & \multirow{2}{4em}{1.3195} & 55.45 & 0.22 & 0.63 & -1.92 & - & - & - & 37.44 & 1.10 & \multirow{2}{4em}{\num{1.84e-2}}\\
&&& 58.12 & 0.27 & 0.76 & -1.92 & \num{3.73e-14} & \num{4.62e-5} & 1.60 & 27.43 & 0.88\\

\hline
\multirow{2}{4cm}{161108A} & \multirow{2}{4em}{105.1} & \multirow{2}{4em}{1.159} & 26.53 & 0.67 & 0.59 & -2.35 & - & - & - & 318.52 & 1.17 & \multirow{2}{4em}{\num{5.69e-5}}\\
&&& 44.48 & 0.66 & 0.71 & -2.35 & \num{1.46e-14} & \num{5.95e-7} & 2.60 & 296.23 & 1.10\\

\hline
\multirow{2}{4cm}{180205A} & \multirow{2}{4em}{15.5} & \multirow{2}{4em}{1.409} & \num{1.20e-5} & 13.30 & -1.89 & -12.67 & - & - & - & 162.59 & 2.80 & \multirow{2}{4em}{\num{7.29e-10}}\\
&&& \num{9.60e-6} & 12.76 & -1.89 & -6.76 & \num{1.29e-14} & \num{1.00e-5} & 1.32 & 117.10 & 2.13\\

\hline
\multirow{2}{4cm}{200829A} & \multirow{2}{4em}{13.04} & \multirow{2}{4em}{1.25} & 0.21 & 1.28 & -0.27 & -1.70 & - & - & - & 816.18 & 1.34 & \multirow{2}{4em}{\num{3.58e-31}}\\
&&& 0.22 & 1.51 & -0.21 & -1.88 & \num{1.84e-14} & \num{4.59e-5} & 2.20 & 671.44 & 1.11\\

\hline
\multirow{2}{4cm}{210411C} & \multirow{2}{4em}{12.8} & \multirow{2}{4em}{2.826} & \num{6.02e-2} & 0.61 & -0.43 & -2.77 & - & - & - & 61.81 & 1.08 & \multirow{2}{4em}{\num{1.33e-2}}\\
&&& 0.10 & 0.55 & -0.34 & -2.49 & \num{7.20e-15} & \num{1.58e-6} & 4.10 & 51.09 & 0.95\\

\hline
\multirow{2}{4cm}{210610B} & \multirow{2}{4em}{69.38} & \multirow{2}{4em}{1.13} & 218.64 & 0.28 & 0.84 & -1.76 & - & - & - & 631.23 & 1.09 & \multirow{2}{4em}{\num{1.50e-8}}\\
&&& 81.50 & 0.29 & 0.67 & -1.76 & \num{2.22e-15} & \num{5.95e-7} & 3.60 & 591.93 & 1.03\\

\hline
\multirow{2}{4cm}{220611A} & \multirow{2}{4em}{57} & \multirow{2}{4em}{2.3608} & 1.80 & 0.47 & \num{2.04e-2} & -2.19 & - & - & - & 291.90 & 1.31 & \multirow{2}{4em}{\num{5.39e-6}}\\
&&& 1.18 & 0.60 & \num{1.63e-2} & -2.41 & \num{2.33e-14} & \num{1.09e-6} & 2.30 & 264.72 & 1.20\\
\hline
\multirow{2}{4cm}{221110A} & \multirow{2}{4em}{8.98} & \multirow{2}{4em}{4.06} & \num{5.06e-4} & 1.64 & -1.00 & -1.53 & - & - & - & 27.50 & 0.76 & \multirow{2}{4em}{0.19}\\
&&& \num{1.12e-3} & 2.18 & -0.80 & -1.83 & \num{2.95e-14} & \num{2.32e-5} & 1.80 & 22.68 & 0.69\\

\hline
\hline
\end{tabular}
}
\footnotesize
\hypertarget{A}{\textsuperscript{$a$} ${\rm sec}^{-1}{\rm cm}^{-2}{\rm keV}^{-2}$}
\hypertarget{$E_0$}{\textsuperscript{$b$} ${\rm keV}$}
\caption{The results of the PBH fitting with oscillation, where the goodness of fit decreases by over 0.05, correspond to Fig.~\ref{fig:PBH_femtolensing1}. The first three columns are the data of the GRB. The fourth column lists the four parameters of the BAND model. The fifth column lists the parameters of the PBH. The sixth and seventh columns represent the goodness of fit and $p$-value, respectively, and are selected using the minimum $\chi^{2}$.}
\label{tab:label1}
\end{table}

\begin{figure}[H]
    \centering
        \includegraphics[width=0.3\linewidth]{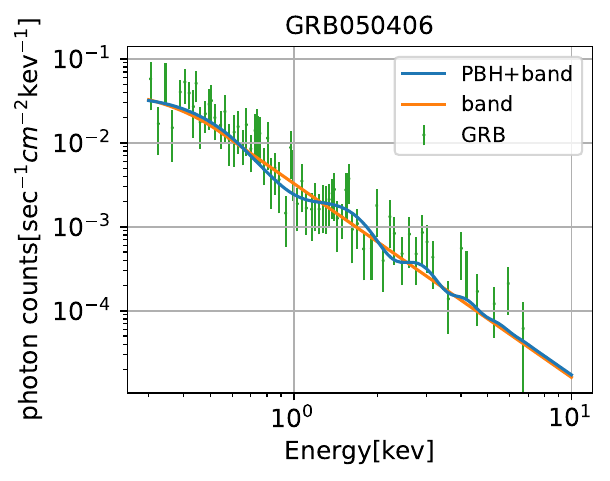}
        \includegraphics[width=0.3\linewidth]{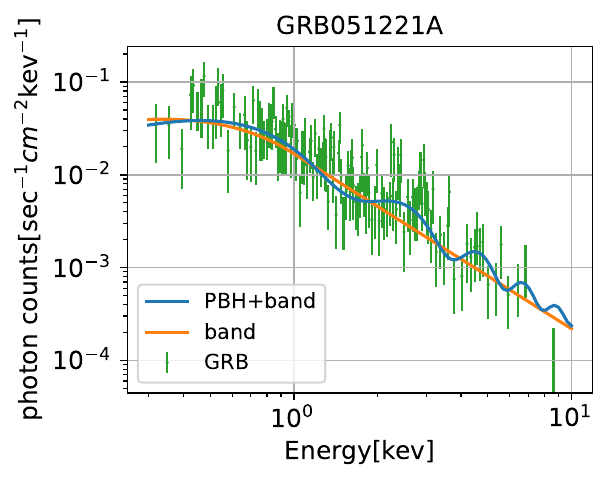}
        \includegraphics[width=0.3\linewidth]{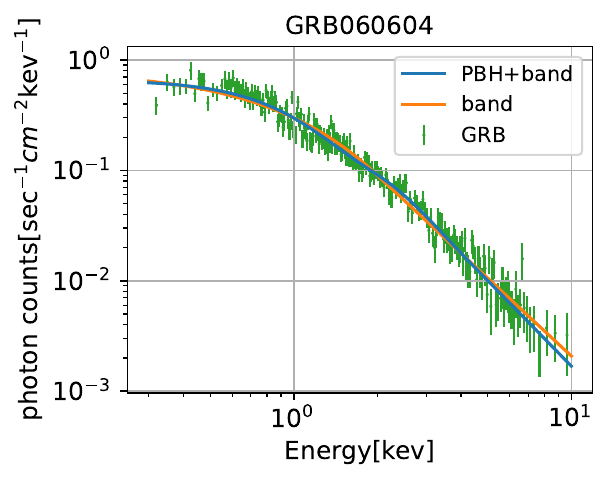}
        \includegraphics[width=0.3\linewidth]{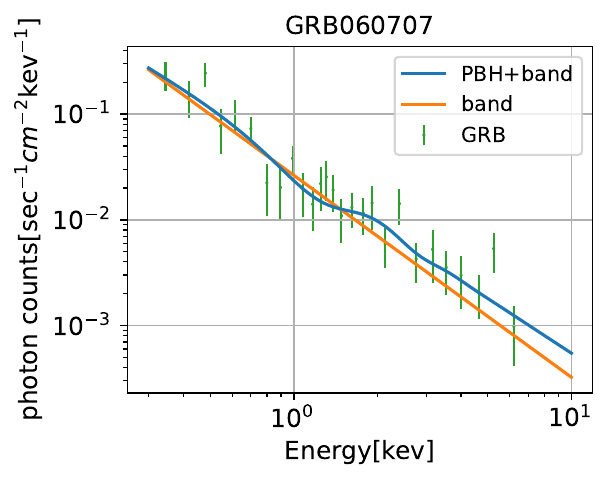}
        \includegraphics[width=0.3\linewidth]{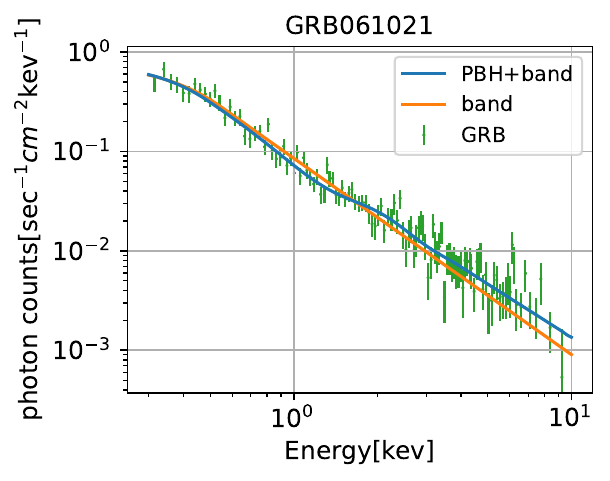}
        \includegraphics[width=0.3\linewidth]{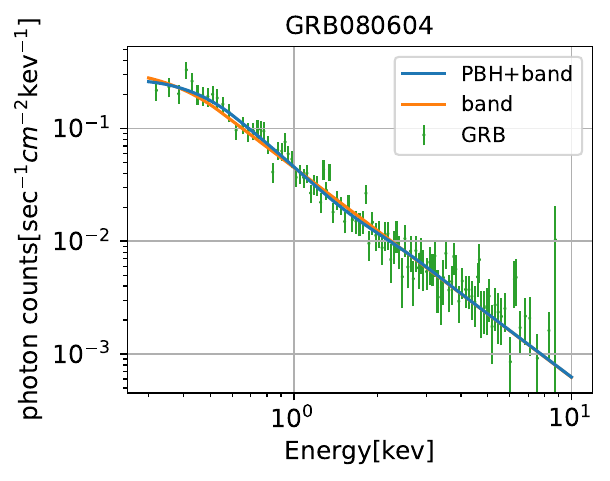}
        \includegraphics[width=0.3\linewidth]{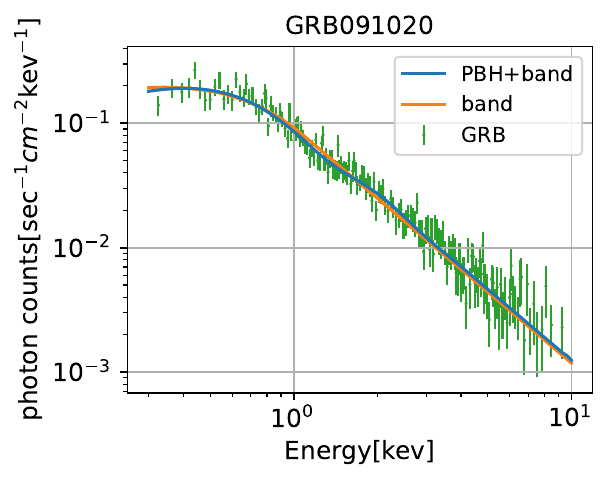}
        \includegraphics[width=0.3\linewidth]{Figures/oscillation_over_0.05/GRB091029.pdf}
        \includegraphics[width=0.3\linewidth]{Figures/oscillation_over_0.05/GRB101219B.pdf}
        \includegraphics[width=0.3\linewidth]{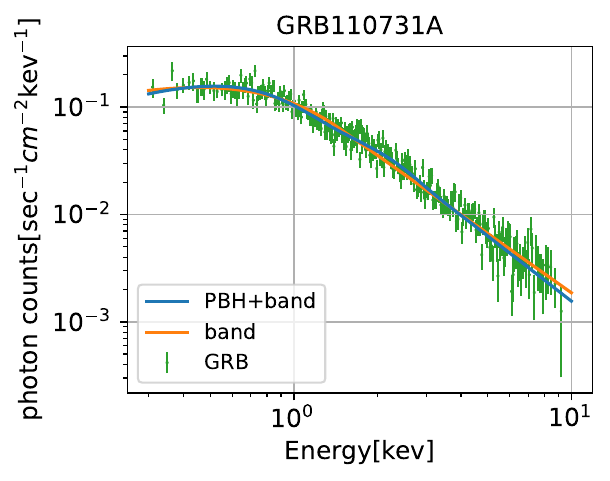}
        \includegraphics[width=0.3\linewidth]{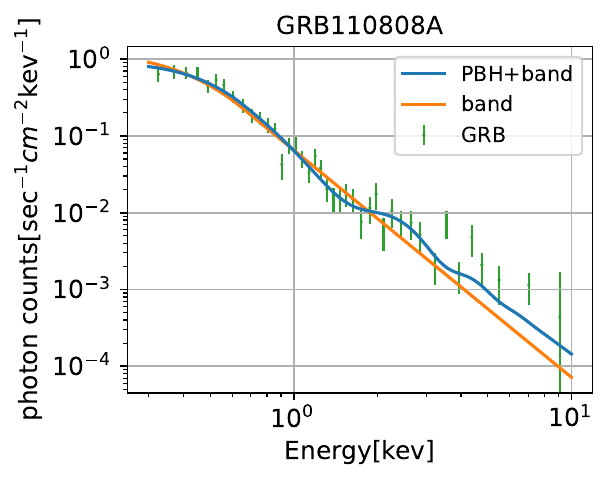}
        \includegraphics[width=0.3\linewidth]{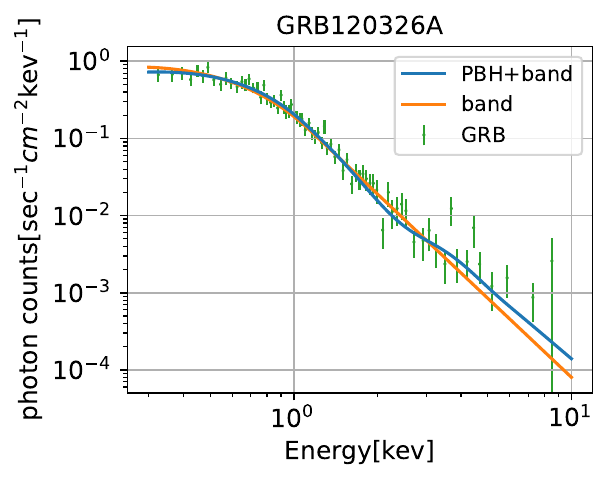}
        \includegraphics[width=0.3\linewidth]{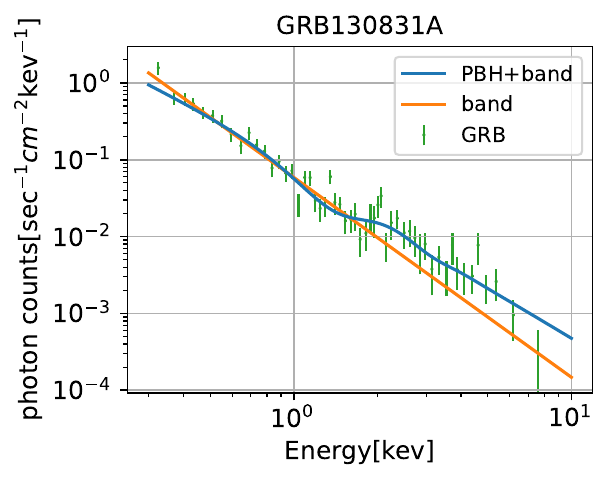}
        \includegraphics[width=0.3\linewidth]{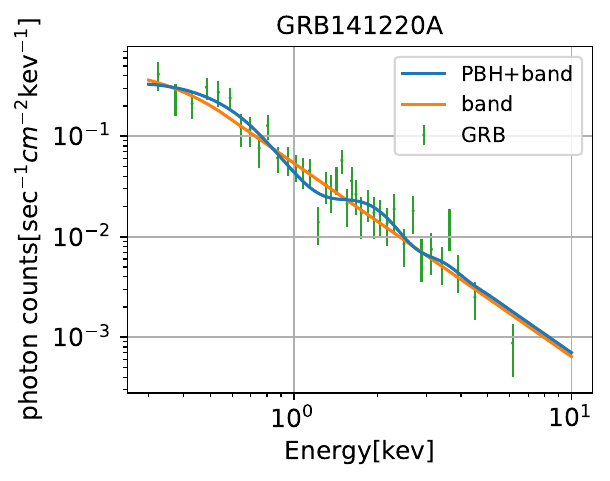}
        \includegraphics[width=0.3\linewidth]{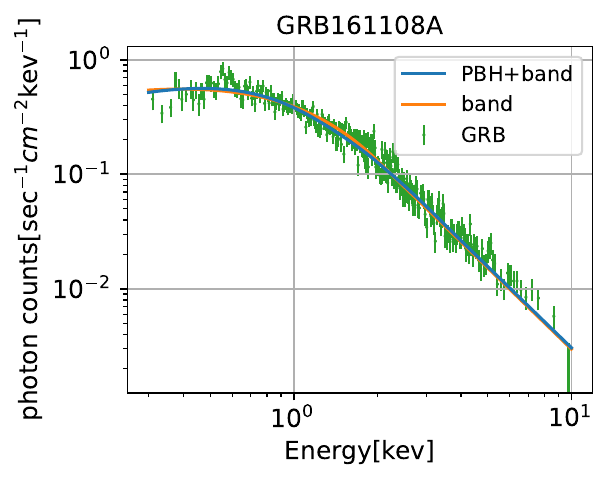}
        \includegraphics[width=0.3\linewidth]{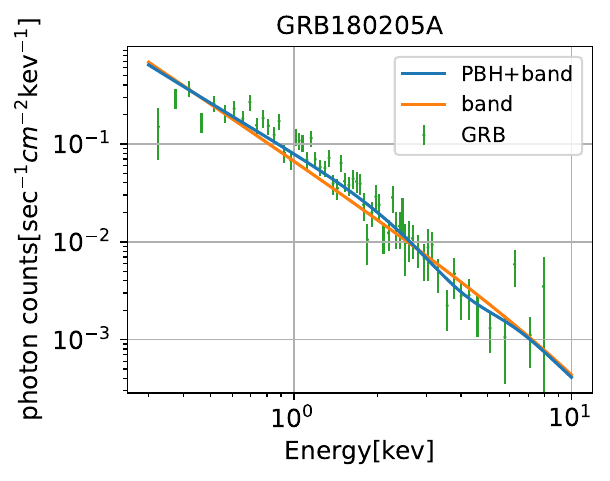}
        \includegraphics[width=0.3\linewidth]{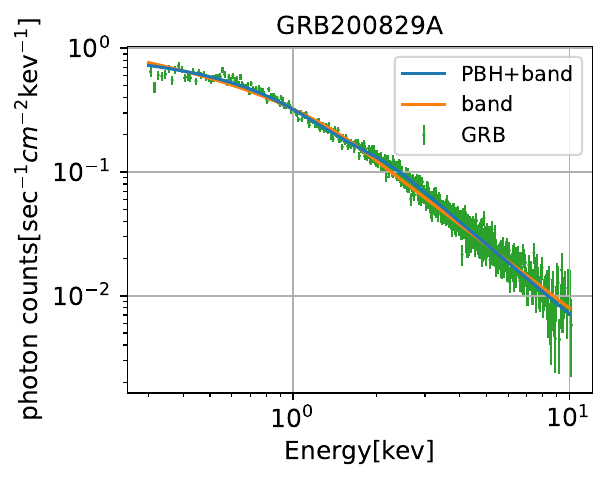}
        \includegraphics[width=0.3\linewidth]{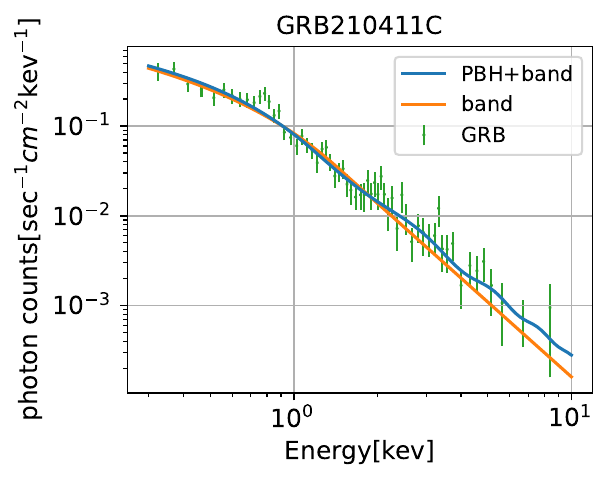}
    \caption{}
    \label{fig:PBH_femtolensing1}
\end{figure}
\begin{figure}[H]\ContinuedFloat
    \centering
        \includegraphics[width=0.3\linewidth]{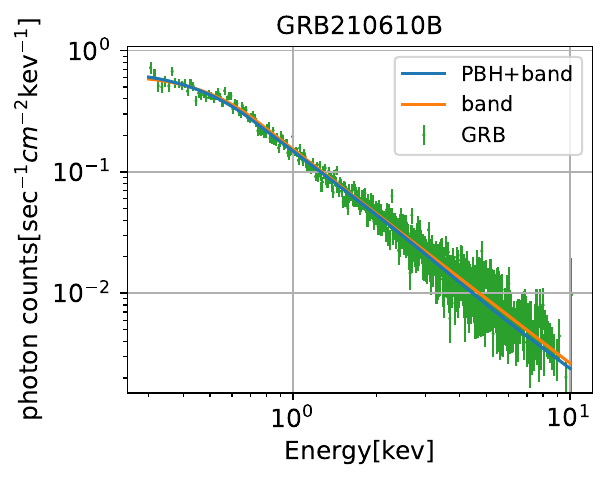}
        \includegraphics[width=0.3\linewidth]{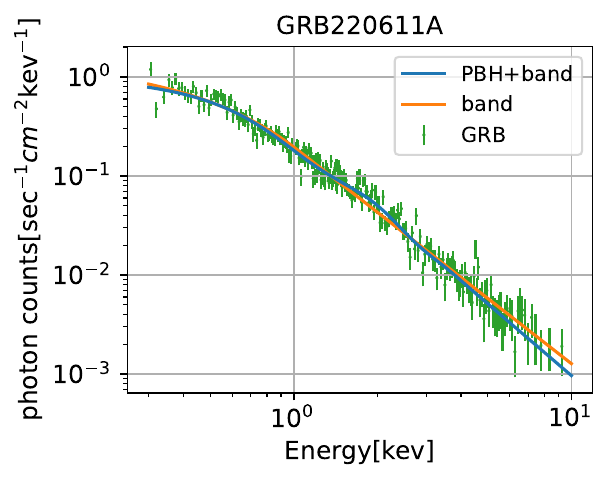}
        \includegraphics[width=0.3\linewidth]{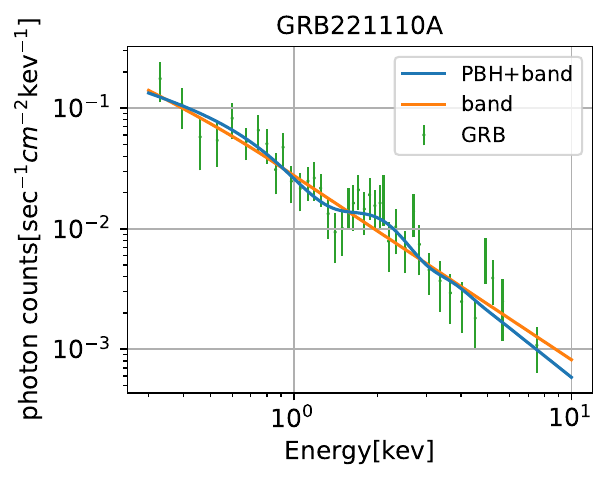}
    \caption{The fitting data with an oscillation pattern, where the goodness of fit decreases by over 0.05, correspond to Table~\ref{tab:label1}. The green error bars represent the GRB data of Swift XRT. The orange curve represents the BAND model only. The blue curve represents the BAND model with lensing effects taken into account.}
    \label{fig:PBH_femtolensing1}
\end{figure}

\begin{table}[H]
\centering
\resizebox{\textwidth}{!}{
\begin{tabular}{c|c|c|cccc|ccc|cc|cc|c}
\hline
\hline
GRB&$T_{90}$ & $z_{S}$ & A\hyperlink{A}{\textsuperscript{$a$}} & $E_0$\hyperlink{$E_0$}{\textsuperscript{$b$}} & $\alpha_1$ & $\alpha_2$ & $M_{\rm PBH}/M_\odot$ & $z_L$ & $y_0$ & $\chi^2_{\rm BAND/PBH}$ & $\chi^2_{\rm BAND/PBH}/{\rm d.o.f}$ & $P$-value~\cite{Cheung:2018ave} \\

\hline
\multirow{2}{4cm}{051109A} & \multirow{2}{4em}{37.2} & \multirow{2}{4em}{2.346} & 50.96 & 0.43 & 1.00 & -1.87 & - & - & - & 108.22 & 0.85 & \multirow{2}{4em}{\num{8.74e-2}}\\
&&& 7.42 & 0.64 & 0.80 & -2.06 & \num{3.73e-14} & \num{3.07e-6} & 1.30 & 101.66 & 0.82\\

\hline
\multirow{2}{4cm}{060908} & \multirow{2}{4em}{19.3} & \multirow{2}{4em}{1.8836} & 0.83 & 0.37 & \num{6.12e-2} & -2.29 & - & - & - & 34.40 & 0.88 & \multirow{2}{4em}{0.37}\\
&&& 0.60 & 0.44 & \num{6.12e-2} & -2.29 & \num{1.46e-14} & \num{1.09e-6} & 2.90 & 31.28 & 0.87\\

\hline
\multirow{2}{4cm}{090809} & \multirow{2}{4em}{5.4} & \multirow{2}{4em}{2.737} & 5.35 & 1.03 & 0.59 & -2.32 & - & - & - & 90.06 & 1.02 & \multirow{2}{4em}{0.16}\\
&&& 2.08 & 1.30 & 0.47 & -2.32 & \num{5.69e-15} & \num{3.07e-6} & 2.70  & 84.95 & 1.00\\

\hline
\multirow{2}{4cm}{130610A} & \multirow{2}{4em}{46.4} & \multirow{2}{4em}{2.092} & \num{3.59e-6} & 14999.02 & -1.96 & -3.95 & - & - & - & 38.78 & 0.95 & \multirow{2}{4em}{0.33}\\
&&& \num{3.43e-6} & 29998.05 & -1.96 & -3.16 & \num{2.33e-14} & \num{5.95e-7} & 2.70 & 35.39 & 0.93\\

\hline
\multirow{2}{4cm}{130701A} & \multirow{2}{4em}{4.38} & \multirow{2}{4em}{1.155} & \num{4.76e-3} & 1.43 & -0.80 & -2.12 & - & - & - & 286.42 & 0.97 & \multirow{2}{4em}{0.21}\\
&&& \num{1.06e-2} & 1.34 & -0.64 & -2.34 & \num{9.10e-15} & \num{4.54e-5} & 3.40 & 281.88 & 0.96\\

\hline
\multirow{2}{4cm}{140301A} & \multirow{2}{4em}{31} & \multirow{2}{4em}{1.416} & 100.62 & 0.31 & 1.00 & -1.91 & - & - & - & 100.10 & 0.93 & \multirow{2}{4em}{0.28}\\
&&& 296.54 & 0.30 & 1.20 & -1.91 & \num{9.10e-15} & \num{5.95e-7} & 4.30 & 96.29 & 0.92\\

\hline
\multirow{2}{4cm}{170604A} & \multirow{2}{4em}{26.7} & \multirow{2}{4em}{1.329} & 0.56 & 1.15 & -0.22 & -2.84 & - & - & - & 436.37 & 1.31 & \multirow{2}{4em}{\num{2.45e-3}}\\
&&& 0.40 & 1.22 & -0.27 & -2.56 & \num{3.56e-15} & \num{5.95e-7} & 3.40 & 422.01 & 1.28\\

\hline
\multirow{2}{4cm}{180115A} & \multirow{2}{4em}{40.9} & \multirow{2}{4em}{2.487} & \num{2.24e-6} & 14999.04 & -2.11 & -4.22 & - & - & - & 45.14 & 1.03 & \multirow{2}{4em}{0.30}\\
&&& \num{4.98e-6} & 48.16 & -1.90 & -3.38 & \num{2.95e-14} & \num{3.56e-6} & 1.50 & 41.47 & 1.01\\

\hline
\multirow{2}{4cm}{180325A} & \multirow{2}{4em}{94.1} & \multirow{2}{4em}{2.25} & 174.72 & 0.42 & 0.96 & -1.34 & - & - & - & 215.32 & 1.01 & \multirow{2}{4em}{\num{4.66e-3}}\\
&&& 353.56 & 0.46 & 1.15 & -1.48 & \num{2.33e-14} & \num{5.95e-7} & 2.10 & 202.33 & 0.96\\

\hline
\multirow{2}{4cm}{190114A} & \multirow{2}{4em}{66.6} & \multirow{2}{4em}{3.3765} & 140.00 & 0.35 & 1.00 & -2.03 & - & - & - & 120.22 & 0.68 & \multirow{2}{4em}{0.31}\\
&&& 213.70 & 0.32 & 1.10 & -1.83 & \num{1.46e-14} & \num{5.95e-7} & 1.60 & 116.64 & 0.67\\

\hline
\multirow{2}{4cm}{210722A} & \multirow{2}{4em}{50.2} & \multirow{2}{4em}{1.145} & \num{1.95e-2} & 0.67 & -0.80 & -2.18 & - & - & - & 207.80 & 0.93 & \multirow{2}{4em}{\num{8.10e-2}}\\
&&& \num{5.22e-2} & 0.57 & -0.64 & -2.18 & \num{9.10e-15} & \num{5.95e-7} & 5.00 & 201.07 & 0.91\\

\hline
\hline
\end{tabular}
}
\footnotesize
\hypertarget{A}{\textsuperscript{$a$} ${\rm sec}^{-1}{\rm cm}^{-2}{\rm keV}^{-2}$}
\hypertarget{$E_0$}{\textsuperscript{$b$} ${\rm keV}$}
\caption{The results of the PBH fitting with oscillation, where the goodness of fit decreases by less than 0.05, correspond to Fig.~\ref{fig:PBH_femtolensing2}. The first three columns are the data of the GRB. The fourth column lists the four parameters of the BAND model. The fifth column lists the parameters of the PBH. The sixth and seventh columns represent the goodness of fit and $p$-value, respectively, and are selected using the minimum $\chi^{2}$.}
\label{tab:label2}
\end{table}  

\begin{figure}[H]
    \centering
        \includegraphics[width=0.3\linewidth]{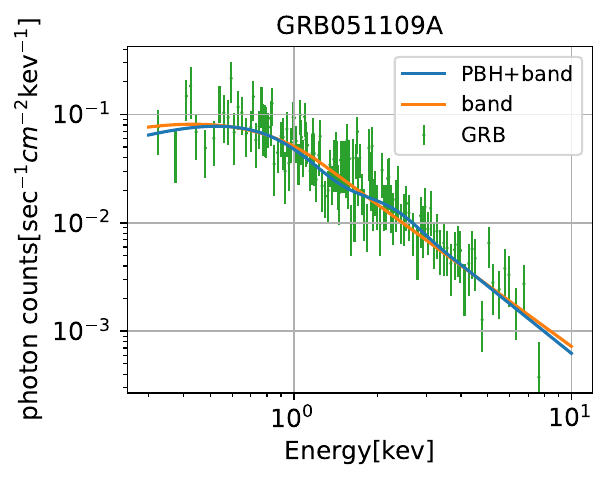}
        \includegraphics[width=0.3\linewidth]{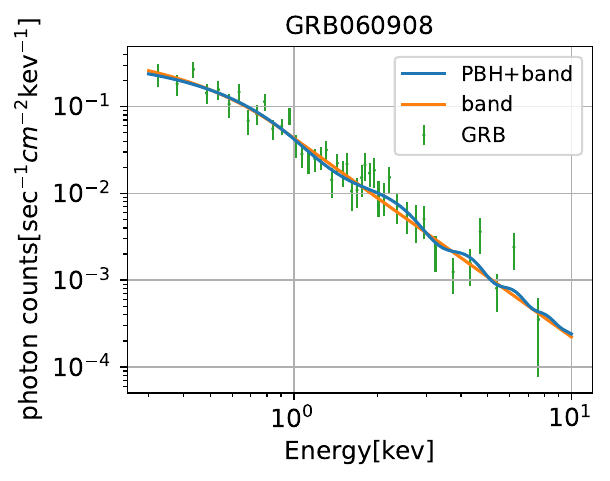}
        \includegraphics[width=0.3\linewidth]{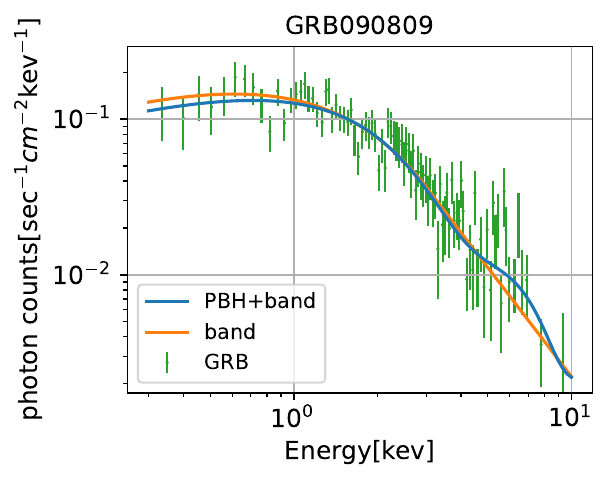}
        \includegraphics[width=0.3\linewidth]{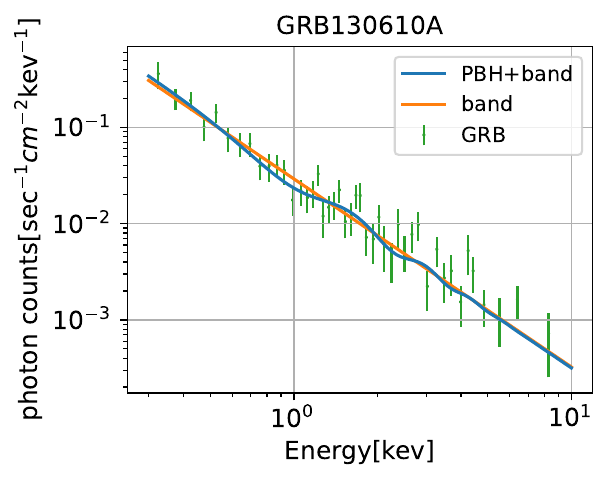}
        \includegraphics[width=0.3\linewidth]{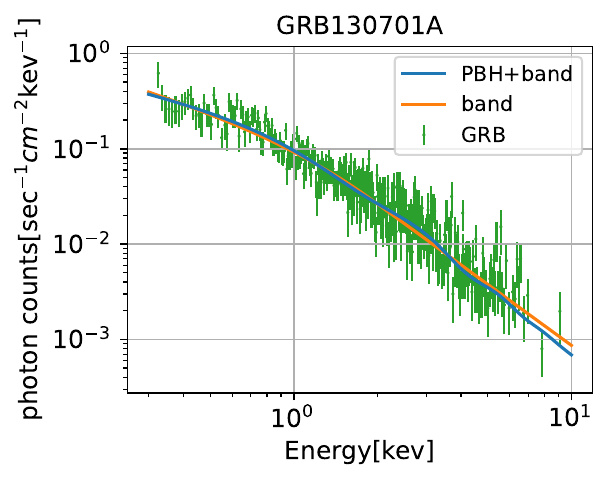}
        \includegraphics[width=0.3\linewidth]{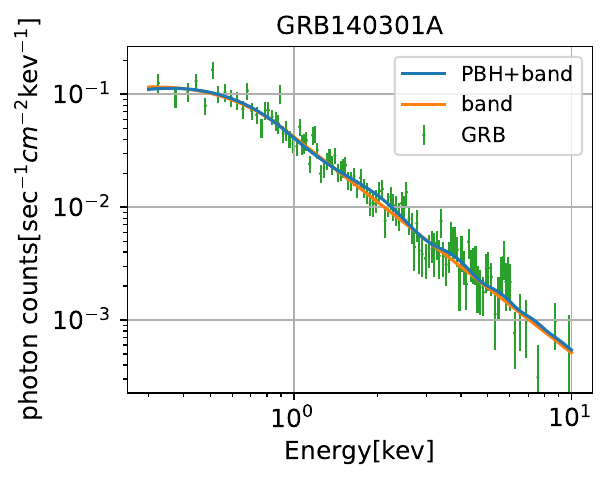}
        \includegraphics[width=0.3\linewidth]{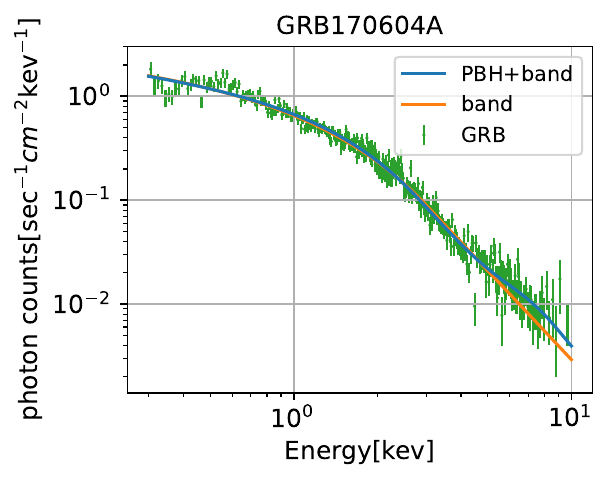}
        \includegraphics[width=0.3\linewidth]{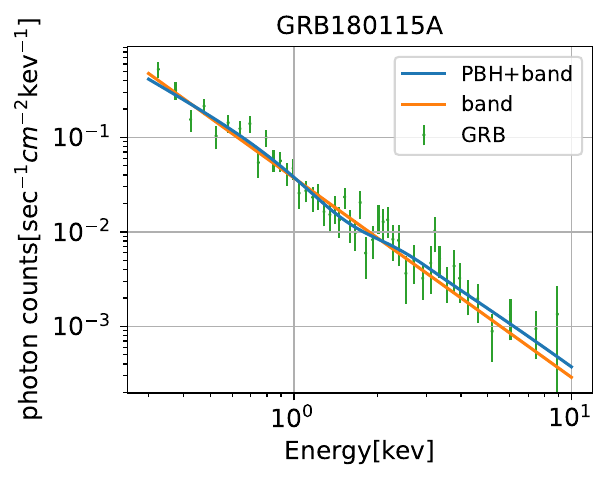}
        \includegraphics[width=0.3\linewidth]{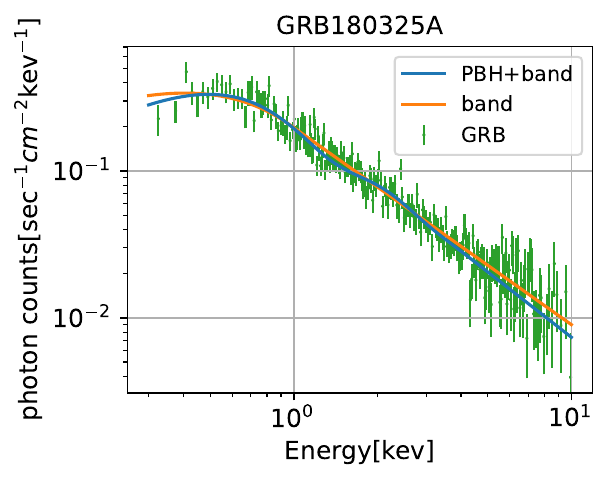}
        \includegraphics[width=0.3\linewidth]{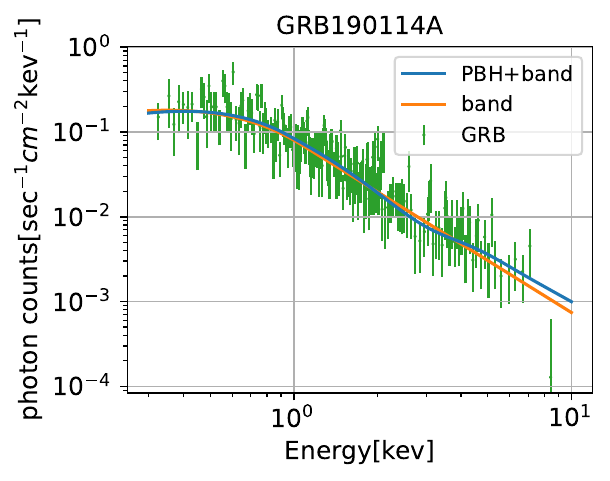}
        \includegraphics[width=0.3\linewidth]{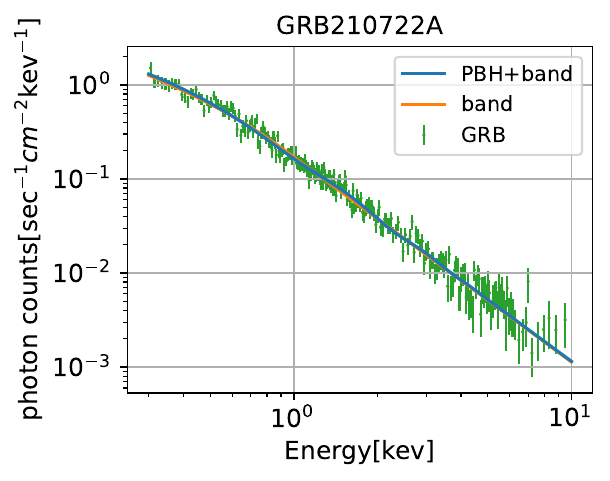}
    \caption{The fitting data with an oscillation pattern, where the goodness of fit decreases by less than 0.05, correspond to Table~\ref{tab:label2}. The green error bars represent the GRB data of Swift XRT. The orange curve represents the BAND model only. The blue curve represents the BAND model with lensing effects taken into account.}
    \label{fig:PBH_femtolensing2}
\end{figure}

\begin{table}[H]
\centering
\resizebox{\textwidth}{!}{
\begin{tabular}{c|c|c|cccc|ccc|cc|cc|c}
\hline
\hline
GRB&$T_{90}$ & $z_{S}$ & A\hyperlink{A}{\textsuperscript{$a$}} & $E_0$\hyperlink{$E_0$}{\textsuperscript{$b$}} & $\alpha_1$ & $\alpha_2$ & $M_{\rm PBH}/M_\odot$ & $z_L$ & $y_0$ & $\chi^2_{\rm BAND/PBH}$ & $\chi^2_{\rm BAND/PBH}/{\rm d.o.f}$ & $P$-value~\cite{Cheung:2018ave} \\

\hline
\multirow{2}{4cm}{050819} & \multirow{2}{4em}{37.7} & \multirow{2}{4em}{2.5043} & 267.76 & 0.16 & 0.88 & -2.28 & - & - & - & 60.19 & 0.85 & \multirow{2}{4em}{0.71}\\
&&& 51.70 & 0.21 & 0.70 & -2.28 & \num{1.46e-14} & \num{1.09e-6} & 2.90 & 58.82 & 0.86\\

\hline
\multirow{2}{4cm}{050915A} & \multirow{2}{4em}{52} & \multirow{2}{4em}{2.5273} & \num{2.84e-2} & 1.17 & -0.35 & -2.14 & - & - & - & 38.22 & 1.06 & \multirow{2}{4em}{0.95}\\
&&& \num{2.27e-2} & 0.84 & -0.31 & -1.71 & \num{1.15e-14} & \num{5.95e-7} & 1.30 & 37.87 & 1.15\\

\hline
\multirow{2}{4cm}{060926} & \multirow{2}{4em}{8} & \multirow{2}{4em}{3.208} & \num{1.03e-2} & 1.03 & \num{2.04e-2} & -1.71 & - & - & - & 47.20 & 0.68 & \multirow{2}{4em}{1.00}\\
&&& \num{1.08e-2} & 0.98 & \num{1.63e-2} & -1.71 & \num{1.46e-14} & \num{6.53e-6} & 4.00 & 48.42 & 0.73\\

\hline
\multirow{2}{4cm}{120118B} & \multirow{2}{4em}{23.26} & \multirow{2}{4em}{2.943} & \num{2.43e-2} & 1.25 & -0.27 & -3.66 & - & - & - & 6.83 & 0.53 & \multirow{2}{4em}{0.95}\\
&&& \num{2.79e-2} & 1.21 & -0.24 & -4.03 & \num{1.46e-14} & \num{2.57e-6} & 3.70 & 6.49 & 0.65\\

\hline
\multirow{2}{4cm}{120724A} & \multirow{2}{4em}{72.8} & \multirow{2}{4em}{1.48} & \num{1.69e-5} & 0.15 & -1.77 & -1.78 & - & - & - & 39.85 & 1.05 & \multirow{2}{4em}{0.55}\\
&&& \num{1.67e-5} & 0.15 & -1.77 & -1.78 & \num{6.25e-15} & \num{1.00e-7} & 3.80 & 37.76 & 1.08\\

\hline
\multirow{2}{4cm}{120805A} & \multirow{2}{4em}{48} & \multirow{2}{4em}{3.1} & 3.02 & 0.96 & 0.84 & -1.00 & - & - & - & 77.97 & 0.70 & \multirow{2}{4em}{0.87}\\
&&& 1.02 & 1.26 & 0.67 & -1.10 & \num{5.69e-15} & \num{5.95e-7} & 3.20 & 77.27 & 0.72\\

\hline
\multirow{2}{4cm}{130603B} & \multirow{2}{4em}{0.18} & \multirow{2}{4em}{0.356} & 2.01 & 0.69 & 0.84 & -1.75 & - & - & - & 60.05 & 0.94 & \multirow{2}{4em}{0.60}\\
&&& 0.64 & 0.81 & 0.67 & -1.58 & \num{5.69e-15} & \num{1.58e-3} & 2.60 & 58.17 & 0.95\\

\hline
\multirow{2}{4cm}{150206A} & \multirow{2}{4em}{83.2} & \multirow{2}{4em}{2.087} & 17.41 & 0.92 & 0.59 & -1.87 & - & - & - & 167.42 & 1.05 & \multirow{2}{4em}{0.37}\\
&&& 7.83 & 1.08 & 0.47 & -1.87 & \num{3.56e-15} & \num{5.95e-7} & 3.90 & 164.28 & 1.05\\

\hline
\multirow{2}{4cm}{150301B} & \multirow{2}{4em}{12.44} & \multirow{2}{4em}{1.5169} & \num{1.51e-3} & 1.52 & -0.76 & -1.52 & - & - & - & 36.13 & 0.88 & \multirow{2}{4em}{0.83}\\
&&& \num{3.05e-3} & 1.51 & -0.60 & -1.52 & \num{1.15e-14} & \num{4.06e-6} & 3.10 & 35.25 & 0.93\\

\hline
\multirow{2}{4cm}{151027B} & \multirow{2}{4em}{80} & \multirow{2}{4em}{4.063} & 45.08 & 0.33 & 0.84 & -2.17 & - & - & - & 26.77 & 0.69 & \multirow{2}{4em}{0.57}\\
&&& 26.88 & 0.40 & 0.84 & -2.39 & \num{2.95e-14} & \num{5.95e-7} & 1.70 & 24.75 & 0.69\\

\hline
\multirow{2}{4cm}{191004B} & \multirow{2}{4em}{37.7} & \multirow{2}{4em}{3.503} & \num{1.67e-3} & 1.59 & -0.96 & -1.76 & - & - & - & 101.31 & 1.04 & \multirow{2}{4em}{0.48}\\
&&& \num{4.69e-4} & 3.11 & -1.15 & -1.76 & \num{4.50e-15} & \num{5.95e-7} & 3.70 & 98.84 & 1.05\\

\hline
\multirow{2}{4cm}{210210A} & \multirow{2}{4em}{6.6} & \multirow{2}{4em}{0.715} & \num{3.37e-6} & 14999.04 & -1.92 & -2.20 & - & - & - & 21.49 & 0.72 & \multirow{2}{4em}{0.67}\\
&&& \num{3.53e-6} & 29998.08 & -1.92 & -1.98 & \num{1.84e-14} & \num{8.02e-6} & 3.40 & 19.96 & 0.74\\

\hline
\multirow{2}{4cm}{231210B} & \multirow{2}{4em}{7.47} & \multirow{2}{4em}{3.13} & \num{4.70e-4} & 1.85 & -1.00 & -2.18 & - & - & - & 19.96 & 0.80 & \multirow{2}{4em}{0.63}\\
&&& \num{2.11e-4} & 3.19 & -1.10 & -2.61 & \num{9.10e-15} & \num{2.57e-6} & 2.70 & 18.23 & 0.83\\

\hline
\multirow{2}{4cm}{241026A} & \multirow{2}{4em}{25.2} & \multirow{2}{4em}{2.79} & 66.98 & 0.30 & 0.92 & -2.15 & - & - & - & 25.52 & 0.85 & \multirow{2}{4em}{0.73}\\
&&& 20.69 & 0.29 & 0.73 & -1.94 & \num{1.15e-14} & \num{2.08e-6} & 1.70 & 24.22 & 0.90\\

\hline
\hline
\end{tabular}
}
\footnotesize
\hypertarget{A}{\textsuperscript{$a$} ${\rm sec}^{-1}{\rm cm}^{-2}{\rm keV}^{-2}$}
\hypertarget{$E_0$}{\textsuperscript{$b$} ${\rm keV}$}
\caption{The fitting data with an oscillation pattern, where the goodness of fit increases, correspond to Fig.~\ref{fig:PBH_femtolensing3}. The first three columns are the data of the GRB. The fourth column lists the four parameters ofthe BAND model. The fifth column lists the parameters of the PBH. The sixth and seventh columns represent the goodness of fit and $p$-value, respectively, and are selected using the minimum $\chi^{2}$.}
\label{tab:label3}
\end{table} 

\begin{figure}[H]
    \centering
        \includegraphics[width=0.3\linewidth]{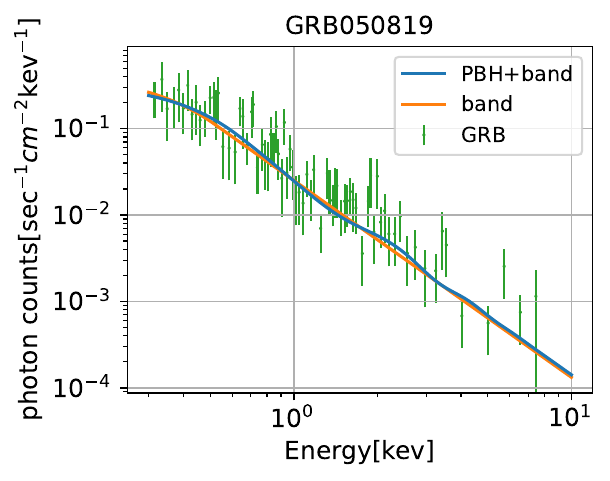}
        \includegraphics[width=0.3\linewidth]{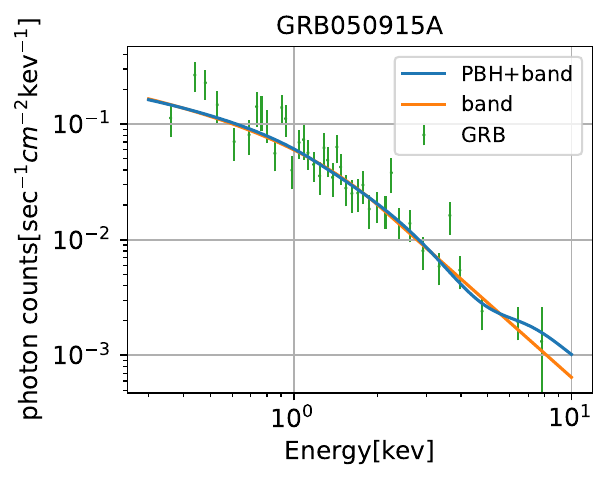}
        \includegraphics[width=0.3\linewidth]{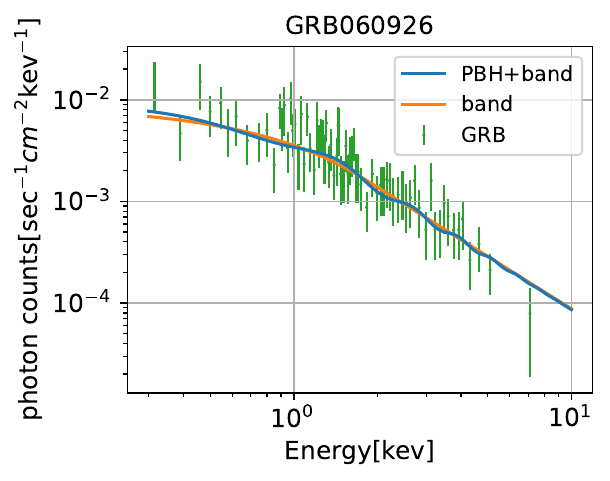}
        \includegraphics[width=0.3\linewidth]{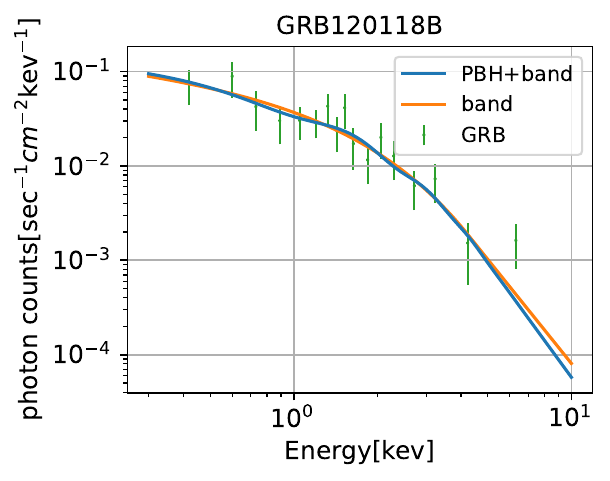}
        \includegraphics[width=0.3\linewidth]{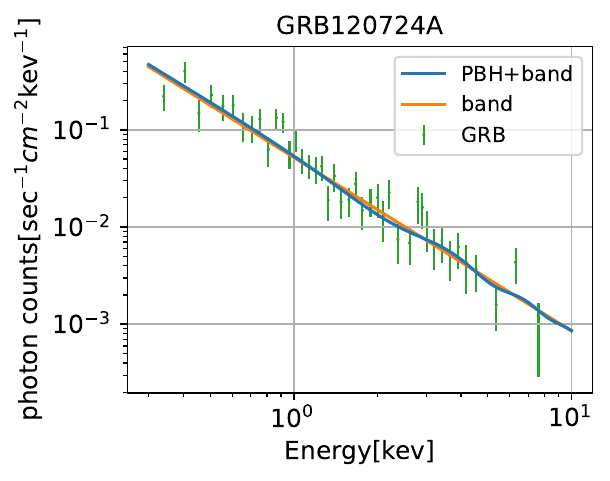}
        \includegraphics[width=0.3\linewidth]{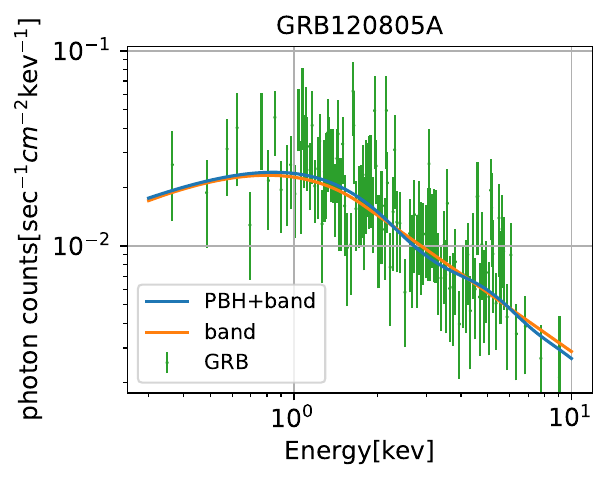}
        \includegraphics[width=0.3\linewidth]{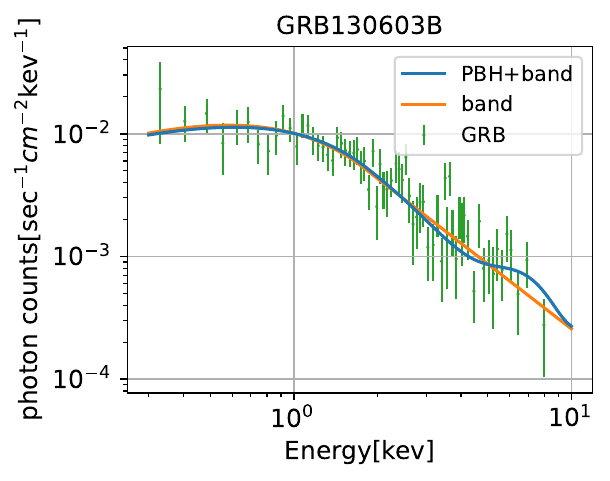}
        \includegraphics[width=0.3\linewidth]{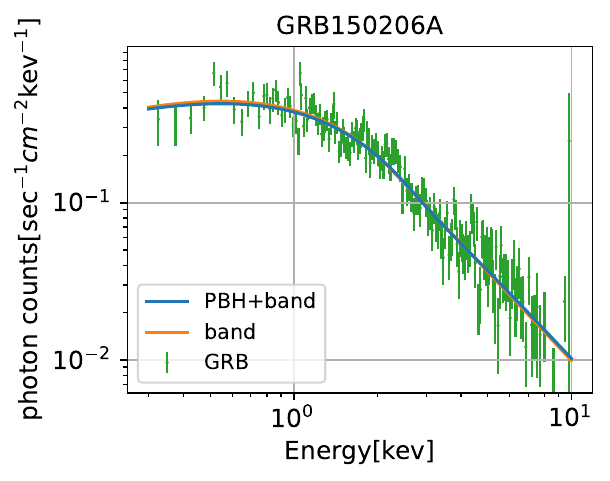}
        \includegraphics[width=0.3\linewidth]{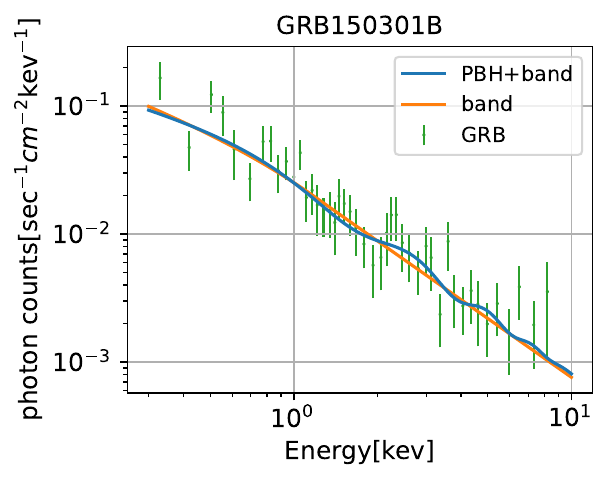}
        \includegraphics[width=0.3\linewidth]{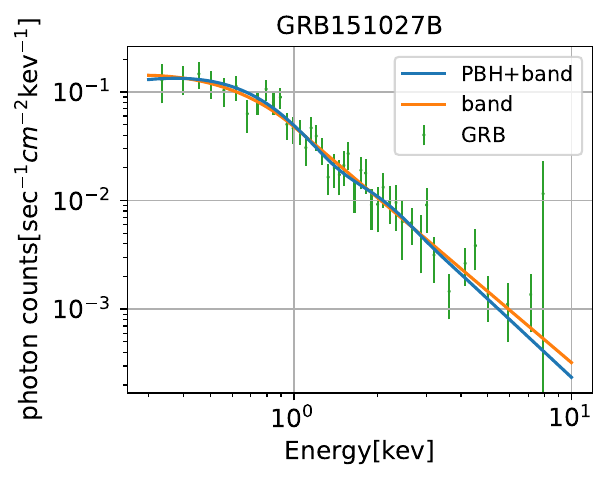}
        \includegraphics[width=0.3\linewidth]{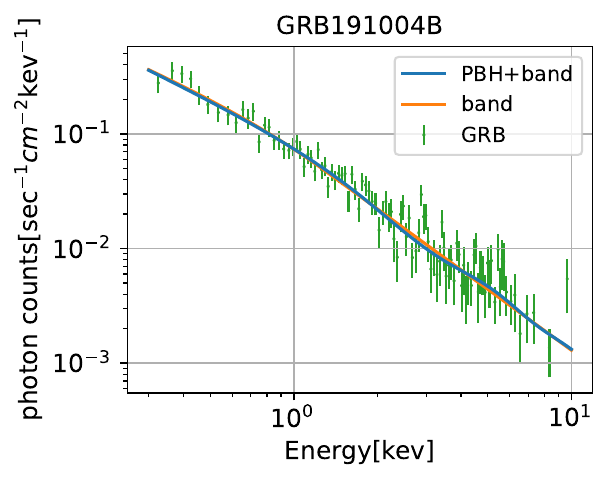}
        \includegraphics[width=0.3\linewidth]{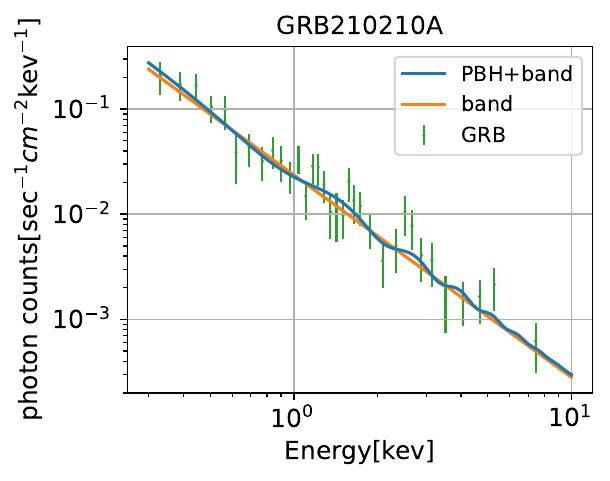}
        \includegraphics[width=0.3\linewidth]{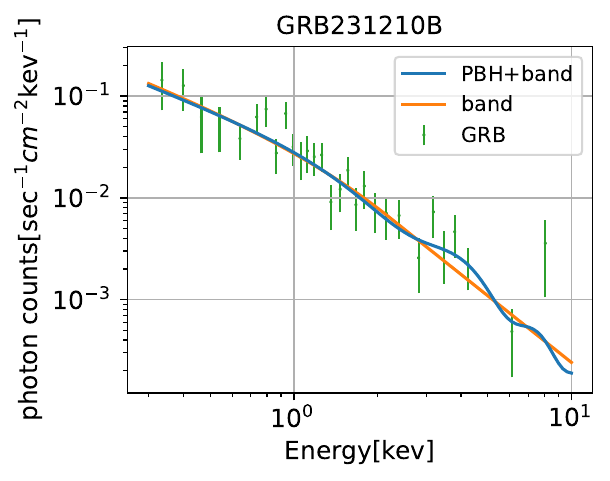}
        \includegraphics[width=0.3\linewidth]{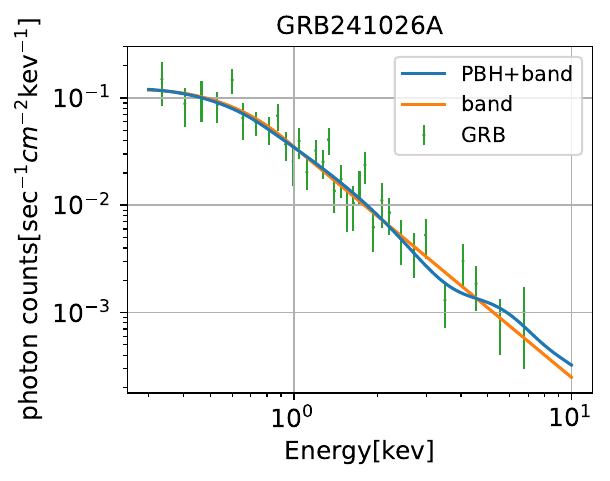}
    \caption{The fitting data with an oscillation pattern, where the goodness of fit increases, correspond to Table~\ref{tab:label3}. The green error bars represent the GRB data of Swift XRT. The orange curve represents the BAND model only. The blue curve represents the BAND model with lensing effects taken into account.}
    \label{fig:PBH_femtolensing3}
\end{figure}

\begin{table}[H]
\centering
\resizebox{\textwidth}{!}{
\begin{tabular}{c|c|c|cccc|ccc|cc|cc|c}
\hline
\hline
GRB&$T_{90}$ & $z_{S}$ & A\hyperlink{A}{\textsuperscript{$a$}} & $E_0$\hyperlink{$E_0$}{\textsuperscript{$b$}} & $\alpha_1$ & $\alpha_2$ & $M_{\rm PBH}/M_\odot$ & $z_L$ & $y_0$ & $\chi^2_{\rm BAND/PBH}$ & $\chi^2_{\rm BAND/PBH}/{\rm d.o.f}$ & $P$-value~\cite{Cheung:2018ave} \\

\hline
\multirow{2}{4cm}{051016B} & \multirow{2}{4em}{4} & \multirow{2}{4em}{0.9364} & 286.74 & 0.24 & 1.00 & -3.86 & - & - & - & 11.49 & 0.82 & \multirow{2}{4em}{0.25}\\
&&& 128.99 & 0.23 & 1.20 & -3.08 & \num{1.93e-13} & \num{6.39e-5} & 0.20 & 7.35 & 0.67\\

\hline
\multirow{2}{4cm}{060512} & \multirow{2}{4em}{8.5} & \multirow{2}{4em}{0.4428} & \num{3.15E-2} & 0.20 & -0.88 & -3.60 & - & - & - & 74.98 & 1.12 & \multirow{2}{4em}{\num{1.57E-2}}\\
&&& \num{7.68E-2} & 0.22 & -0.70 & -4.32 & \num{5.18e-17} & \num{3.77e-7} & 0.1 & 64.61 & 1.01\\

\hline
\multirow{2}{4cm}{061121} & \multirow{2}{4em}{81.3} & \multirow{2}{4em}{1.314} & 109.16 & 0.49 & 1.00 & -1.59 & - & - & - & 296.35 & 1.14 & \multirow{2}{4em}{\num{1.76e-6}}\\
&&& 94.81 & 0.58 & 1.10 & -1.75 & \num{3.73e-14} & \num{1.09e-6} & 1.20 & 266.85 & 1.04\\

\hline
\multirow{2}{4cm}{061222A} & \multirow{2}{4em}{71.4} & \multirow{2}{4em}{2.088} & 43.28 & 0.92 & 1.00 & -2.21 & - & - & - & 437.80 & 1.58 & \multirow{2}{4em}{\num{7.44e-31}}\\
&&& 21.39 & 0.80 & 1.20 & -1.99 & \num{1.21e-13} & \num{5.95e-7} & 0.20 & 294.53 & 1.07\\

\hline
\multirow{2}{4cm}{061222B} & \multirow{2}{4em}{40} & \multirow{2}{4em}{3.355} & 19.94 & 0.69 & 1.00 & -2.52 & - & - & - & 164.60 & 2.25 & \multirow{2}{4em}{\num{3.84e-11}}\\
&&& 6.16 & 0.76 & 1.20 & -2.52 & \num{1.53e-13} & \num{5.95e-7} & 0.20 & 113.11 & 1.62\\

\hline
\multirow{2}{4cm}{070508} & \multirow{2}{4em}{20.9} & \multirow{2}{4em}{0.82} & 14.01 & 0.96 & 0.96 & -2.00 & - & - & - & 622.00 & 1.67 & \multirow{2}{4em}{\num{1.10e-44}}\\
&&& 11.08 & 0.99 & 1.15 & -2.20 & \num{5.96e-14} & \num{6.53e-6} & 0.50 & 414.67 & 1.12\\

\hline
\multirow{2}{4cm}{070724A} & \multirow{2}{4em}{0.4} & \multirow{2}{4em}{0.457} & 15.34 & 0.91 & 1.00 & -1.62 & - & - & - & 76.60 & 1.14 & \multirow{2}{4em}{\num{4.67e-3}}\\
&&& 2.77 & 0.98 & 1.20 & -1.95 & \num{3.91e-13} & \num{2.54e-3} & 0.10 & 63.62 & 0.99\\

\hline
\multirow{2}{4cm}{080605} & \multirow{2}{4em}{20} & \multirow{2}{4em}{1.6398} & 4.05 & 0.84 & 0.55 & -1.67 & - & - & - & 303.98 & 0.98 & \multirow{2}{4em}{\num{3.88e-5}}\\
&&& 4.66 & 0.42 & 0.66 & -1.67 & \num{5.96e-14} & \num{9.01e-6} & 0.10 & 280.89 & 0.91\\

\hline
\multirow{2}{4cm}{080607} & \multirow{2}{4em}{79} & \multirow{2}{4em}{3.036} & 127.19 & 0.60 & 0.96 & -2.05 & - & - & - & 307.62 & 1.31 & \multirow{2}{4em}{\num{1.23e-12}}\\
&&& 160.87 & 0.37 & 1.15 & -1.85 & \num{7.54e-14} & \num{5.95e-7} & 0.10 & 249.12 & 1.08\\

\hline
\multirow{2}{4cm}{080916A} & \multirow{2}{4em}{60} & \multirow{2}{4em}{0.689} & 69.63 & 0.67 & 1.00 & -2.22 & - & - & - & 227.06 & 1.67 & \multirow{2}{4em}{\num{3.35e-19}}\\
&&& 37.80 & 0.67 & 1.20 & -2.22 & \num{1.21e-13} & \num{5.95e-7} & 0.30 & 137.92 & 1.04\\

\hline
\multirow{2}{4cm}{081007} & \multirow{2}{4em}{10} & \multirow{2}{4em}{0.5295} & 62.02 & 0.50 & 1.00 & -2.53 & - & - & - & 107.98 & 2.04 & \multirow{2}{4em}{\num{3.30e-11}}\\
&&& 9.92 & 0.56 & 1.20 & -2.78 & \num{3.91e-13} & \num{2.57e-6} & 0.10 & 56.18 & 1.12\\

\hline
\multirow{2}{4cm}{081118} & \multirow{2}{4em}{67} & \multirow{2}{4em}{2.58} & 853.89 & 0.19 & 1.00 & -2.57 & - & - & - & 39.95 & 1.05 & \multirow{2}{4em}{0.26}\\
&&& 2085.25 & 0.16 & 1.20 & -2.31 & \num{4.71e-14} & \num{1.58e-6} & 0.90 & 35.93 & 1.03\\

\hline
\multirow{2}{4cm}{081221} & \multirow{2}{4em}{34} & \multirow{2}{4em}{2.26} & 91.18 & 0.67 & 1.00 & -2.31 & - & - & - & 676.05 & 2.09 & \multirow{2}{4em}{\num{1.59e-58}}\\
&&& 49.50 & 0.68 & 1.20 & -2.31 & \num{1.21e-13} & \num{2.08e-6} & 0.30 & 404.72 & 1.26\\

\hline
\multirow{2}{4cm}{090812} & \multirow{2}{4em}{66.7} & \multirow{2}{4em}{2.452} & 10.56 & 0.51 & 0.43 & -1.93 & - & - & - & 407.81 & 1.21 & \multirow{2}{4em}{\num{2.25e-14}}\\
&&& 6.42 & 0.40 & 0.34 & -2.31 & \num{1.46e-14} & \num{5.95e-7} & 0.30 & 341.17 & 1.02\\

\hline
\multirow{2}{4cm}{090926B} & \multirow{2}{4em}{109.7} & \multirow{2}{4em}{1.24} & 26.44 & 0.83 & 1.00 & -1.71 & - & - & - & 398.39 & 1.90 & \multirow{2}{4em}{\num{2.01e-34}}\\
&&& 17.32 & 0.66 & 1.20 & -1.54 & \num{9.54e-14} & \num{5.95e-7} & 0.10 & 238.57 & 1.15\\

\hline
\multirow{2}{4cm}{100424A} & \multirow{2}{4em}{104} & \multirow{2}{4em}{2.465} & 0.24 & 2.04 & 0.22 & -1.51 & - & - & - & 210.04 & 1.01 & \multirow{2}{4em}{\num{8.46e-6}}\\
&&& \num{9.78e-2} & 2.22 & 0.27 & -1.66 & \num{7.54e-14} & \num{5.95e-7} & 0.50 & 183.79 & 0.90\\

\hline
\multirow{2}{4cm}{100425A} & \multirow{2}{4em}{37} & \multirow{2}{4em}{1.755} & 314.32 & 0.31 & 0.88 & -3.74 & - & - & - & 90.57 & 1.01 & \multirow{2}{4em}{\num{1.30e-2}}\\
&&& 37.93 & 0.29 & 0.79 & -3.37 & \num{1.53e-13} & \num{5.05e-6} & 0.10 & 79.79 & 0.92\\

\hline
\multirow{2}{4cm}{100615A} & \multirow{2}{4em}{39} & \multirow{2}{4em}{1.398} & 8.50 & 1.59 & 1.00 & -3.86 & - & - & - & 277.18 & 2.89 & \multirow{2}{4em}{\num{1.34e-32}}\\
&&& 5.07 & 1.48 & 1.20 & -4.24 & \num{5.96e-14} & \num{5.95e-7} & 0.30 & 125.81 & 1.35\\

\hline
\multirow{2}{4cm}{100621A} & \multirow{2}{4em}{63.6} & \multirow{2}{4em}{0.542} & 135.07 & 0.92 & 1.00 & -2.79 & - & - & - & 2071.92 & 4.14 & \multirow{2}{4em}{\num{9.82e-231}}\\
&&& 80.55 & 0.76 & 1.20 & -2.51 & \num{7.54e-14} & \num{5.95e-7} & 0.10 & 1006.17 & 2.02\\

\hline
\multirow{2}{4cm}{100906A} & \multirow{2}{4em}{114.4} & \multirow{2}{4em}{1.727} & 145.37 & 0.96 & 1.00 & -1.73 & - & - & - & 359.93 & 1.31 & \multirow{2}{4em}{\num{1.74e-17}}\\
&&& 95.23 & 0.81 & 1.20 & -1.56 & \num{9.54e-14} & \num{5.95e-7} & 0.20  & 278.78 & 1.02\\

\hline
\multirow{2}{4cm}{110213A} & \multirow{2}{4em}{48} & \multirow{2}{4em}{1.46} & 108.04 & 0.44 & 1.00 & -2.71 & - & - & - & 58.29 & 1.88 & \multirow{2}{4em}{\num{3.00e-5}}\\
&&& 64.43 & 0.43 & 1.20 & -2.44 & \num{1.21e-13} & \num{5.95e-7} & 0.30 & 34.67 & 1.24\\

\hline
\multirow{2}{4cm}{110715A} & \multirow{2}{4em}{13} & \multirow{2}{4em}{0.82} & 8.52 & 1.15 & 0.96 & -2.14 & - & - & - & 317.77 & 1.71 & \multirow{2}{4em}{\num{7.29e-22}}\\
&&& 0.85 & 1.22 & 1.06 & -2.36 & \num{2.44e-13} & \num{4.06e-6} & 0.10 & 216.24 & 1.18\\

\hline
\multirow{2}{4cm}{121128A} & \multirow{2}{4em}{23.3} & \multirow{2}{4em}{2.2} & 53.87 & 0.62 & 1.00 & -2.54 & - & - & - & 97.38 & 1.04 & \multirow{2}{4em}{\num{3.59e-3}}\\
&&& 56.46 & 0.40 & 1.10 & -2.03 & \num{4.71e-14} & \num{1.58e-6} & 0.30 & 83.83 & 0.92\\

\hline
\multirow{2}{4cm}{130505A} & \multirow{2}{4em}{88} & \multirow{2}{4em}{2.27} & \num{1.32e-2} & 2.45 & -0.88 & -1.85 & - & - & - & 402.35 & 0.92 & \multirow{2}{4em}{\num{2.93e-6}}\\
&&& \num{3.07e-3} & 4.52 & -1.05 & -2.22 & \num{2.95e-14} & \num{1.58e-6} & 1.10 & 373.91 & 0.86\\

\hline
\multirow{2}{4cm}{130604A} & \multirow{2}{4em}{37.7} & \multirow{2}{4em}{1.06} & 25.60 & 0.69 & 0.92 & -1.86 & - & - & - & 127.57 & 1.29 & \multirow{2}{4em}{\num{2.56e-6}}\\
&&& 24.43 & 0.78 & 1.10 & -2.23 & \num{5.96e-14} & \num{3.07e-6} & 0.70 & 98.85 & 1.03\\

\hline
\multirow{2}{4cm}{131103A} & \multirow{2}{4em}{17.3} & \multirow{2}{4em}{0.599} & 40.98 & 0.62 & 1.00 & -2.27 & - & - & - & 62.26 & 1.17 & \multirow{2}{4em}{\num{4.30e-5}}\\
&&& 8.69 & 0.64 & 1.10 & -2.04 & \num{1.21e-13} & \num{2.08e-6} & 0.20 & 39.39 & 0.79\\

\hline
\multirow{2}{4cm}{140419A} & \multirow{2}{4em}{94.7} & \multirow{2}{4em}{3.956} & \num{2.27e-3} & 2.21 & -1.00 & -2.10 & - & - & - & 374.91 & 1.14 & \multirow{2}{4em}{\num{4.58e-20}}\\
&&& \num{7.35e-3} & 1.02 & -0.80 & -1.68 & \num{9.10e-15} & \num{5.95e-7} & 1.80 & 281.74 & 0.86\\

\hline
\multirow{2}{4cm}{140703A} & \multirow{2}{4em}{67.1} & \multirow{2}{4em}{3.14} & 274.71 & 0.48 & 1.00 & -3.01 & - & - & - & 215.49 & 1.38 & \multirow{2}{4em}{\num{1.65e-12}}\\
&&& 27.47 & 0.47 & 1.00 & -3.01 & \num{2.44e-13} & \num{5.95e-7} & 0.10 & 157.59 & 1.03\\

\hline
\multirow{2}{4cm}{140907A} & \multirow{2}{4em}{79.2} & \multirow{2}{4em}{1.21} & 15.84 & 0.55 & 0.96 & -1.90 & - & - & - & 36.77 & 0.99 & \multirow{2}{4em}{\num{9.33e-2}}\\
&&& 29.19 & 0.43 & 1.15 & -1.52 & \num{3.73e-14} & \num{5.95e-7} & 0.70 & 30.36 & 0.89\\

\hline
\multirow{2}{4cm}{150314A} & \multirow{2}{4em}{14.79} & \multirow{2}{4em}{1.758} & 111.92 & 0.56 & 0.96 & -1.72 & - & - & - & 624.70 & 1.42 & \multirow{2}{4em}{\num{5.95e-21}}\\
&&& 88.49 & 0.31 & 1.06 & -1.72 & \num{9.54e-14} & \num{9.01e-6} & 0.10  & 527.41 & 1.20\\

\hline
\multirow{2}{4cm}{150403A} & \multirow{2}{4em}{40.9} & \multirow{2}{4em}{2.06} & 0.22 & 1.28 & -0.22 & -1.84 & - & - & - & 739.94 & 1.29 & \multirow{2}{4em}{\num{1.80e-30}}\\
&&& 0.20 & 1.45 & -0.18 & -1.84 & \num{2.33e-14} & \num{6.04e-6} & 1.60  & 598.45 & 1.05\\

\hline
\multirow{2}{4cm}{151021A} & \multirow{2}{4em}{110.2} & \multirow{2}{4em}{2.33} & 349.65 & 0.49 & 0.92 & -1.74 & - & - & - & 343.83 & 1.20 & \multirow{2}{4em}{\num{9.74e-7}}\\
&&& 89.40 & 0.37 & 0.73 & -1.57 & \num{4.71e-14} & \num{1.09e-6} & 0.20 & 313.11 & 1.10\\

\hline
\multirow{2}{4cm}{160327A} & \multirow{2}{4em}{28} & \multirow{2}{4em}{4.99} & \num{9.99e-6} & 14998.93 & -1.95 & -12.67 & - & - & - & 79.35 & 1.24 & \multirow{2}{4em}{\num{4.02e-6}}\\
&&& \num{5.96e-6} & 612.30 & -1.75 & -10.14 & \num{1.53e-13} & \num{2.08e-6} & 0.30 & 51.56 & 0.85\\

\hline
\multirow{2}{4cm}{170113A} & \multirow{2}{4em}{20.66} & \multirow{2}{4em}{1.968} & 20.31 & 0.58 & 0.71 & -1.74 & - & - & - & 200.10 & 1.18 & \multirow{2}{4em}{\num{2.14e-4}}\\
&&& 6.89 & 0.80 & 0.64 & -2.09 & \num{3.73e-14} & \num{1.49e-5} & 1.20 & 180.59 & 1.08\\

\hline
\hline
\end{tabular}
}
\footnotesize
\hypertarget{A}{\textsuperscript{$a$} ${\rm sec}^{-1}{\rm cm}^{-2}{\rm keV}^{-2}$}
\hypertarget{$E_0$}{\textsuperscript{$b$} ${\rm keV}$}
\caption{}
\label{tab:label4}
\end{table}

\begin{table}[H]
\ContinuedFloat
\centering
\resizebox{\textwidth}{!}{
\begin{tabular}{c|c|c|cccc|ccc|cc|cc|c}
\hline
\hline
GRB&$T_{90}$ & $z_{S}$ & A\hyperlink{A}{\textsuperscript{$a$}} & $E_0$\hyperlink{$E_0$}{\textsuperscript{$b$}} & $\alpha_1$ & $\alpha_2$ & $M_{\rm PBH}/M_\odot$ & $z_L$ & $y_0$ & $\chi^2_{\rm BAND/PBH}$ & $\chi^2_{\rm BAND/PBH}/{\rm d.o.f}$ & $P$-value~\cite{Cheung:2018ave} \\

\hline
\multirow{2}{4cm}{180728A} & \multirow{2}{4em}{8.68} & \multirow{2}{4em}{0.117} & 19.80 & 0.79 & 1.00 & -1.83 & - & - & - & 520.02 & 1.28 & \multirow{2}{4em}{\num{2.94e-18}}\\
&&& 4.39 & 0.83 & 1.10 & -1.83 & \num{1.53e-13} & \num{3.56e-6} & 0.20 & 435.27 & 1.08\\

\hline
\multirow{2}{4cm}{190106A} & \multirow{2}{4em}{76.8} & \multirow{2}{4em}{1.86} & \num{2.95e-4} & 15.00 & -1.45 & -9.20 & - & - & - & 464.69 & 2.37 & \multirow{2}{4em}{\num{1.11e-60}}\\
&&& \num{2.13e-4} & 4.15 & -1.16 & -7.36 & \num{1.53e-13} & \num{1.09e-6} & 0.10 & 183.39 & 0.95\\

\hline
\multirow{2}{4cm}{191221B} & \multirow{2}{4em}{48} & \multirow{2}{4em}{1.19} & 561.89 & 0.28 & 1.00 & -1.86 & - & - & - & 501.32 & 1.23 & \multirow{2}{4em}{\num{5.83e-26}}\\
&&& 1035.05 & 0.23 & 1.10 & -2.04 & \num{1.46e-14} & \num{4.55e-6} & 0.60 & 380.75 & 0.94\\

\hline
\multirow{2}{4cm}{201104B} & \multirow{2}{4em}{8.66} & \multirow{2}{4em}{1.954} & 0.63 & 0.76 & 0.27 & -1.65 & - & - & - & 146.85 & 1.10 & \multirow{2}{4em}{\num{1.79e-4}}\\
&&& 0.31 & 0.97 & 0.29 & -1.98 & \num{4.71e-14} & \num{1.08e-4} & 0.90 & 126.96 & 0.98\\

\hline
\multirow{2}{4cm}{201216C} & \multirow{2}{4em}{48} & \multirow{2}{4em}{1.1} & 8.63 & 0.92 & 1.00 & -2.57 & - & - & - & 241.18 & 0.96 & \multirow{2}{4em}{\num{5.61e-12}}\\
&&& 7.50 & 0.99 & 1.20 & -2.83 & \num{3.73e-14} & \num{5.95e-7} & 0.60 & 185.76 & 0.75\\

\hline
\multirow{2}{4cm}{210619B} & \multirow{2}{4em}{60.9} & \multirow{2}{4em}{1.937} & 61.24 & 0.67 & 0.92 & -1.74 & - & - & - & 773.47 & 1.41 & \multirow{2}{4em}{\num{2.26e-41}}\\
&&& 93.48 & 0.35 & 1.10 & -1.74 & \num{7.54e-14} & \num{5.95e-7} & 0.10 & 581.48 & 1.06\\

\hline
\multirow{2}{4cm}{210731A} & \multirow{2}{4em}{22.51} & \multirow{2}{4em}{1.2525} & 582.22 & 0.24 & 0.96 & -2.85 & - & - & - & 85.09 & 1.49 & \multirow{2}{4em}{\num{4.79e-6}}\\
&&& 77.32 & 0.22 & 1.06 & -2.28 & \num{3.91e-13} & \num{1.09e-6} & 0.10 & 57.66 & 1.07\\

\hline
\multirow{2}{4cm}{220117A} & \multirow{2}{4em}{49.81} & \multirow{2}{4em}{4.961} & 0.70 & 0.65 & -0.10 & -2.74 & - & - & - & 127.93 & 1.08 & \multirow{2}{4em}{\num{2.07e-2}}\\
&&& 0.67 & 0.46 & \num{-9.18e-2} & -2.47 & \num{2.33e-14} & \num{5.95e-7} & 0.70 & 118.16 & 1.03\\

\hline
\hline
\end{tabular}
}
\footnotesize
\hypertarget{A}{\textsuperscript{$a$} ${\rm sec}^{-1}{\rm cm}^{-2}{\rm keV}^{-2}$}
\hypertarget{$E_0$}{\textsuperscript{$b$} ${\rm keV}$}
\caption{The results of the PBH fitting without oscillation, where the goodness of fit decreases by over 0.05, correspond to Fig.~\ref{fig:PBH_femtolensing4}. The first three columns are the data of the GRB. The fourth column lists the four parameters of the BAND model. The fifth column lists the parameters of the PBH. The sixth and seventh columns represent the goodness of fit and $p$-value, respectively, and are selected using the minimum $\chi^{2}$.}
\label{tab:label4}
\end{table}  

\begin{figure}[H]
    \centering
        \includegraphics[width=0.3\linewidth]{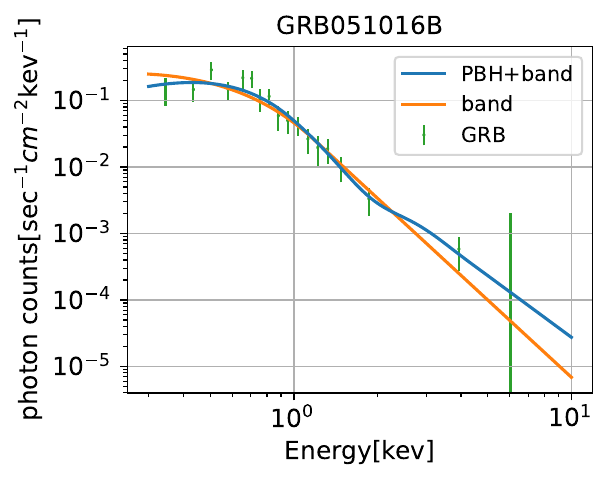}
        \includegraphics[width=0.3\linewidth]{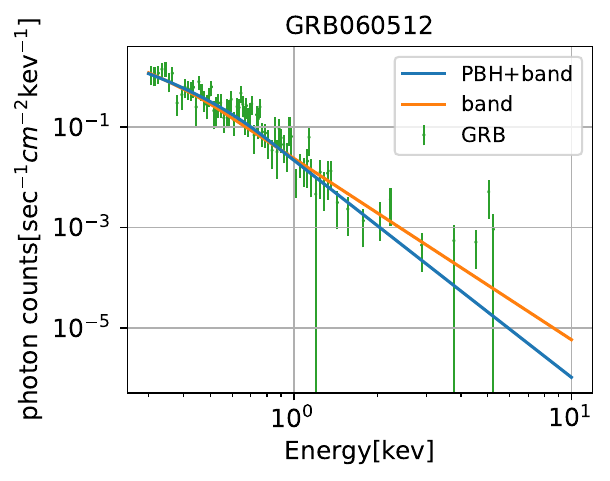}
        \includegraphics[width=0.3\linewidth]{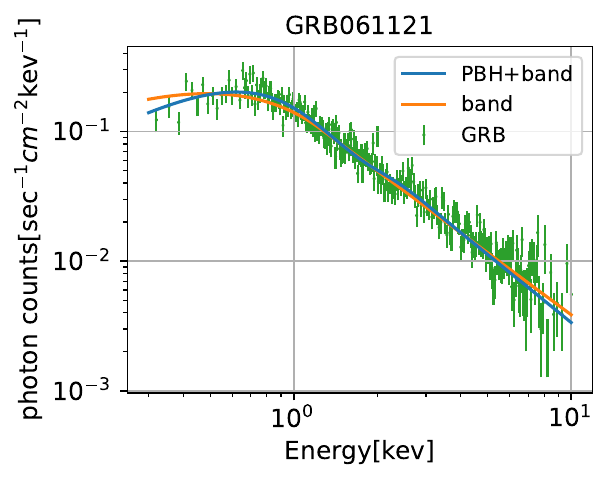}
        \includegraphics[width=0.3\linewidth]{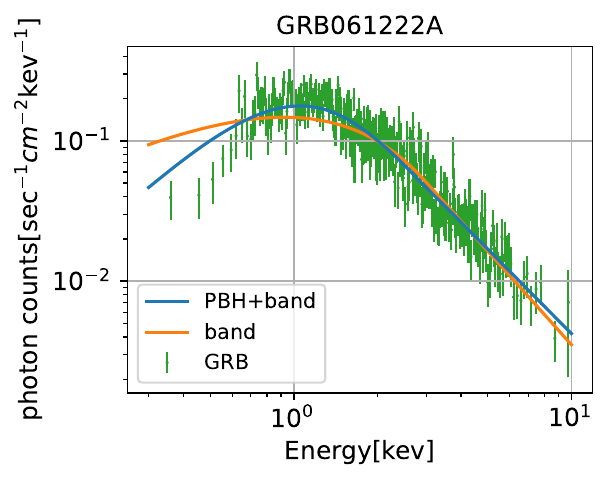}
        \includegraphics[width=0.3\linewidth]{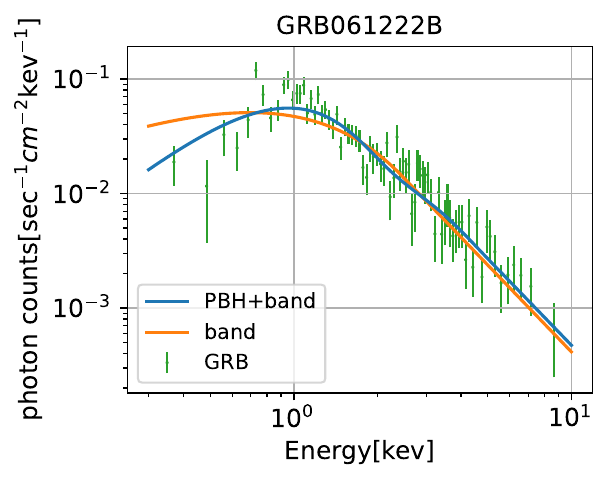}
        \includegraphics[width=0.3\linewidth]{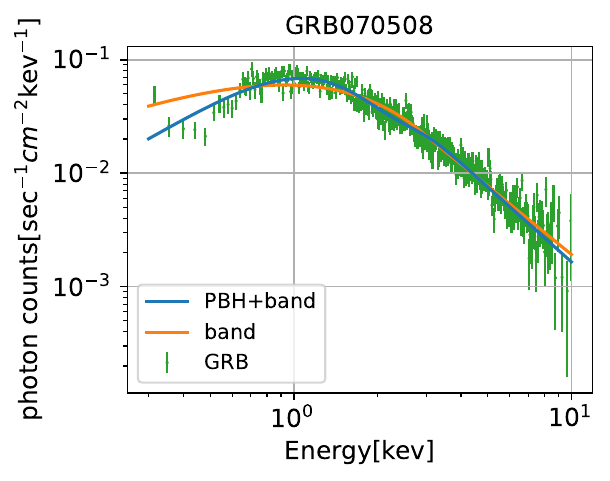}
        \includegraphics[width=0.3\linewidth]{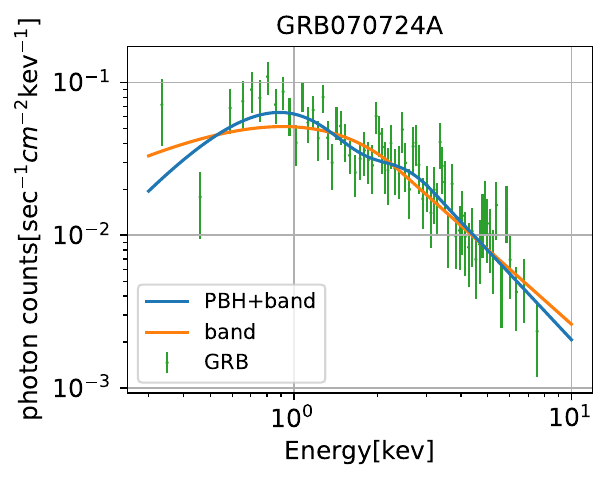}
        \includegraphics[width=0.3\linewidth]{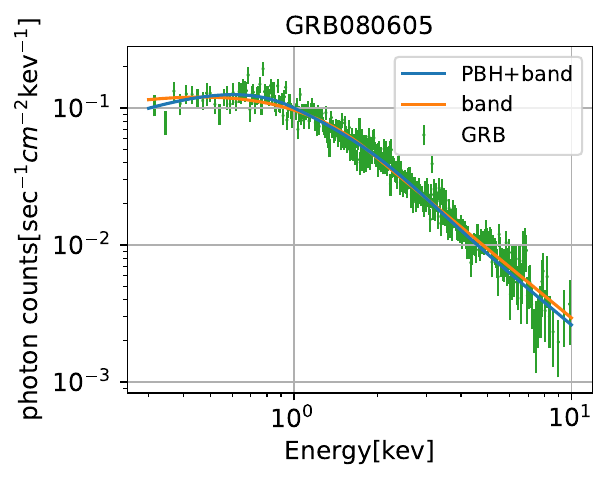}
        \includegraphics[width=0.3\linewidth]{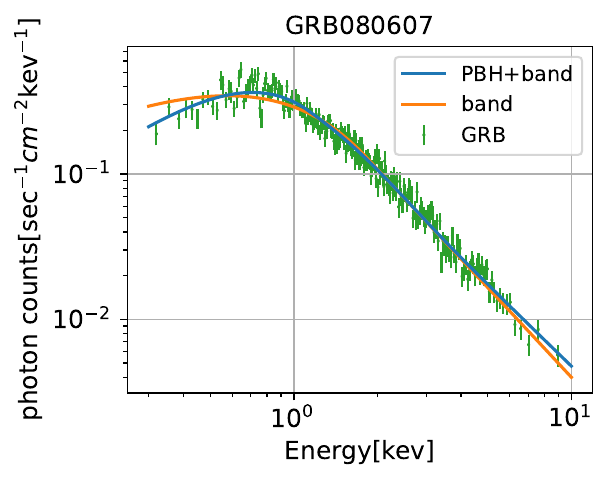}
        \includegraphics[width=0.3\linewidth]{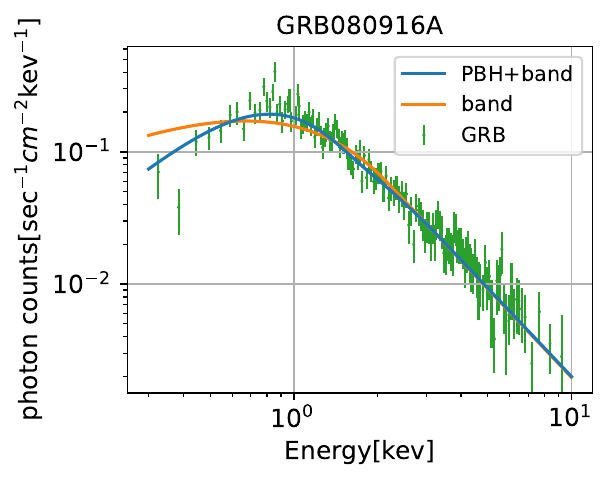}
        \includegraphics[width=0.3\linewidth]{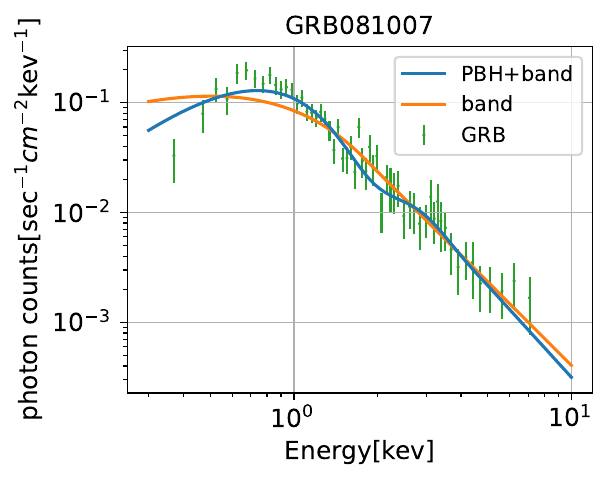}
        \includegraphics[width=0.3\linewidth]{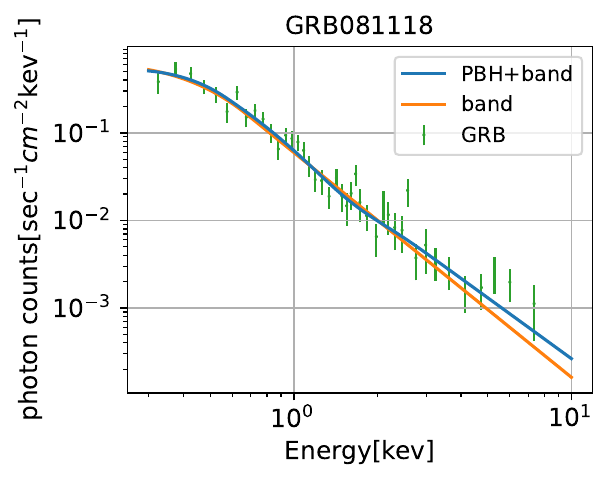}
        \includegraphics[width=0.3\linewidth]{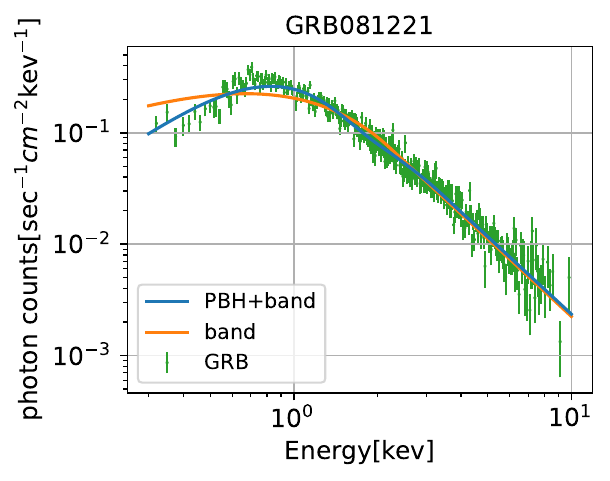}
        \includegraphics[width=0.3\linewidth]{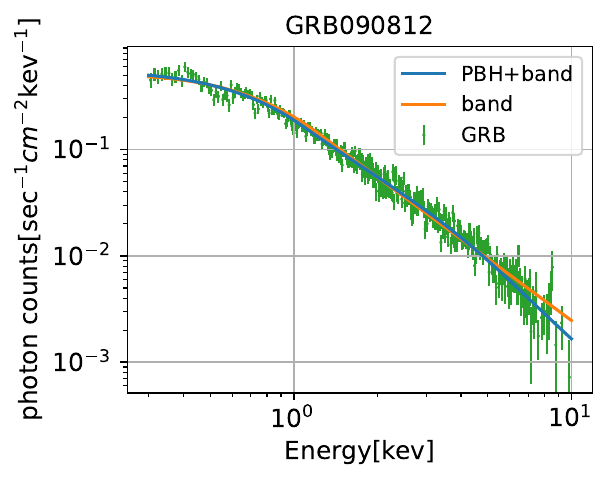}
        \includegraphics[width=0.3\linewidth]{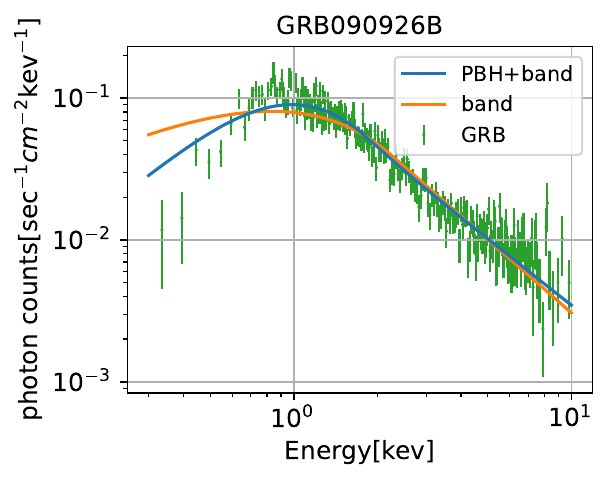}
        \includegraphics[width=0.3\linewidth]{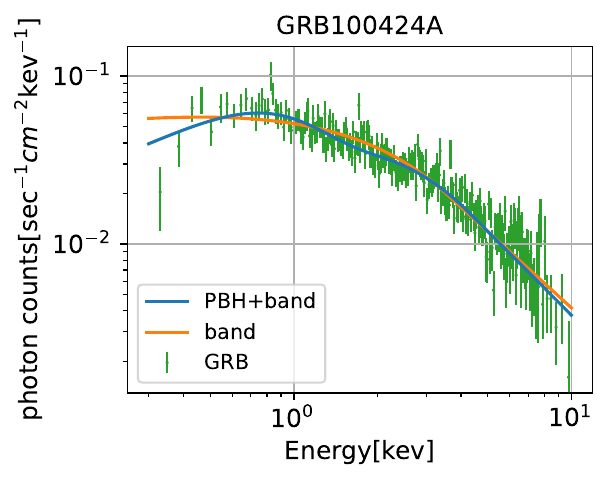}
        \includegraphics[width=0.3\linewidth]{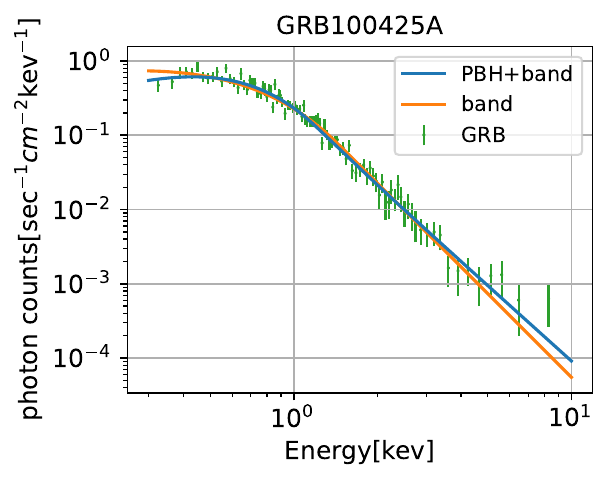}
        \includegraphics[width=0.3\linewidth]{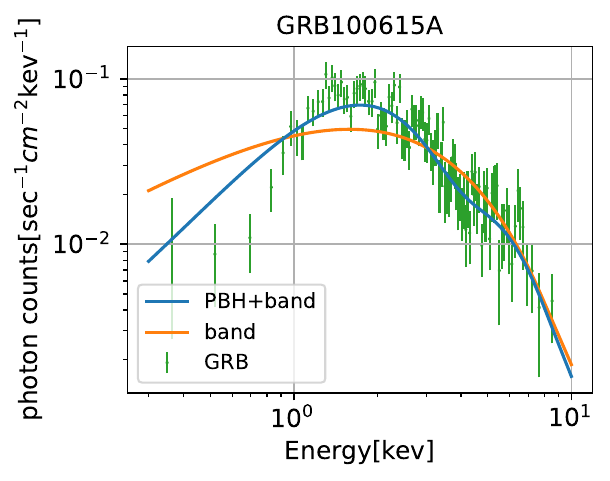}
    \caption{} 
\end{figure}
\begin{figure}[H]
\ContinuedFloat
    \centering
        \includegraphics[width=0.3\linewidth]{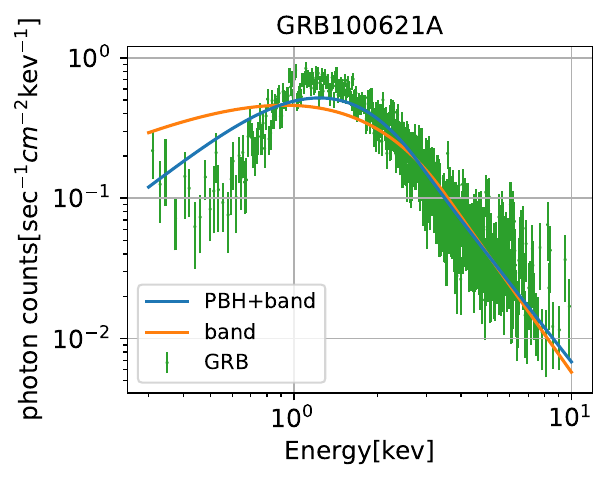}
        \includegraphics[width=0.3\linewidth]{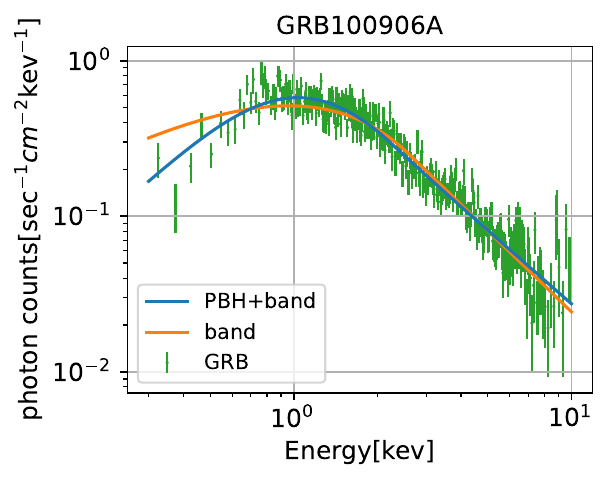}
        \includegraphics[width=0.3\linewidth]{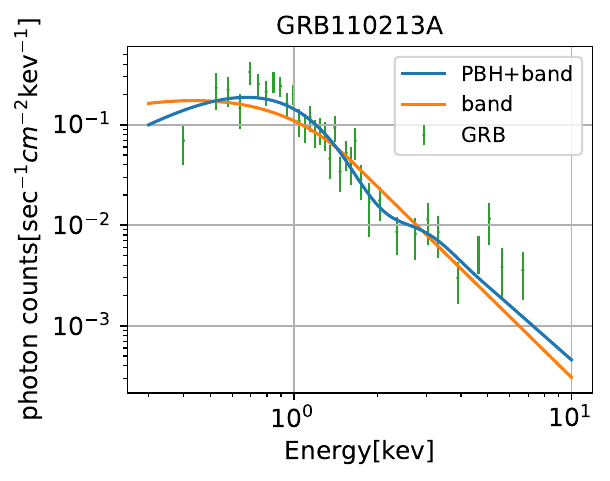} 
        \includegraphics[width=0.3\linewidth]{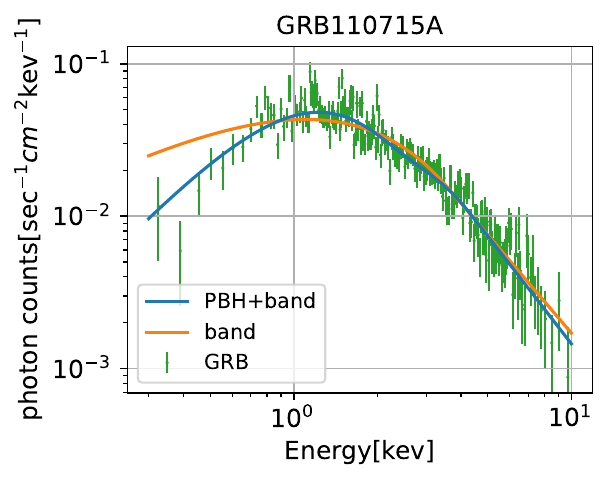}
        \includegraphics[width=0.3\linewidth]{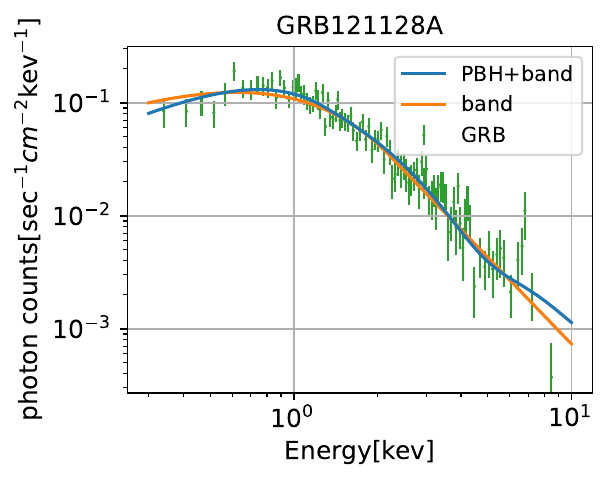}
        \includegraphics[width=0.3\linewidth]{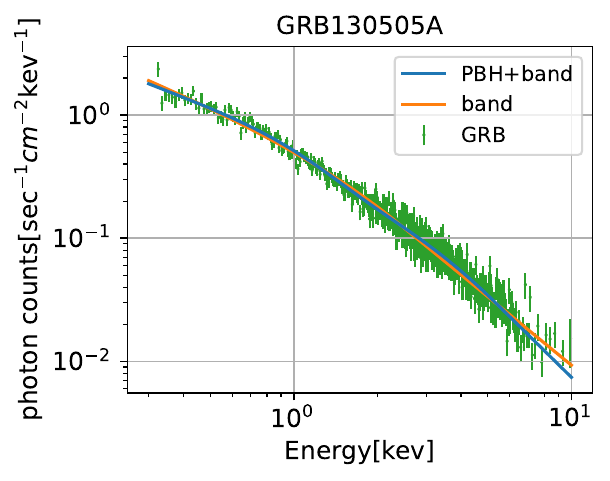}
        \includegraphics[width=0.3\linewidth]{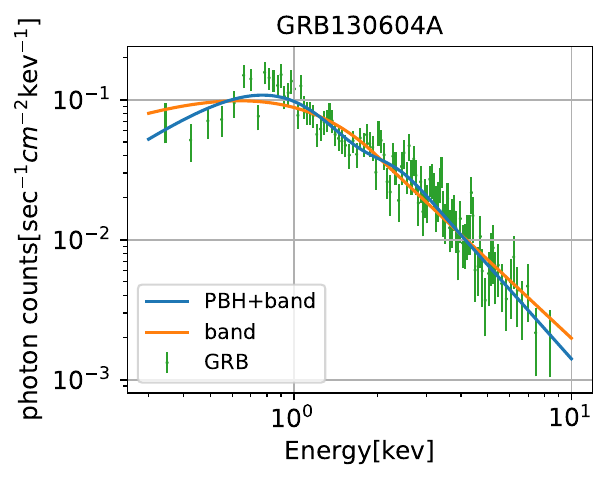}
        \includegraphics[width=0.3\linewidth]{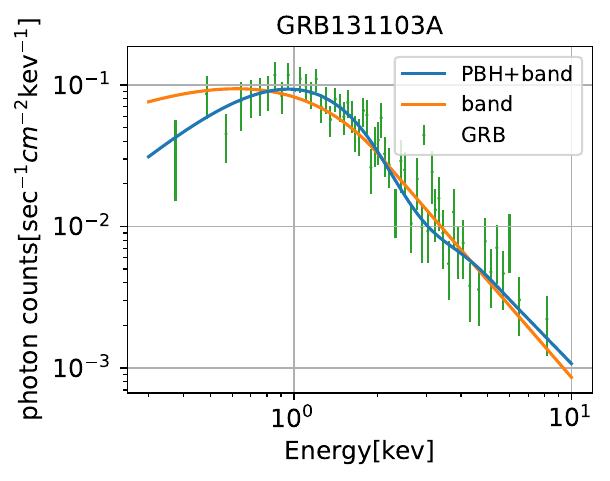}
        \includegraphics[width=0.3\linewidth]{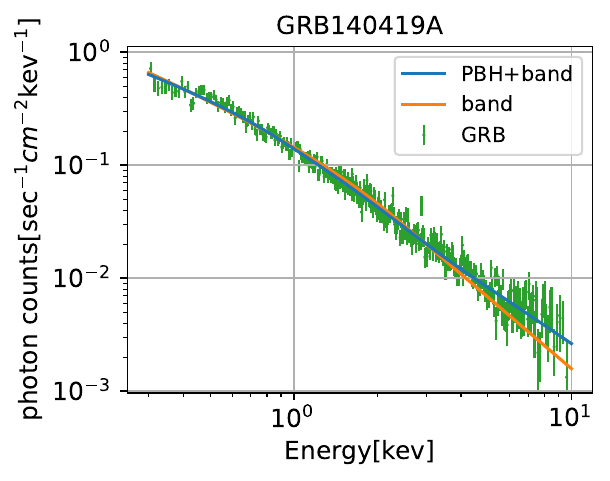}
        \includegraphics[width=0.3\linewidth]{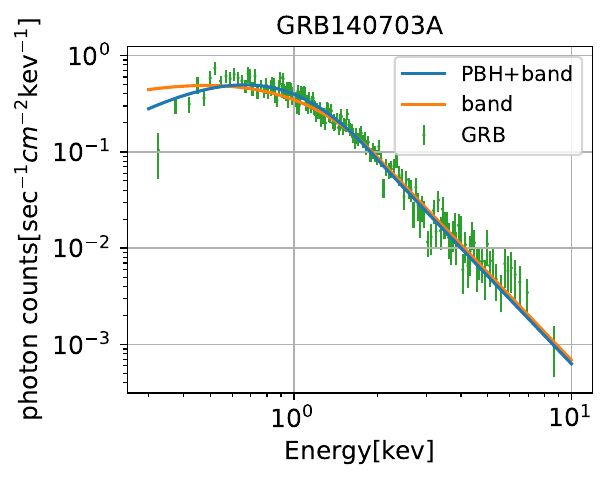}
        \includegraphics[width=0.3\linewidth]{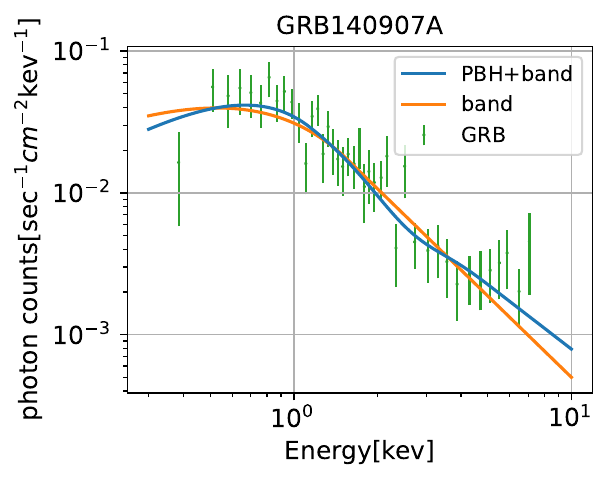}
        \includegraphics[width=0.3\linewidth]{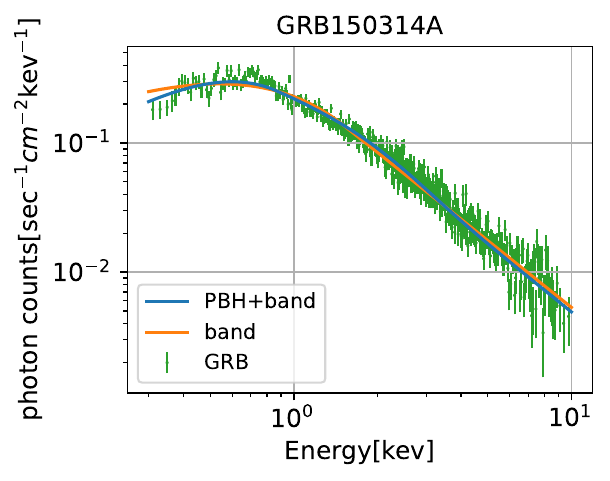}
        \includegraphics[width=0.3\linewidth]{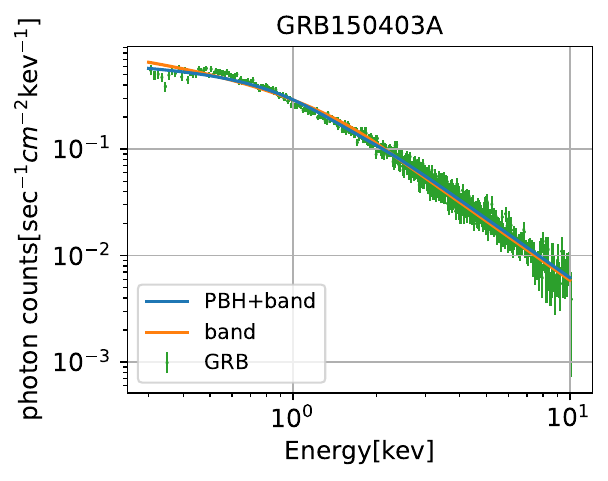}
        \includegraphics[width=0.3\linewidth]{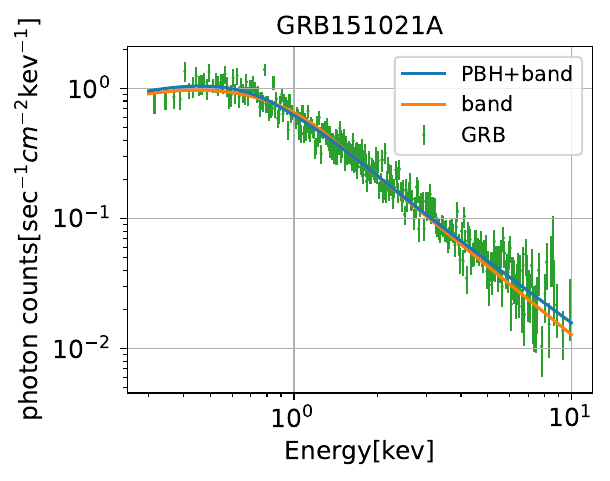}
        \includegraphics[width=0.3\linewidth]{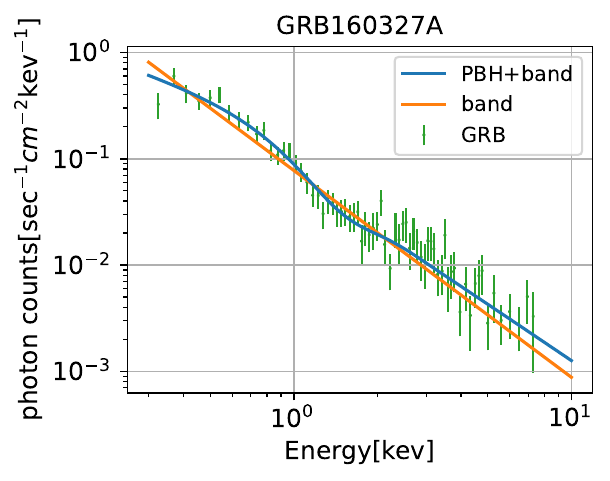}
        \includegraphics[width=0.3\linewidth]{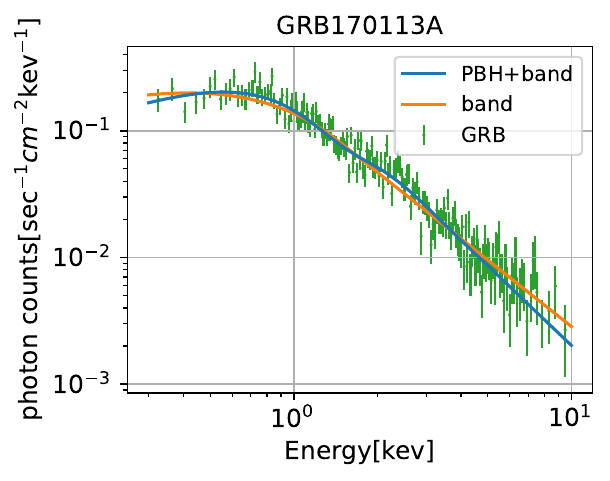}
        \includegraphics[width=0.3\linewidth]{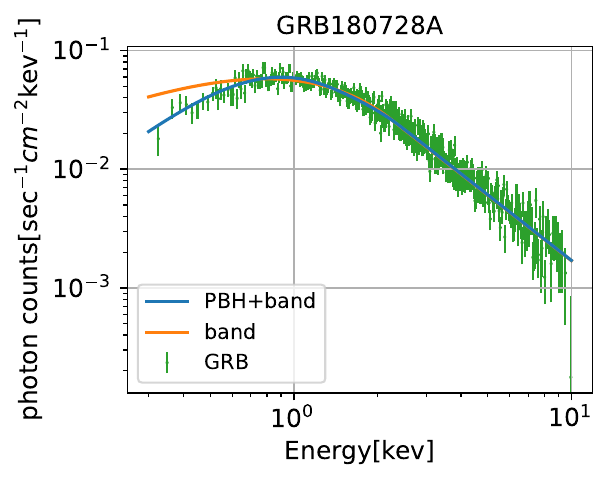}
        \includegraphics[width=0.3\linewidth]{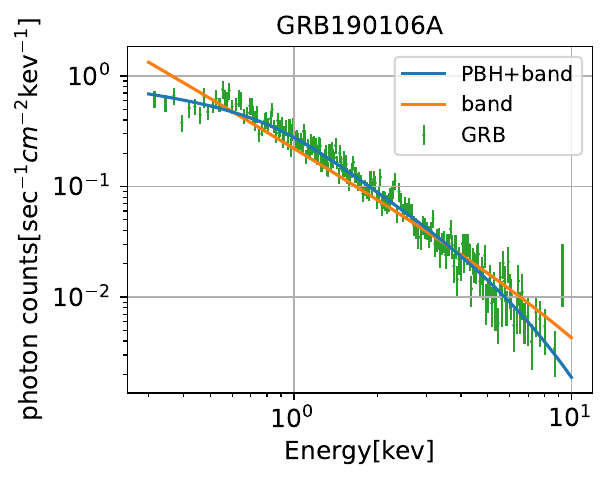}
    \caption{} 
    \label{fig:PBH_femtolensing4}
\end{figure}
\begin{figure}[H]
\ContinuedFloat
    \centering
        \includegraphics[width=0.3\linewidth]{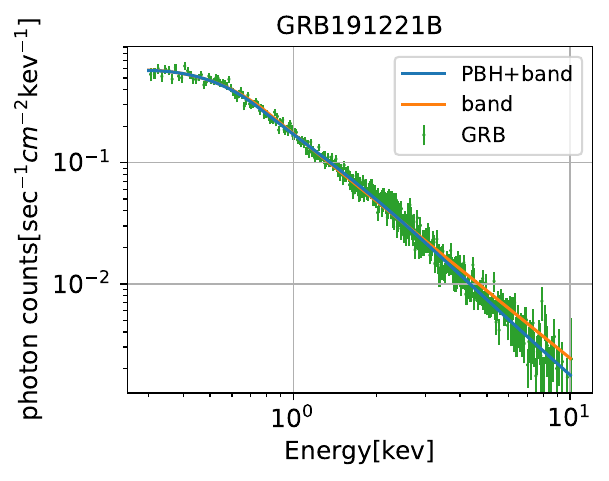}
        \includegraphics[width=0.3\linewidth]{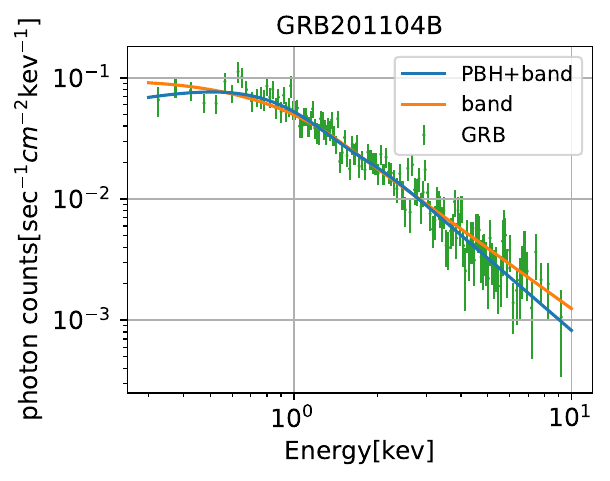}
        \includegraphics[width=0.3\linewidth]{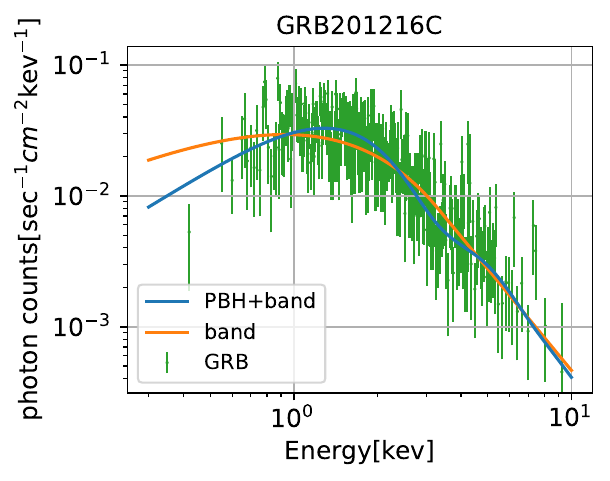}
        \includegraphics[width=0.3\linewidth]{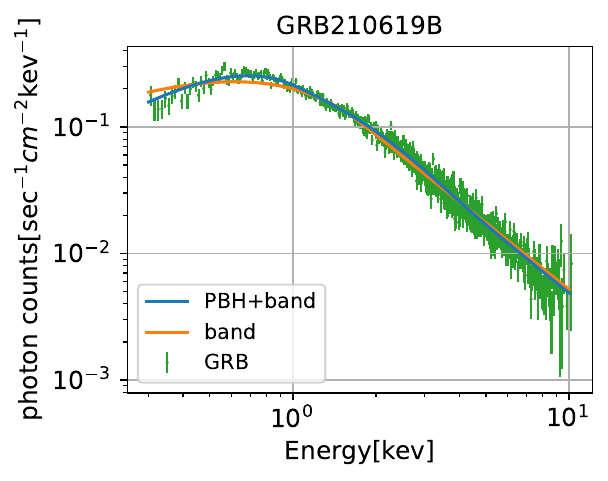}
        \includegraphics[width=0.3\linewidth]{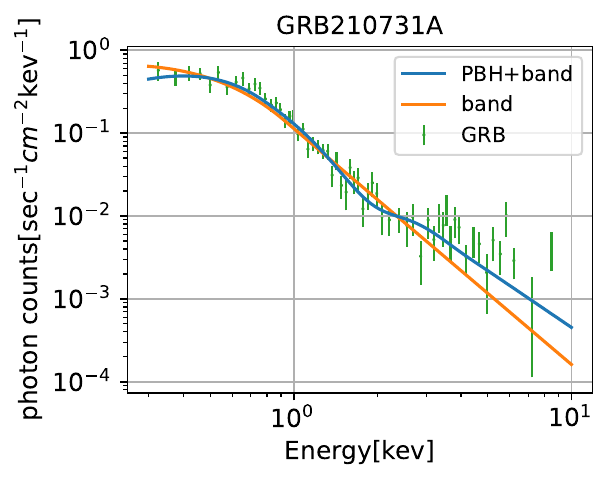}
        \includegraphics[width=0.3\linewidth]{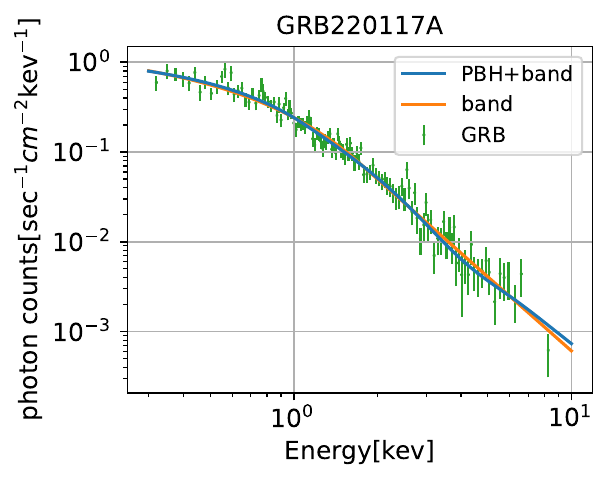}
    \caption{The fitting data without an oscillation pattern, where the goodness of fit decreases by over 0.05, correspond to Table~\ref{tab:label4}.  The green error bars represent the GRB data of Swift XRT. The orange curve represents the BAND model only. The blue curve represents the BAND model with lensing effects taken into account.}
    \label{fig:PBH_femtolensing4}
\end{figure}

\begin{table}[H]
\centering
\resizebox{\textwidth}{!}{
\begin{tabular}{c|c|c|cccc|ccc|cc|cc|c}
\hline
\hline
GRB&$T_{90}$ & $z_{S}$ & A\hyperlink{A}{\textsuperscript{$a$}} & $E_0$\hyperlink{$E_0$}{\textsuperscript{$b$}} & $\alpha_1$ & $\alpha_2$ & $M_{\rm PBH}/M_\odot$ & $z_L$ & $y_0$ & $\chi^2_{\rm BAND/PBH}$ & $\chi^2_{\rm BAND/PBH}/{\rm d.o.f}$ & $P$-value~\cite{Cheung:2018ave} \\

\hline
\multirow{2}{4cm}{050401} & \multirow{2}{4em}{33.3} & \multirow{2}{4em}{2.9} & 9.14 & 0.48 & 0.80 & -1.84 & - & - & - & 382.42 & 1.09 & \multirow{2}{4em}{\num{2.71e-2}}\\
&&& 20.32 & 0.33 & 0.96 & -1.84 & \num{2.95e-14} & \num{1.00e-5} & 0.10 & 373.24 & 1.07\\

\hline
\multirow{2}{4cm}{060607A} & \multirow{2}{4em}{102.2} & \multirow{2}{4em}{3.082} & \num{2.10e-2} & 1.03 & -0.63 & -1.71 & - & - & - & 443.23 & 1.03 & \multirow{2}{4em}{\num{2.76e-2}}\\
&&& \num{4.66e-2} & 0.70 & -0.51 & -1.89 & \num{5.69e-15} & \num{5.95e-7} & 0.50 & 434.10 & 1.02\\

\hline
\multirow{2}{4cm}{061110A} & \multirow{2}{4em}{40.7} & \multirow{2}{4em}{0.758} & 956.50 & 0.28 & 0.96 & -2.82 & - & - & - & 246.40 & 1.12 & \multirow{2}{4em}{\num{8.76e-3}}\\
&&& 756.21 & 0.28 & 0.96 & -2.53 & \num{1.46e-14} & \num{2.57e-6} & 1.80 & 234.77 & 1.08\\

\hline
\multirow{2}{4cm}{080805} & \multirow{2}{4em}{78} & \multirow{2}{4em}{1.505} & 49.40 & 0.60 & 0.92 & -1.21 & - & - & - & 404.45 & 0.94 & \multirow{2}{4em}{0.10}\\
&&& 159.93 & 0.44 & 1.10 & -1.34 & \num{7.20e-15} & \num{1.58e-6} & 0.10 & 398.25 & 0.94\\

\hline
\multirow{2}{4cm}{090418A} & \multirow{2}{4em}{56} & \multirow{2}{4em}{1.608} & 51.67 & 0.36 & 0.92 & -1.73 & - & - & - & 44.40 & 0.73 & \multirow{2}{4em}{0.28}\\
&&& 33.85 & 0.29 & 1.10 & -1.73 & \num{1.53e-13} & \num{2.57e-6} & 0.10 & 40.56 & 0.70\\

\hline
\multirow{2}{4cm}{100302A} & \multirow{2}{4em}{17.9} & \multirow{2}{4em}{4.813} & 0.23 & 0.54 & -0.31 & -2.89 & - & - & - & 165.72 & 0.91 & \multirow{2}{4em}{0.38}\\
&&& \num{6.97e-2} & 0.32 & -0.34 & -2.89 & \num{1.21e-13} & \num{3.56e-6} & 0.10 & 162.65 & 0.91\\

\hline
\multirow{2}{4cm}{110818A} & \multirow{2}{4em}{103} & \multirow{2}{4em}{3.36} & 67.49 & 0.25 & 0.84 & -2.06 & - & - & - & 78.84 & 0.87 & \multirow{2}{4em}{0.31}\\
&&& 22.90 & 0.25 & 0.75 & -1.85 & \num{4.71e-14} & \num{5.95e-7} & 0.90 & 75.29 & 0.86\\

\hline
\multirow{2}{4cm}{121024A} & \multirow{2}{4em}{69} & \multirow{2}{4em}{2.298} & 211.46 & 0.33 & 1.00 & -2.01 & - & - & - & 94.15 & 0.79 & \multirow{2}{4em}{0.16}\\
&&& 684.60 & 0.27 & 1.20 & -2.21 & \num{9.10e-15} & \num{5.95e-7} & 0.80 & 89.03 & 0.77\\

\hline
\multirow{2}{4cm}{130427B} & \multirow{2}{4em}{27} & \multirow{2}{4em}{2.78} & 216.70 & 0.28 & 1.00 & -1.70 & - & - & - & 191.34 & 0.92 & \multirow{2}{4em}{0.36}\\
&&& 399.17 & 0.26 & 1.10 & -2.04 & \num{4.50e-15} & \num{5.95e-7} & 0.10 & 188.10 & 0.91\\

\hline
\multirow{2}{4cm}{140518A} & \multirow{2}{4em}{60.5} & \multirow{2}{4em}{4.707} & \num{5.80e-3} & 1.97 & -0.67 & -1.87 & - & - & - & 197.97 & 0.90 & \multirow{2}{4em}{0.36}\\
&&& \num{2.37e-3} & 1.04 & -0.81 & -1.49 & \num{2.33e-14} & \num{5.95e-7} & 0.80 & 194.79 & 0.89\\

\hline
\multirow{2}{4cm}{151111A} & \multirow{2}{4em}{76.93} & \multirow{2}{4em}{3.5} & 0.27 & 1.12 & \num{2.04e-2} & -1.34 & - & - & - & 180.31 & 0.93 & \multirow{2}{4em}{\num{1.74e-2}}\\
&&& 0.19 & 1.36 & \num{2.24e-2} & -1.34 & \num{2.33e-14} & \num{1.09e-6} & 1.70 & 170.17 & 0.90\\

\hline
\multirow{2}{4cm}{180314A} & \multirow{2}{4em}{51.2} & \multirow{2}{4em}{1.445} & 232.39 & 0.31 & 1.00 & -1.96 & - & - & - & 255.17 & 0.91 & \multirow{2}{4em}{\num{7.01e-3}}\\
&&& 114.84 & 0.25 & 1.20 & -1.96 & \num{2.44e-13} & \num{1.58e-6} & 0.10 & 243.05 & 0.88\\

\hline
\multirow{2}{4cm}{181010A} & \multirow{2}{4em}{16.4} & \multirow{2}{4em}{1.39} & 27.67 & 0.51 & 1.00 & -2.02 & - & - & - & 48.60 & 1.10 & \multirow{2}{4em}{0.21}\\
&&& 4.86 & 0.53 & 1.20 & -2.02 & \num{3.91e-13} & \num{6.04e-6} & 0.10 & 44.06 & 1.07\\

\hline
\hline
\end{tabular}
}
\footnotesize
\hypertarget{A}{\textsuperscript{$a$} ${\rm sec}^{-1}{\rm cm}^{-2}{\rm keV}^{-2}$}
\hypertarget{$E_0$}{\textsuperscript{$b$} ${\rm keV}$}
\caption{The results of the PBH fitting without oscillation, where the goodness of fit decreases by less than 0.05, correspond to Fig.~\ref{fig:PBH_femtolensing5}. The first three columns are the data of the GRB. The fourth column lists the four parameters of the BAND model. The fifth column lists the parameters of the PBH. The sixth and seventh columns represent the goodness of fit and $p$-value, respectively, and are selected using the minimum $\chi^{2}$.}
\label{tab:label5}
\end{table}  

\begin{figure}[H]
    \centering
        \includegraphics[width=0.3\linewidth]{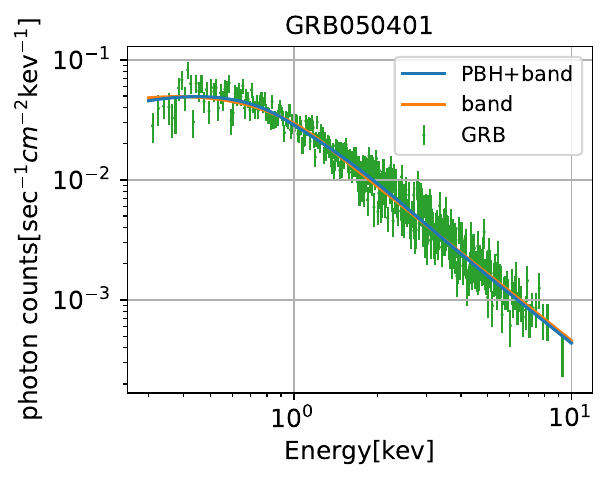}
        \includegraphics[width=0.3\linewidth]{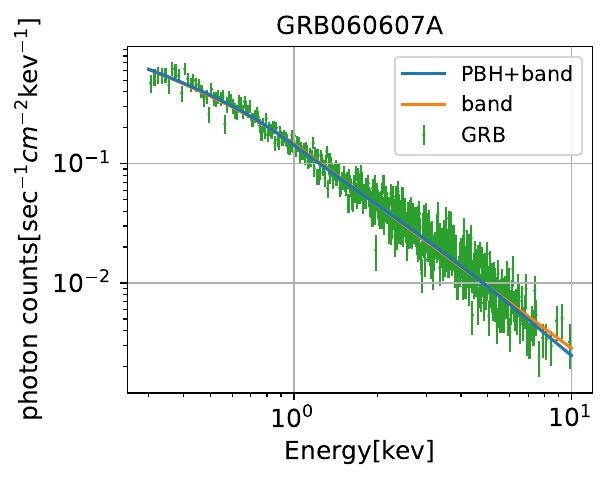}
        \includegraphics[width=0.3\linewidth]{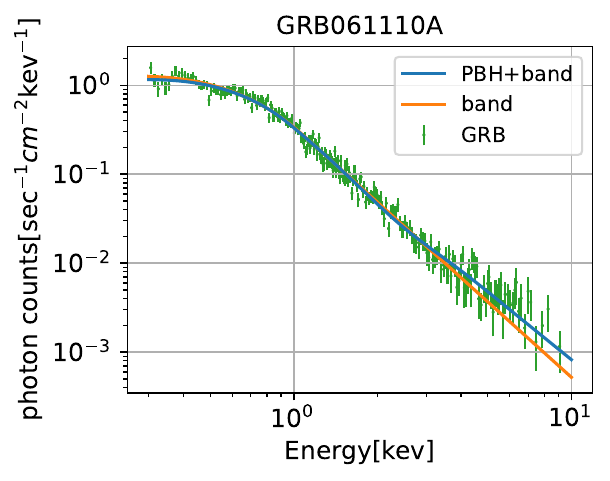}
        \includegraphics[width=0.3\linewidth]{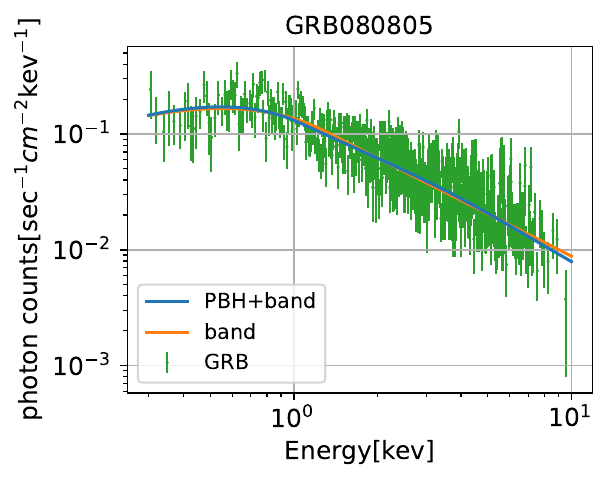}
        \includegraphics[width=0.3\linewidth]{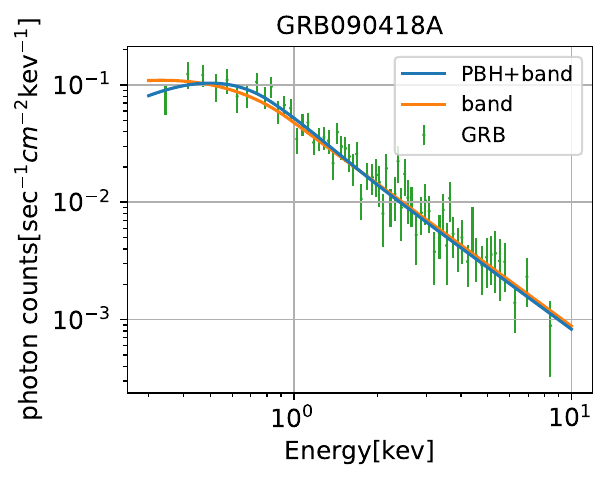}
        \includegraphics[width=0.3\linewidth]{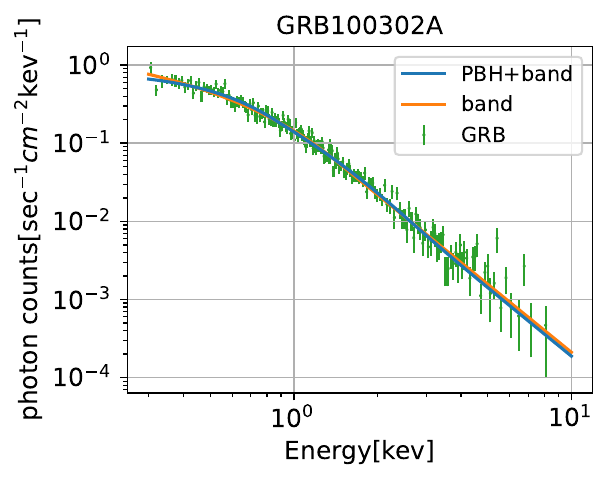}
        \includegraphics[width=0.3\linewidth]{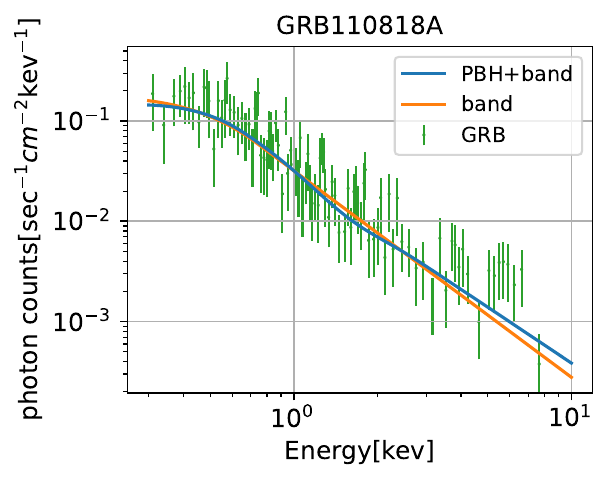}
        \includegraphics[width=0.3\linewidth]{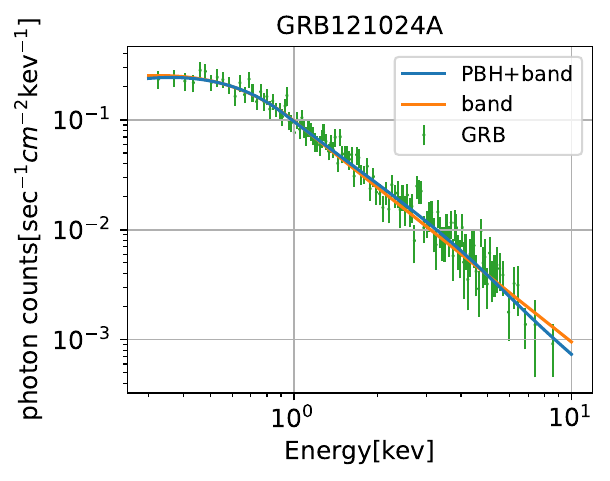}
        \includegraphics[width=0.3\linewidth]{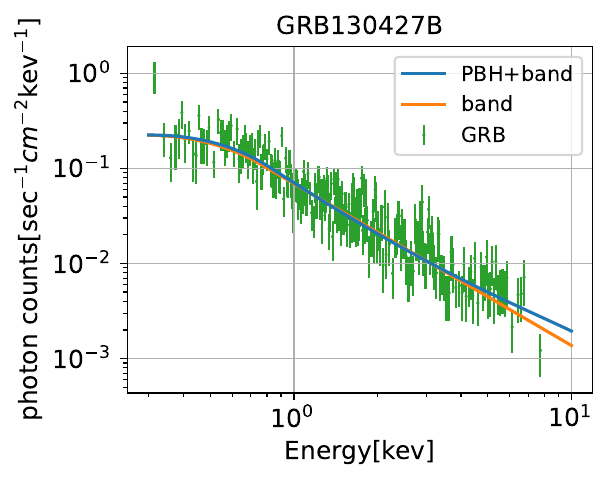}
        \includegraphics[width=0.3\linewidth]{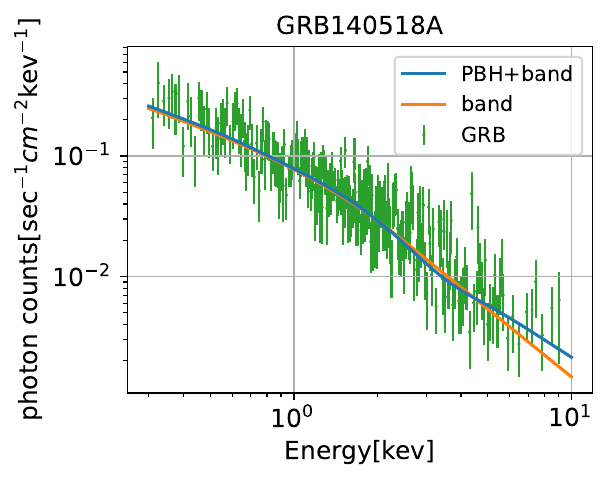}
        \includegraphics[width=0.3\linewidth]{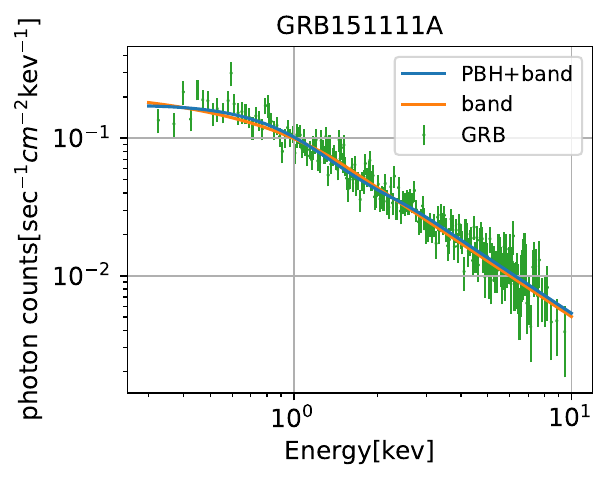}
        \includegraphics[width=0.3\linewidth]{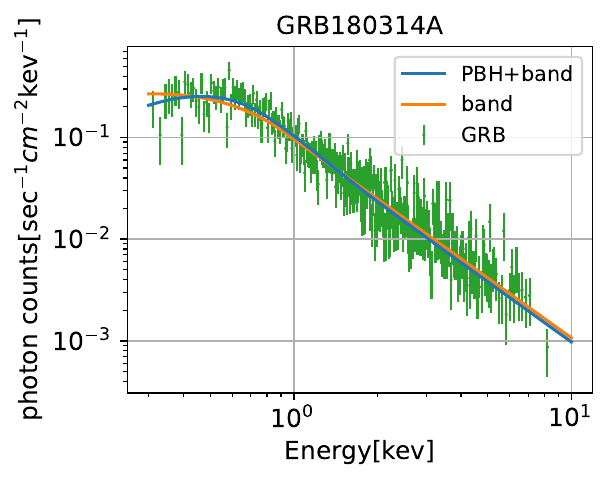}
        \includegraphics[width=0.3\linewidth]{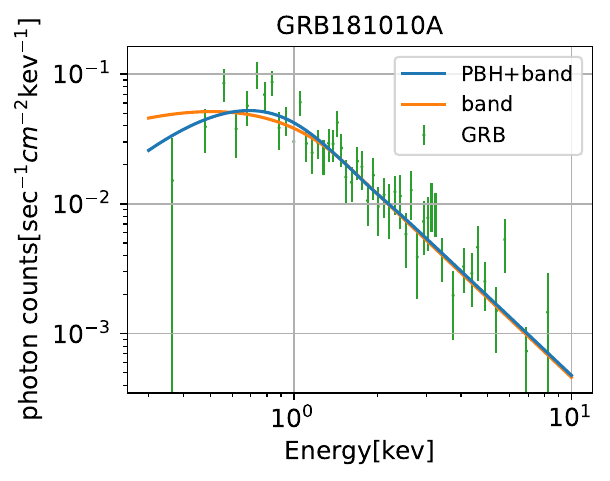}
    \caption{The fitting data without an oscillation pattern, where the goodness of fit decrease by less than 0.05, correspond to Table~\ref{tab:label5}.The green error bars represent the GRB data of Swift XRT. The orange curve represents the BAND model only. The blue curve represents the BAND model with lensing effects taken into account.}
    \label{fig:PBH_femtolensing5}
\end{figure}

\begin{table}[H]
\centering
\resizebox{\textwidth}{!}{
\begin{tabular}{c|c|c|cccc|ccc|cc|cc|c}
\hline
\hline
GRB&$T_{90}$ & $z_{S}$ & A\hyperlink{A}{\textsuperscript{$a$}} & $E_0$\hyperlink{$E_0$}{\textsuperscript{$b$}} & $\alpha_1$ & $\alpha_2$ & $M_{\rm PBH}/M_\odot$ & $z_L$ & $y_0$ & $\chi^2_{\rm BAND/PBH}$ & $\chi^2_{\rm BAND/PBH}/{\rm d.o.f}$ & $P$-value~\cite{Cheung:2018ave} \\

\hline
\multirow{2}{4cm}{050315} & \multirow{2}{4em}{95.6} & \multirow{2}{4em}{1.949} & 1.48 & 0.43 & 0.96 & -2.00 & - & - & - & 33.01 & 0.66 & \multirow{2}{4em}{0.98}\\
&&& 0.67 & 0.44 & 0.86 & -1.80 & \num{1.46e-14} & \num{5.95e-7} & 1.60 & 32.82 & 0.70\\

\hline
\multirow{2}{4cm}{070714B} & \multirow{2}{4em}{64} & \multirow{2}{4em}{0.92} & 0.45 & 1.43 & 0.31 & -1.37 & - & - & - & 112.20 & 0.97 & \multirow{2}{4em}{0.96}\\
&&& 0.63 & 1.00 & 0.37 & -1.64 & \num{7.20e-15} & \num{1.09e-6} & 0.10 & 111.92 & 0.99\\

\hline
\multirow{2}{4cm}{071122} & \multirow{2}{4em}{68.7} & \multirow{2}{4em}{1.14} & 66.78 & 0.33 & 0.92 & -1.91 & - & - & - & 87.17 & 1.36 & \multirow{2}{4em}{0.40}\\
&&& 196.81 & 0.26 & 1.10 & -1.72 & \num{1.46e-14} & \num{1.09e-6} & 1.40 & 84.22 & 1.38\\

\hline
\multirow{2}{4cm}{090510} & \multirow{2}{4em}{0.3} & \multirow{2}{4em}{0.903} & 2.98 & 0.45 & 0.51 & -1.67 & - & - & - & 64.19 & 0.99 & \multirow{2}{4em}{0.61}\\
&&& 2.14 & 0.39 & 0.46 & -2.00 & \num{9.10e-15} & \num{7.75e-3} & 0.70  & 62.37 & 1.01\\

\hline
\multirow{2}{4cm}{120729A} & \multirow{2}{4em}{71.5} & \multirow{2}{4em}{0.8} & 5.89 & 0.83 & 0.76 & -1.29 & - & - & - & 94.06 & 0.78 & \multirow{2}{4em}{0.97}\\
&&& 5.62 & 0.77 & 0.76 & -1.54 & \num{3.56e-15} & \num{5.95e-7} & 0.70 & 93.83 & 0.80\\

\hline
\hline
\end{tabular}
}
\footnotesize
\hypertarget{A}{\textsuperscript{$a$} ${\rm sec}^{-1}{\rm cm}^{-2}{\rm keV}^{-2}$}
\hypertarget{$E_0$}{\textsuperscript{$b$} ${\rm keV}$}
\caption{The results of the PBH fitting without oscillation, where the goodness of fit increases, correspond to Fig.~\ref{fig:PBH_femtolensing6}. The first three columns are the data of the GRB. The fourth column lists the four parameters of the BAND model. The fifth column lists the parameters of the PBH. The sixth and seventh columns represent the goodness of fit and $p$-value, respectively, and are selected using the minimum $\chi^{2}$.}
\label{tab:label6}
\end{table}  

\begin{figure}[H]
    \centering
        \includegraphics[width=0.3\linewidth]{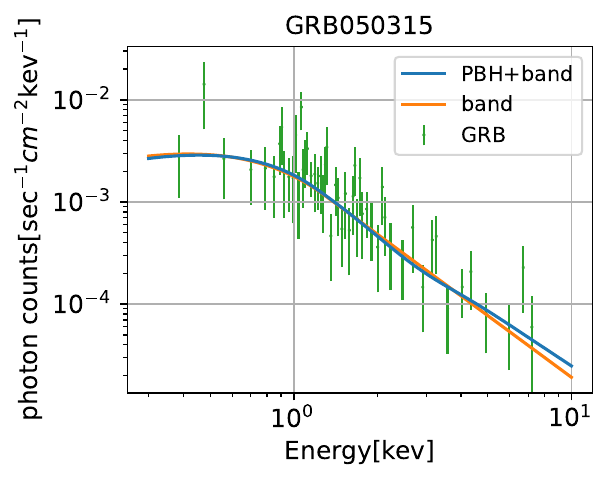}
        \includegraphics[width=0.3\linewidth]{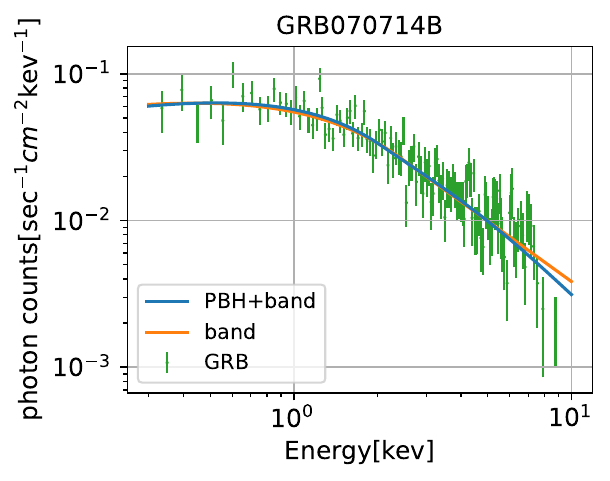}
        \includegraphics[width=0.3\linewidth]{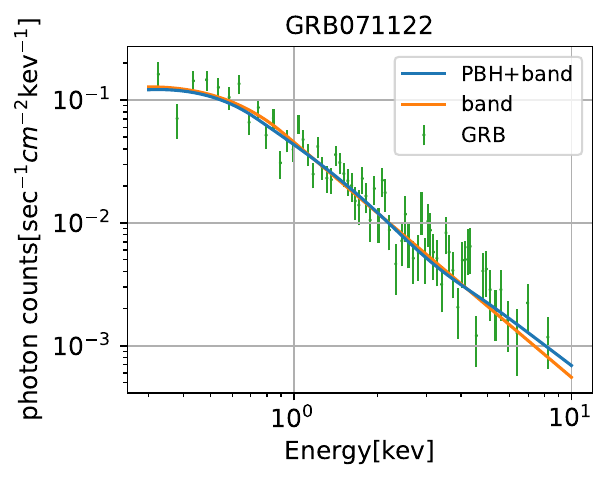}
        \includegraphics[width=0.3\linewidth]{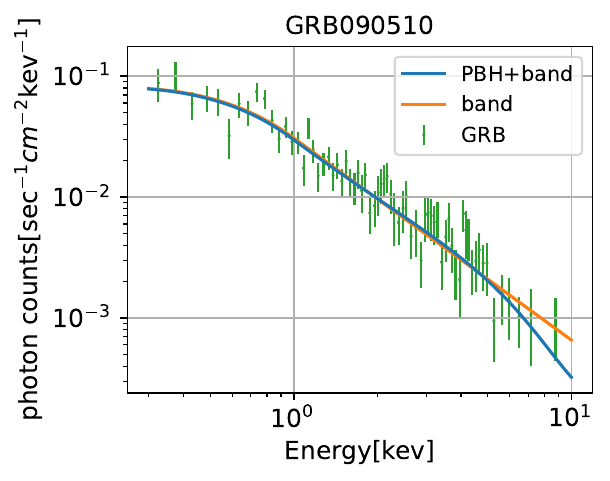}
        \includegraphics[width=0.3\linewidth]{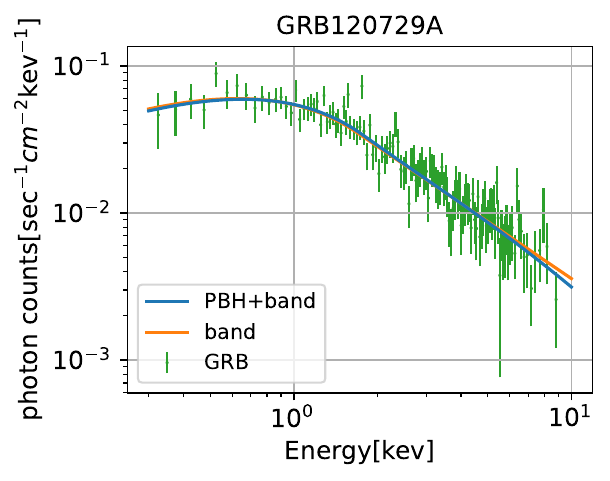}
    \caption{The fitting data without an oscillation pattern, where the goodness of fit increase, correspond to Table~\ref{tab:label6}. The green error bars represent the GRB data of Swift XRT. The orange curve represents the BAND model only. The blue curve represents the BAND model with lensing effects taken into account.}
    \label{fig:PBH_femtolensing6}
\end{figure}

\end{document}